\documentclass[sigconf]{acmart}
\usepackage{graphicx}
\usepackage{subfigure}
\usepackage{paralist}
\usepackage{multirow}
\usepackage{upgreek}
\usepackage{enumitem}
\usepackage{algorithm}
\usepackage{algorithmic}
\usepackage{balance}
\setlist[itemize]{leftmargin=*,topsep=0em}

\usepackage{makecell}

\usepackage{amsmath}

\newtheorem{definition}{Definition}

\DeclareMathOperator*{\argmin}{argmin}
\DeclareMathOperator*{\argmax}{argmax}

\AtBeginDocument{%
  }

\copyrightyear{2024}
\acmYear{2024}
\setcopyright{acmlicensed}\acmConference[SIGIR '24]{Proceedings of the 47th International ACM SIGIR Conference on Research and Development in Information Retrieval}{July 14--18, 2024}{Washington, DC, USA}
\acmBooktitle{Proceedings of the 47th International ACM SIGIR Conference on Research and Development in Information Retrieval (SIGIR '24), July 14--18, 2024, Washington, DC, USA}
\acmDOI{10.1145/3626772.3657709}
\acmISBN{979-8-4007-0431-4/24/07}

\begin{document}

\title{Adaptive Fair Representation Learning for Personalized Fairness in Recommendations via Information Alignment}

%
%


\author{Xinyu Zhu}
\authornote{Xinyu Zhu and Lilin Zhang share the first authorship.}
\affiliation{%
\department{School of Computer Science}
\institution{Sichuan University}
\city{Chengdu}
 \country{China}
}
\email{zhuxinyu@stu.scu.edu.cn}

\author{Lilin Zhang}
\authornotemark[1]
\affiliation{%
\department{School of Computer Science}
\institution{Sichuan University}
\city{Chengdu}
 \country{China}
}
\email{zhanglilin@stu.scu.edu.cn}

\author{Ning Yang}
\authornote{Ning Yang is the corresponding author.}
\affiliation{%
\department{School of Computer Science}
\institution{Sichuan University}
\city{Chengdu}
 \country{China}
}
\email{yangning@scu.edu.cn}

\renewcommand{\shortauthors}{Xinyu Zhu, Lilin Zhang, \& Ning Yang}

\begin{abstract}
Personalized fairness in recommendations has been attracting increasing attention from researchers. The existing works often treat a fairness requirement, represented as a collection of sensitive attributes, as a hyper-parameter, and pursue extreme fairness by completely removing information of sensitive attributes from the learned fair embedding, which suffer from two challenges: huge training cost incurred by the explosion of attribute combinations, and the suboptimal trade-off between fairness and accuracy. In this paper, we propose a novel Adaptive Fair Representation Learning (AFRL) model, which achieves a real personalized fairness due to its advantage of training only one model to adaptively serve different fairness requirements during inference phase. Particularly,  AFRL treats fairness requirements as inputs and can learn an attribute-specific embedding for each attribute from the unfair user embedding, which endows AFRL with the adaptability during inference phase to determine the non-sensitive attributes under the guidance of the user's unique fairness requirement. To achieve a better trade-off between fairness and accuracy in recommendations, AFRL conducts a novel Information Alignment to exactly preserve discriminative information of non-sensitive attributes and incorporate a debiased collaborative embedding into the fair embedding to capture attribute-independent collaborative signals, without loss of fairness. Finally, the extensive experiments conducted on real datasets together with the sound theoretical analysis demonstrate the superiority of AFRL. The codes and datasets are available on https://github.com/zhuxinyu2700/AFRL.
 \end{abstract}



\maketitle

\section{Introduction}
In the era of exponential information growth, recommender systems play a critical role in alleviating the problem of information overloading \cite{he2017neural, mnih2007probabilistic, kang2018self}. Despite the great success of recommender systems, existing studies have shown that machine learning based recommendation models often suffer from unfair recommendations due to the bias in training data \cite{kleinberg2016inherent, pessach2022review, dwork2012fairness, xiao2017fairness}. In recent years, many efforts have been made for fair recommendations from various perspectives \cite{beutel2019fairness, fu2020fairness, ZhuZiwei2021FaNI,ge2021towards}, among which user-side fairness attracts special attention due to its significance to the improvement of user experience.

A range of approaches have been proposed for user-side fair recommendations, following various technical lines including Pareto optimization \cite{lin2019pareto}, adversarial training \cite{li2021towards,wu2021fairness,wu2022selective}, and disentangled representation learning \cite{ zhao2023fair, park2021learning, gong2020jointly, sarhan2020fairness}. The common idea of the existing works is to filter out the information of user sensitive attributes from the learned fair user embeddings, so that recommendations can be made independently of these sensitive attributes. However, the traditional approaches often assume that all users share the common sensitive attributes and consider the user fairness requirement, represented by a collection of user sensitive attributes, as hyper-parameter specified in advance, which limits the model to serving only one fairness requirement. In real world, however, it is the fact that different users are sensitive to different attributes and have \textit{personalized fairness} requirements \cite{li2021towards}. For example, some users may want the recommendations to be made for them without consideration of their racial information, while some others want the recommendations to be made without gender bias. 

Recently, a few methods have been proposed for personalized fairness in recommendations. For example, to meet diverse fairness requirements, Li \textit{et al}. \cite{li2021towards} propose to train a filter for each possible combination of sensitive attributes. Wu \textit{et al}. \cite{wu2022selective}  propose a model PFRec, which builds a set of prompt-based bias eliminators and adapters with customized attribute-specific prompts to learn fair embeddings for different attribute combinations. Creager \textit{et al}. \cite{creager2019flexibly} propose FFVAE, a disentangled representation learning model that separates representations into sensitive and non-sensitive subspaces. It addresses personalized fairness by excluding from the learned fair embeddings the relevant semantic factors corresponding to different sensitive attributes specified by user fairness requirements. However, the problem of personalized fairness in recommendations is still far from being well solved partly due to the following challenges:

(1) \textbf{Unacceptable training cost caused by exponential combinations of attributes} Essentially, most existing methods for personalized fairness still treat a fairness requirement as a hyper-parameter, which makes them have to train and store a model for each combination of attributes in a brute force way \cite{li2021towards,wu2022selective}. Such inflexibility incurs a huge training cost due to the explosion of attribute combinations, which leads to the impracticability of the existing methods.

(2) \textbf{Compromise on recommendation accuracy} It is well-recognized by previous works that gains in fairness necessarily come at the cost of losses in accuracy \cite{gupta2021controllable, kearns2019empirical}. However, in order to completely erase the sensitive attribute information from the fair embeddings, the existing approaches also remove information that overlaps with non-sensitive attributes \cite{li2021towards,wu2022selective,creager2019flexibly}. This inevitably causes the loss of the discriminative information of the non-sensitive attributes and results in a suboptimal trade-off between accuracy and fairness. Therefore, it is worth exploring better solutions to achieve personalized fairness with less compromise on recommendation accuracy.

In this paper, to address the above challenges, we propose a novel fair model called Adaptive Fair Representation Learning (AFRL) for personalized fairness recommendation. Particularly, to overcome the challenge of the unacceptable training cost incurred by the explosion of attribute combinations, AFRL treats a user fairness requirement as input, rather than hyper-parameter, by which AFRL can adaptively generate the fair embeddings for different users by fusing the corresponding embeddings of non-sensitive attributes specified by the fairness requirement. The generation of fair embeddings is dynamically controlled by the fairness requirements, which endows AFRL with the adaptability during the inference phase to the exponential combinations of attributes. 

To strike a better trade-off between fairness and accuracy, we propose an Information Alignment Module (IAlignM) for AFRL. This module is designed to enable the learned fair embeddings to retain discriminative information from non-sensitive attributes and unbiased collaborative signals from the user's interaction history. In contrast to existing approaches, where removing sensitive attributes results in the loss of discriminative information from non-sensitive attributes, IAlignM adopts a novel strategy. It learns attribute-specific embeddings from the original user embeddings for each attribute, ensuring the precise alignment of information between the embeddings and their respective attributes. Here \textit{information alignment} refers to preserving and only preserving the discriminative information related to the attributes within the corresponding attribute-specific embeddings. For this purpose, we employ a bilevel optimization approach to ensure that an attribute-specific embedding exactly encodes the information of its corresponding attribute, with a theoretical guarantee rooted in information bottleneck principle. At the same time, IAlignM will further improve the recommendation accuracy by learning a debiased collaborative embedding to explicitly capture into the fair user embeddings the collaborative signals where bias to any user attribute is filtered out to avoid the loss of fairness.

It is worth mentioning that AFRL is recommendation model-agnostic, which means it can serve as a plugin for an off-the-shelf recommendation model. In this paper, we will test our AFRL for non-sequential and sequential recommendation tasks. In summary, our contributions can be outlined as follows:
\begin{itemize}
  \item We propose a novel fair model called Adaptive Fair Representation Learning (AFRL) which achieves a real personalized fairness in recommendations due to its advantage of training only one model to adaptively serve different fairness requirements during inference phase. 
  
\item To achieve a better trade-off between fairness and accuracy in recommendations, we propose an Information Alignment Module (IAlignM). IAlignM reduces the loss of recommendation accuracy by exactly preserving discriminative information of non-sensitive attributes and incorporating a debiased collaborative embedding into the fair embedding to capture attribute-independent collaborative signals.
   
  \item At last, the extensive experiments conducted on real datasets together with the theoretical analysis demonstrate the superiority of AFRL. 
 \end{itemize}
 
  \begin{figure*}[t]
    \centering
      \includegraphics[width=0.8\linewidth]{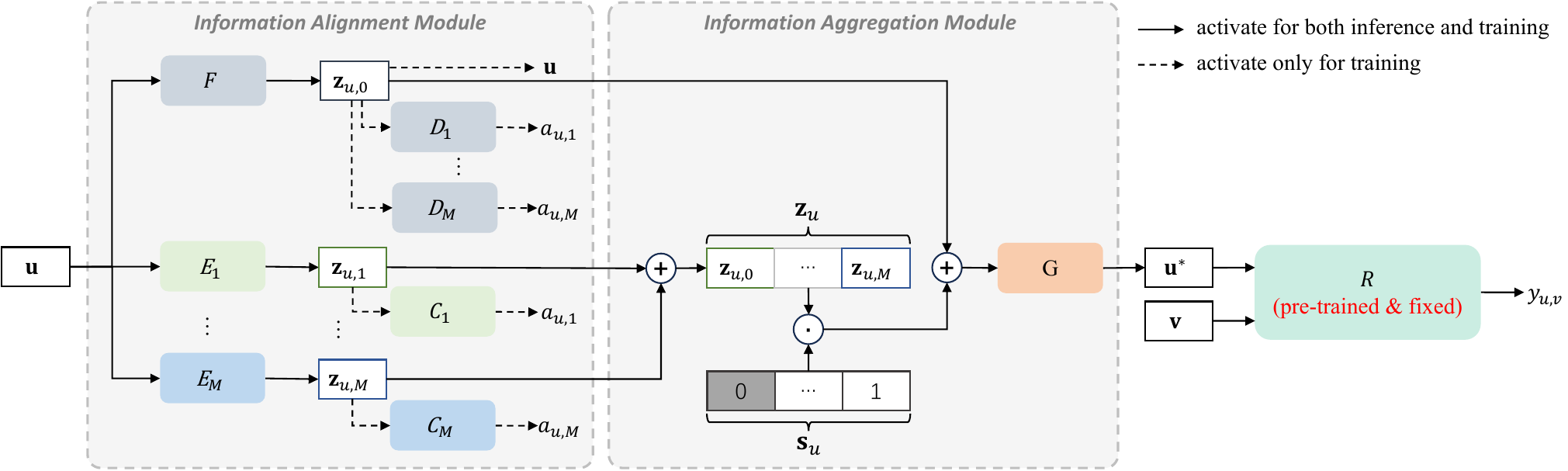}
    \caption{ The overview of AFRL. AFRL contains Information Alignment Module (IAlignM) and Information Aggregation Module (IAggM). IAlignM generates $M$ attribute-specific embeddings $\{\mathbf{z}_{u,i}\}$ ($1 \le i \le M$) and a debiased collaborative embedding $\mathbf{z}_{u,0}$ from the user embedding $\mathbf{u}$, using attribute encoders $\{E_{i}\}$ ($1 \le i \le M$) and a debiased collaborative signal encoder $F$, respectively. IAggM aggregates these embeddings with respect to the personalized fairness requirement $\mathbf{s}_u$ to produce the final fair user embedding $\mathbf{u}^{*}$ for downstream recommendation.}
      \label{Fig:model}
  \end{figure*}

\section{Preliminaries}
\subsection{Basic Notations and Definitions}
We denote user set by $\mathcal{U}$ and item set by $\mathcal{V}$. Let $\mathcal{O} \in \{ 0, 1\}^{|\mathcal{U}| \times |\mathcal{V}|}$ be the interaction matrix, where a cell $o_{u,v} = 1$ indicates the observed interaction of user $u \in \mathcal{U}$ with item $v \in \mathcal{V}$, and the $u$th row $\mathcal{O}_{u,*}$ represents the historical user-item interaction set of user $u$. Let $\mathcal{A}$ be the user attribute matrix of $|\mathcal{U}| \times M$, where $M$ is the number of the attributes of a user, and the cell at $u$th row and $i$th column, $a_{u, i}$, is the $i$th attribute of user $u$.

\subsection{Formulation of the Target Problem}
Let $y_{u, v} = R(\mathbf{u}, \mathbf{v}) $ be a given recommendation model, which infers the probability $y_{u, v}$ that user $u \in \mathcal{U} $ will interact with item $v \in \mathcal{V}$ based on the user embedding $\mathbf{u} \in \mathbb{R}^{d \times 1}$ and the item embedding $\mathbf{v} \in \mathbb{R}^{d \times 1} $, where $d$ is the embedding dimensionality. Let $\mathbf{s}_u \in \{ 0, 1\}^{Md \times 1}$ representing the fairness requirement of user $u$, where the $i$th piece $\mathbf{s}_{u,[1+(i-1)d : i\times d]} = 0$ means the $i$th attribute is a sensitive attribute of user $u$, otherwise a non-sensitive attribute. The target problem can be formulated as follow:

Given an unfair recommendation model $R$, the original user embeddings $ \{ \mathbf{u} \} $ and item embeddings $ \{ \mathbf{v} \} $ generated by $R$, the historical interaction matrix $\mathcal{O}$, and the user attribute matrix $\mathcal{A}$, we want to train a fair model $FairR$ to generate the fair user embedding $\mathbf{u}^{*}$ for user $u$ based on $u$'s original embedding $\mathbf{u}$ and fairness requirement $ \mathbf{s}_u$, i.e., $\mathbf{u}^{*} = FairR(\mathbf{u}, \mathbf{s}_u)$.

 \subsection{Counterfactual Fairness}
In this paper, the recommendations made based on $\mathbf{u}^{*}$ are supposed to be independent
of the user’s sensitive attributes specified by $\mathbf{s}_u$, which is a kind of counterfactual fairness \cite{kusner2017counterfactual}. Intuitively, counterfactual fairness requires the recommendations made in the imaginary counterfactual world, where only the user's sensitive attributes are modified, should be the same as in the real world. The formal definition of counterfactual fairness in recommender systems is provided as follow:
 \begin{definition}[Counterfactually fair recommendation \cite{li2021towards}]
 	A recommender model is counterfactually fair if for any possible user with non-sensitive attributes $A_n=a_n$ and sensitive attributes $A_s=a_s$:
\begin{displaymath}
  P\left(Y_{a_{s}}\middle| A_n=a_n,A_s=a_s\right)=P\left(Y_{a_s^\prime}|A_n=a_n,A_s=a_s\right)
\end{displaymath}
for all $Y$ and for any value $a_s$ attainable by $A_s$, where $Y$ denotes the generated recommendation results. 
\label{Def:fairness}
\end{definition}


\section{The Proposed Model} 
\subsection{Overview}
Fig. \ref{Fig:model} gives an overview of the architecture of AFRL. As shown in Fig. \ref{Fig:model}, AFRL takes as inputs the original user embedding $\mathbf{u}$ and the personalized fairness requirement $\mathbf{s}_u$, and produces the fair embedding $\mathbf{u}^{*}$. We can see that AFRL consists of two modules, the Information Alignment Module (IAlignM) and the Information Aggregation Module (IAggM). IAlignM is responsible for learning $M$ attribute-specific embeddings $\{ \mathbf{z}_{u, i} \}$ and one debiased collaborative embedding $\mathbf{z}_{u, 0}$ from the original user embedding $\mathbf{u}$ via $M$ \textbf{attribute encoders} $\{E_{i}\}$ ($1 \le i \le M$) and a \textbf{debiased collaborative signal encoder} $F$, respectively. In particular, to ensure the \textbf{information alignment} between $\mathbf{z}_{u, i}$ and the $i$th attribute $a_{u, i}$ of user $u$, IAlignM will maximize the mutual information between $\mathbf{z}_{u, i}$ and $a_{u, i}$, and minimize the mutual information between $\mathbf{z}_{u, i}$ and $\mathbf{u}$, using a bilevel optimization for attribute encoder $E_i$ together with the auxiliary attribute classifier $C_i$. At the same time, to generate the debiased collaborative embedding $\mathbf{z}_{u, 0}$, IAlignM will ensure $F$ to remove the correlation between $\mathbf{z}_{u, 0}$ and the user attributes via an adversarial training together with $M$ attribute discriminators $\{D_i\}$ ($1 \le i \le M$), and encode the collaborative signals by maximizing the ability of $\mathbf{z}_{u, 0}$ to reconstruct the original user embedding $\mathbf{u}$.

Once the attribute-specific embeddings and the debiased collaborative embedding are produced, IAggM will mask the embeddings of sensitive attributes with respect to $\mathbf{s}_u$ and fuse the embeddings of non-sensitive attributes with the debiased collaborative embedding using an MLP $G$ to generate the fair embedding $\mathbf{u}^{*}$. To make $\mathbf{u}^{*}$ legal for the given recommendation model $R$, the training of IAggM will be supervised by the historical interactions $\mathcal{O}$ through $R$, which locates $\mathbf{u}^{*}$ in the same embedding space as the original user embedding $\mathbf{u}$. Finally, $\mathbf{u}^*$ and the item embedding $\mathbf{v}$ are fed to $R$ to make a fair prediction of $y_{u,v}$.

\subsection{Information Alignment Module}
\label{Sec:IAlignM}
As mentioned before, IAlignM learns attribute-specific embeddings $\{ \mathbf{z}_{u, 1}, $ $\cdots, \mathbf{z}_{u, M}  \}$ and debiased collaborative embedding $\mathbf{z}_{u, 0}$ from the original user embedding $\mathbf{u}$. Since $\mathbf{u}$ is generated by the given recommendation model $R$, it is reasonable to assume that $\mathbf{u}$ has encoded the attribute information and the collaborative signals of user $u$ in an entangling and biased way.

\begin{figure}[t]
  \centering
  \subfigure[Larger $\beta$]{
    \begin{minipage}[t]{0.3\linewidth}
      \centering
      \label{Fig:big_beta}
      \includegraphics[width=\linewidth]{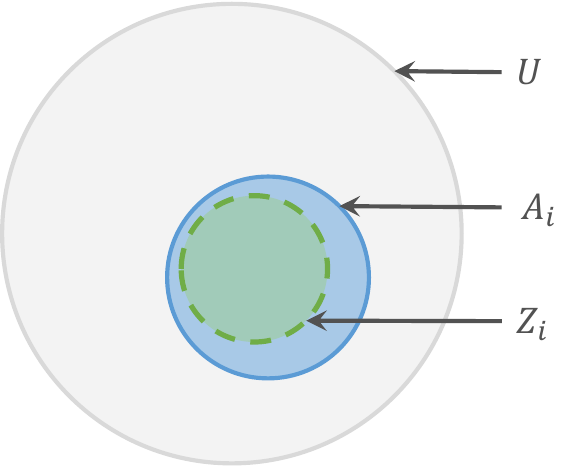}
    \end{minipage}%
  }%
  \subfigure [Smaller $\beta$]{
    \begin{minipage}[t]{0.3\linewidth}
      \centering
      \label{Fig:small_beta}
      \includegraphics[width=\linewidth]{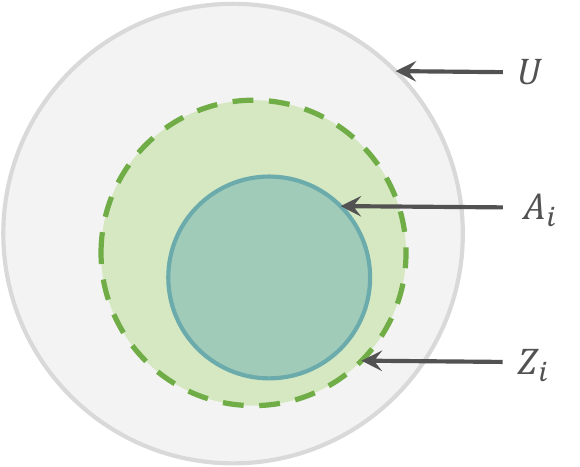}
    \end{minipage}%
  }%
   \subfigure [Ideal $\beta$]{
    \begin{minipage}[t]{0.3\linewidth}
      \centering
      \label{Fig:idea_beta}
      \includegraphics[width=\linewidth]{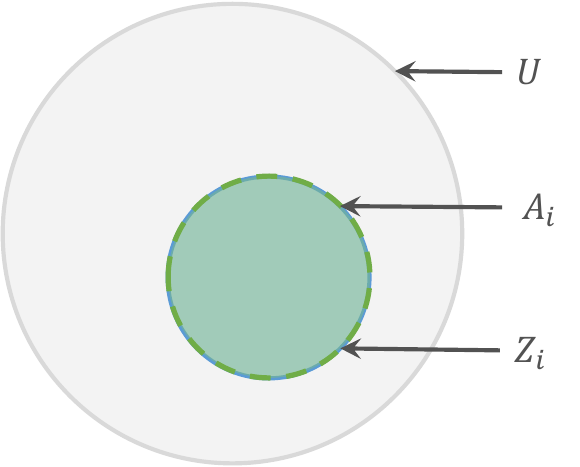}
    \end{minipage}%
  }%
  \centering
  \caption{Illustration of the attribute-specific embedding learning for Information Alignment, where (a) and (b) are the cases of information misalignment, while (c) is the optimal case of information alignment.}
  \label{Fig:ialignm}
\end{figure}

\subsubsection{Attribute-Specific Embedding}
\label{Sec:AE}
The attribute-specific embedding $\mathbf{z}_{u, i}$ of the $i$th attribute of user $u$ is extracted by the attribute encoder $E_i$ from the original user embedding $ \mathbf{u} $, $1 \le i \le M$, i.e.,
\begin{equation}
\mathbf{z}_{u,i} = E_i(\mathbf{u}).
\end{equation}
To achieve information alignment, our objective is to make $E_i$ able to encode and only encode the complete information of the $i$th attribute of users. Since $E_i$ is shared by all users, the output of $E_i$ can be represented by a stochastic variable $Z_i$ with various values $\{ \mathbf{z}_{u, i} | u \in \mathcal{U} \}$. Meanwhile, let $A_i$ be the stochastic variable representing the $i$th column of the user attribute matrix $\mathcal{A}$, of which the values are $\{ a_{u, i}| u \in \mathcal{U} \}$, and let $U$ be the stochastic variable representing different user embeddings $\{ \mathbf{u} \}$. Obviously, $Z_i = E_i(U)$. In the light of the information bottleneck principle \cite{shwartz2017opening}, the optimization objective of $E_i$ is to maximize the mutual information between $Z_i$ and $A_i$, and simultaneously minimize that between $Z_i$ and $U$, i.e.,
\begin{equation}
\label{Eq:loss_ei}
\min_{E_i} -I(Z_i;A_i)+{\beta}I(Z_i;U),
\end{equation}
where $I(\cdot, \cdot)$ is mutual information function, $\beta$ is a balance factor to control the amount of information from $U$ encoded by $Z_i$. 

Fig. \ref{Fig:ialignm} illustrates the insight of the attribute-specific embedding learning for information alignment defined by Equation (\ref{Eq:loss_ei}), where cycle areas represent the information content of corresponding variables. We can see that the amount of the information encoded in $Z_i$ can be regularized by $\beta$. Particularly, as shown in Figs. \ref{Fig:big_beta} and \ref{Fig:small_beta}, the smaller the $\beta$, the more information about $U$ covered by $Z_i$. Ideally, by regularizing $\beta$, the optimal solution $E_i$ to Equation (\ref{Eq:loss_ei}) will cause that the area of $Z_i$ exactly overlaps with the area of $A_i$, as shown in Fig. \ref{Fig:idea_beta}. At this ideal time, (1) $Z_i$ covers the information of $A_i$, which makes the discriminative information of $A_i$ be reserved completely in $Z_i$; (2) $Z_i$ only covers the information of $A_i$, which is favorable to fairness since the correlation between $Z_i$ and other attributes is reduced to minimum. In sharp contrast, the existing works pursue extreme fairness by completely removing the information of sensitive attributes, which will result in information about $A_i$ not being preserved intact if it is a non-sensitive attribute, because the parts of it that overlap with sensitive information are also removed. As we will see later in the experiments, exactly preserving the information of non-sensitive attributes, including the parts overlapping with the information of sensitive attributes, will increase the discriminability of the fair embeddings without loss of fairness, which helps AFRL achieve a better trade-off between fairness and accuracy.


However, the evaluation of mutual information is difficult. To overcome this challenge, we adopt different strategies to approximately maximize $I(Z_i;A_i)$ and minimize $I(Z_i;U)$, respectively.

(1) \textbf{The minimization of $\boldsymbol{I(Z_i;U)}$}
At first, the definition of $I(Z_i;U)$ is 
\begin{equation}
\label{Eq:MI_ZiU}
\begin{aligned}
I(Z_i;U)=\mathbb{E}_{(\mathbf{z}_{u,i}, \mathbf{u}) \sim p(Z_i, U)} \bigg[ \log{\frac{p(\mathbf{z}_{u,i} | \mathbf{u})}{p(\mathbf{z}_{u,i})}} \bigg].
\end{aligned}
\end{equation}
As previous works did \cite{liang2018variational, liu2021mitigating}, we first assume $p(\mathbf{z}_{u,i})$ $\sim$ $\mathcal{N}(0, \mathbf{I})$. To make Equation (\ref{Eq:MI_ZiU}) tractable, our idea is to view the output of $E_i$ as a sample from a special Gaussian distribution with 0 variance, i.e., $p(\mathbf{z}_{u,i} | \mathbf{u})$ $\sim$ $\mathcal{N}(\boldsymbol{\mu}_u,0)$. Then the optimal $E_i$ is
\begin{equation}
\begin{aligned}
\argmin_{E_i} I(Z_i;U) =\mathbb{E}_{\mathbf{u} \sim p(U)} \frac{1}{2} { \Vert \boldsymbol{\mu}_u} \Vert_2^2,
\end{aligned}
\label{Eq:upper_bound}
\end{equation}
where $ \boldsymbol{\mu}_u = \mathbf{z}_{u,i} = E_i(\mathbf{u}) $ and the expectation can be evaluated via Monte Carlo method.

(2) \textbf{The maximization of $\boldsymbol{I(Z_i;A_i)}$}
The definition of $I(Z_i;A_i)$ is 
\begin{equation}
I(Z_i;A_i)=\mathbb{E}_{(\mathbf{z}_{u,i}, a_{u,i}) \sim p(Z_i, A_i)} \bigg[\log{\frac{p(a_{u,i} | \mathbf{z}_{u,i})}{p(a_{u,i})}} \bigg],
\end{equation}
where $a_{u,i}$ represents the value of the $i$th attribute of a user $u$, and $p(a_{u,i})$ and $p(a_{u,i} | \mathbf{z}_{u,i})$ are the priori and posterior probabilities of $a_{u,i}$, respectively. Since $p(a_{u,i} | \mathbf{z}_{u,i})$ is intractable, we introduce the predicted probability $q(a_{u,i} | \mathbf{z}_{u,i}; C_i)$ offered by the classifier $C_i$:
\begin{equation}
\begin{aligned}
I(Z_i;A_i)
&=\mathbb{E}_{(\mathbf{z}_{u,i}, a_{u,i}) \sim p(Z_i, A_i)} \bigg[\log{\frac{p(a_{u,i} | \mathbf{z}_{u,i})q(a_{u,i} | \mathbf{z}_{u,i}; C_i)}{p(a_{u,i})q(a_{u,i} | \mathbf{z}_{u,i}; C_i)}} \bigg]\\
&= \Phi(Z_i, C_i) + \Delta(Z_i, C_i),
\end{aligned}
\label{Eq:ZiAi}
\end{equation}
where 
\begin{equation}
\begin{aligned}
\Phi(Z_i, C_i) &= \mathbb{E}_{(\mathbf{z}_{u,i}, a_{u,i}) \sim p(Z_i, A_i)} \bigg[\log{\frac{q(a_{u,i} | \mathbf{z}_{u,i}; C_i)}{p(a_{u,i})}} \bigg],\\
\Delta(Z_i, C_i) &= \mathbb{E}_{\mathbf{z}_{u,i} \sim p(Z_i)} \big[ \text{KL} \big(p(a_{u,i} | \mathbf{z}_{u,i}) \Vert {q(a_{u,i} | \mathbf{z}_{u,i}; C_i) \big)} \big].
\end{aligned}
\end{equation}
To maximize $I(Z_i;A_i)$, we borrow the idea of EM algorithm to iteratively update $C_i$ and $Z_i$, i.e., at $t+1$ time step,
\begin{equation}
C_i^{t+1} = \argmin_{C_i^{t}}\Delta(Z_i^{t}, C_i^{t}),
\label{Eq:Ci1}
\end{equation}
\begin{equation}
Z_i^{t+1} = \argmax_{Z_i^{t}}\Phi(Z_i^{t}, C_i^{t+1}).
\label{Eq:Zi1}
\end{equation}
By some simple derivations, it is easy to know Equation (\ref{Eq:Ci1}) is equivalent to
\begin{equation}
C_i^{t+1} = \argmin_{C_i}\mathbb{E}_{(\mathbf{z}_{u,i}, a_{u,i}) \sim p(Z_i^t,A_i)}[-\log{q(a_{u,i} | \mathbf{z}_{u,i}; C_i)}],\
\label{Eq:Ci}
\end{equation}
and Equation (\ref{Eq:Zi1}) is equivalent to
\begin{equation}
Z_i^{t+1} = \argmin_{Z_i^t}\mathbb{E}_{(\mathbf{z}_{u,i}, a_{u,i}) \sim p(Z_i^t,A_i)}[-\log{q(a_{u,i} | \mathbf{z}_{u,i}; C_i^{t+1})}].
\label{Eq:Zi}
\end{equation}
By combining Equations (\ref{Eq:upper_bound}),(\ref{Eq:Ci}) and (\ref{Eq:Zi}), and considering $Z_i$ is updated through $E_i$ since $Z_i = E_i(U)$, we can obtain the following bilevel optimization objective of the attribute encoders:
\begin{equation}
\begin{aligned}
\min_{E_i}\min_{C_i} & \mathbb{E}_{\mathbf{u} \sim p(U)} \frac{1}{2} { \Vert \mathbf{z}_{u, i}} \Vert_2^2 \\ 
&+\beta\mathbb{E}_{(\mathbf{z}_{u,i}, a_{u,i}) \sim p(Z_i, A_i)}[-\log{q(a_{u,i} | \mathbf{z}_{u,i}; C_i)}].
\end{aligned}
\label{Eq:Ei1}
\end{equation}
At the same time, note that the sampling process $(\mathbf{z}_{u,i}, a_{u,i}) \sim p(Z_i, A_i)$ is fulfilled by first sampling a user's embedding $\mathbf{u} \sim p(U)$ and then obtaining $(\mathbf{z}_{u,i} = E_i(\mathbf{u}), a_{u, i})$. Hence the sampling in the second term in Equation (\ref{Eq:Ei1}) can be equivalently changed to be the same as that in the first term, which leads to the following equivalent tractable bilevel optimization:
\begin{equation}
\min_{E_i}\min_{C_i}\mathbb{E}_{\mathbf{u} \sim p(U)} \bigg[\frac{1}{2} { \Vert \mathbf{z}_{u, i}} \Vert_2^2 -\beta\log{q(a_{u,i} | \mathbf{z}_{u,i}; C_i)} \bigg].
\label{Eq:Ei2}
\end{equation}



\subsubsection{Debiased Collaborative Embedding}
As mentioned before, the original unfair user embedding $\mathbf{u}$ provided by the given unfair recommendation model $R$ has encoded the collaborative signals biased to user attributes. IAlignM employs a debiased collaborative signal encoder $F$ to generate the debiased collaborative embedding $\mathbf{z}_{u, 0}$ to capture from $\mathbf{u}$ the collaborative signals independent of user attributes. For this purpose, we optimize $F$ with two objectives: (1) minimizing the discriminability of $\mathbf{z}_{u,0}$ to user attributes for fairness, and (2) maximizing the information of $\mathbf{u}$ encoded in $\mathbf{z}_{u,0}$ to preserve the collaborative signals as much as possible for recommendation accuracy. 

To achieve the objective (1), we introduce a discriminator $D_i$ for $i$th attribute, $1 \le i \le M$, and adjust $F$ to fool $D_i$ with the following adversarial training between $F$ and $D_i$:
\begin{equation}
\min_{F}\max_{D_i} - \mathbb{E}_{ \mathbf{u} \sim p(U) } \sum_{i=1}^{M} { \left[ - \log{ q ( a_{u,i} | \mathbf{z}_{u,0} ; D_i ) } \right] },
\label{Eq:adv}
\end{equation}
where $q ( a_{u,i} | \mathbf{z}_{u,0} ; D_i )$ is the probability that $\mathbf{z}_{u,0}$ is classified as $a_{u,i}$ by $D_i$.

To achieve the objective (2), we minimize the following reconstruction loss:
\begin{equation}
\min_{F}\mathbb{E}_{\mathbf{u} \sim p(U) } \Vert  \mathbf{z}_{u,0}-\mathbf{u} \Vert_2^2 . \\
\label{Eq:dec}
\end{equation}

By combining Equations (\ref{Eq:adv}) and (\ref{Eq:dec}), we obtain the following optimization objective of $F$:
\begin{equation}
\begin{aligned}
\min_{F} \max_{D_i} \mathbb{E}_{ \mathbf{u} \sim p(U) } \Vert \mathbf{z_{u,0} - \mathbf{u}} \Vert_2^2 
 - \lambda \sum_{i=1}^{M} { \left[ -\log{ q ( a_{u,i} | \mathbf{z}_{u,0} ; D_i ) } \right] },
\label{Eq:Di}
\end{aligned}
\end{equation}
where $\lambda$ controls the trade-off between recommendation accuracy and fairness. 

\subsection{Information Aggregation Module}
After obtaining attribute-specific embeddings $\{\mathbf{z}_{u,i}\}$ ($1 \leq i \leq M$) and the debiased collaborative embedding $\mathbf{z}_{u,0}$, the Information Aggregation Module (IAggM), which is implemented as an MLP $G$, combines these embeddings based on the personalized fairness requirement $\mathbf{s}_u$ to produce the final fair embedding $\mathbf{u}^{*}$ as follow:
\begin{equation}
 \mathbf{u}^{*} = G\bigg\{\mathbf{z}_{u,0}\oplus \big[(  \mathbf{z}_{u,1} \oplus  \ldots \oplus \mathbf{z}_{u,M}) \odot \mathbf{s}_{u}\big] \bigg\},
 \label{Eq:G}
\end{equation}
where $\oplus$ represents concatenation, $\odot$ represents element-wise product, and $\mathbf{s}_u \in \{ 0, 1\}^{Md \times 1}$ represents the fairness requirement of user $u$. The $i$th part $\mathbf{s}_{u,[1+(i-1)d : i\times d]} = 0$ means the $i$th attribute is a sensitive attribute of user $u$, otherwise a non-sensitive attribute.

To ensure $ \mathbf{u}^{*} $ to be compatibility with the original recommendation model $R$, $G$ will be adjusted with respect to the supervision offered by $R$ as follow:
\begin{equation}
\min_{G}  -\sum_{(u,v^+,v^-)}{\ln{\sigma(y_{u,v^+}-y_{u,v^-})}},
\label{Eq:Lrec}
\end{equation}
where $v^+$ and $v^-$ are a positive example and a negative example of $u$, respectively, $\sigma(\cdot)$ is the sigmoid function, and $y_{u, v} = R( \mathbf{u}^{*}, \mathbf{v})$. Note that Equation (\ref{Eq:Lrec}) is a function of $G$ rather than $R$, which means it is $G$ that will be adjusted during the optimization of Equation (\ref{Eq:Lrec}) with fixed $R$. Algorithm \ref{Alg:train} gives the training process of AFRL.

\renewcommand{\algorithmicrequire}{\textbf{Input:}}
\renewcommand{\algorithmicensure}{\textbf{Output:}}

\begin{algorithm}[t]
\caption{Training of AFRL}
\label{Alg:train}
\begin{algorithmic}[1]
\REQUIRE ~~ 
user-item interaction matrix $\mathcal{O}$, user attribute matrxi $\mathcal{A}$, unfair recommendation model $R$, and the set of original user embeddings $U$ and item embeddings $V$ generated by $R$; 
\ENSURE ~~ 
$F$, $\{E_i\}$ ($1 \le i \le M$), and $G$;
\STATE Randomly initialize $F$, $\{D_i\}$, $\{E_i\}$, $\{C_i\}$ and $G$, $1 \le i \le M$;
\FOR {$1$ to $T$}
\FOR {each $\mathbf{u} \in U$}
\STATE obtain $\mathbf{z}_{u,0}=F(\mathbf{u})$ and $\{ \mathbf{z}_{u,i}=E_i(\mathbf{u})\}$;
\STATE randomly sample an $\mathbf{s}_u$ from a uniform distribution over different attribute combinations;
\STATE obtain $\mathbf{u}^*$ by Equation (\ref{Eq:G});
\STATE update $\{C_i\}$ by inner minimization of Equation (\ref{Eq:Ei2});
\STATE update $\{E_i\}$ by outer minimization of Equation (\ref{Eq:Ei2});
\STATE update $\{D_i\}$ by inner maximization of Equation (\ref{Eq:Di});
\STATE update $F$ by outer minimization of Equation (\ref{Eq:Di});
\STATE update G by Equation (\ref{Eq:Lrec}).
\ENDFOR
\ENDFOR
\end{algorithmic}
\end{algorithm}

\subsection{Theoretical Justification}
\subsubsection{Convergence Analysis}
Now we analyze the convergence of the optimization of Equation (\ref{Eq:Ei2}), which is the tractable approximation of Equation (\ref{Eq:loss_ei}) to realize the information alignment. At first, it is easy to see that $I(Z_i;U)$ will achieve its minimum as defined in Equation (\ref{Eq:upper_bound}). So, we only justify the iterative optimization defined Equations (\ref{Eq:Ci1}), (\ref{Eq:Zi1}) will lead to the maximum of $I(Z_i;A_i)$, which equivalently to prove $I (Z_i^{t+1}; A_i) - I (Z_i^t; A_i) \ge 0$. At first, according to Equation (\ref{Eq:ZiAi}), we can obtain
\begin{equation}
\begin{aligned}
& I ( Z_i^{t+1}, A_i ) - I ( Z_i^{t} , A_i )\\
= &\underbrace{\Phi(Z_i^{t+1}, C_i) - \Phi(Z_i^{t}, C_i) }_{A} + \underbrace{ \Delta(Z_i^{t+1}, C_i) -  \Delta(Z_i^{t}, C_i) }_{B} . 
\end{aligned}
\end{equation}
By taking a close look at Equation (\ref{Eq:ZiAi}), we can observe that $I(Z_i, A_i )$ is independent of $C_i$. Based on this observation, we can set $C_i$ to the optimal $C_i^{t+1} = \argmin_{C_i^{t}}\Delta(Z_i^{t}, C_i^{t})$ (Equation (\ref{Eq:Ci1})) without changing $I(Z_i, A_i )$, which gives:
\begin{equation}
\begin{aligned}
& I ( Z_i^{t+1}, A_i ) - I ( Z_i^{t} , A_i )\\
= &\underbrace{\Phi(Z_i^{t+1}, C^{t+1}_i) - \Phi(Z_i^{t}, C^{t+1}_i) }_{A} + \underbrace{ \Delta(Z_i^{t+1}, C^{t+1}_i) -  \Delta(Z_i^{t}, C^{t+1}_i) }_{B} . 
\end{aligned}
\end{equation}
Since $\Delta$ is a KL-divergence, the optimal $C_i^{t+1}$ makes $\Delta(Z_i^{t}$, $C^{t+1}_i)$ $ = 0$, and hence $B = \Delta(Z_i^{t+1}, C^{t+1}_i)\ge 0$. At the same time, since $Z_i^{t+1}$ maximizes $\Phi(Z_i^{t}, C^{t+1}_i)$ (see Equation (\ref{Eq:Zi1})), $\Phi(Z_i^{t+1}$, $C^{t+1}_i)$ $\ge\Phi(Z_i^{t}, C^{t+1}_i) $, i.e., $A \ge 0$. Thus $I ( Z_i^{t+1}, A_i ) - I ( Z_i^{t} , A_i ) \ge 0$ holds.

\subsubsection{Informativeness Analysis}

Now we show that AFRL completely extracts the information of a non-sensitive attribute into the fair embedding. Without loss of generality, let $A_n$, $Z_n$, $U$, and $U^{*}$ be the stochastic variables representing a non-sensitive attribute, the embedding of $A_n$, the original user embedding, and the fair user embedding, respectively, $1 \le i \le M$. Obviously, the information encoded by $U$ can be divided into two parts:  $H( U ) = I ( U ; A_n) + H( U | A_n ) $, where $I ( U ; A_n)$ is the part relevant to $A_n$. We want to prove $U^{*}$ completely encodes $I ( U ; A_n)$, which is equal to proving $I(A_n ; U | U^*) = 0$. 

In terms of the theory of information bottleneck \cite{shwartz2017opening}, AFRL can be framed as a Markov Chain $A_n \rightarrow U \rightarrow Z_n \rightarrow U^* $. Suppose the attribute encoder $E_n$ is the optimal solution to the bilevel optimization defined in Equation (\ref{Eq:Ei2}), which means $Z_n= E_n(U)$ completely encodes the $A_n$'s information from $U$, and equivalently, $I(A_n ; U| Z_n) = 0$. Considering $I(A_n ; U | Z_n) = I(A_n; U) - I(U ; A_n; Z_n) =  I(A_n ; U) - I(A_n; Z_n) $, we have
\begin{equation}
I(A_n ; U) = I(A_n; Z_n). 
\label{Eq:AiU}
\end{equation}
Let $Z_s$ be the stochastic variable representing the embeddings of sensitive attributes, and due to the Markov Chain $Z_s \leftarrow U \rightarrow Z_n$, $Z_n$ and $Z_s$ are conditional independent, i.e., $Z_n \perp Z_s | U$. Meanwhile, the variable of debiased collaborative embedding, $Z_0$, is optimized to be independent of any $Z_n$. Therefore, removing $Z_s$ from and incorporating $Z_0$ to $U^*$ will not affect the mutual information between $Z_n$ and $U^*$, $I(Z_n ; U^*)$. When $E_n$ and $G$ are both optimal,  $I(Z_n ; U^*)$ achieves maximum, which results in 
\begin{equation}
H( Z_n | U^*) = 0. 
\label{Eq:ZiU*}
\end{equation}
Based on Equations (\ref{Eq:AiU}) and (\ref{Eq:ZiU*}), we have the following derivations:
\begin{equation}
\begin{aligned}
I(A_n ; U | U^*) =& I(A_n; U) - I(A_n ; U ; U^*) \\
=& I(A_n; Z_n ) - \big( I(A_n ; U^*) - I(A_n ; U^* | U) \big) \\
=& I(A_n; Z_n ) - I(A_n ; U^*) \\
=& I(A_n; Z_n ) - \big( I(A_n ; U^*| Z_n) + I(A ; U^* ; Z_n) \big) \\
=& I(A_n; Z_n ) - I(A_n ; U^* ; Z_n) \\
=& I(A_n; Z_n ) - I(Z_n ; U^*) + I(Z_n ; U^* | A_n)\\
=& \big( H(Z_n) - H(Z_n | A_n) \big) - \big( H(Z_n) - H(Z_n| U^*) \big) \\
& + \big( H(Z_n | A_n) - H(Z_n | U^*, A_n) \big)\\
=& H(Z_n| U^*) - H(Z_n | U^*, A_n) \\
=&- H(Z_n | U^*, A_n) \ge 0.
\end{aligned}
\end{equation}
Note that during the above derivations, $I(A_n ; U^* | U)$ and $I(A_n ; U^*| Z_n)$ are both equal to 0 because of the Markov Chain $A_n \rightarrow U \rightarrow Z_n \rightarrow U^* $. At the same time, since entropy is non-negative, we also have $-H(Z_n | U^*, A_n) \le 0$, and thus $H(Z_n | U^*, A_n) = 0$, and consequently $I(A_n ; U | U^*) = 0$.

\section{experiments}
The experiments are designed to address the following research questions:
\begin{itemize}
  \item \textbf{RQ1} How does AFRL perform as compared to the state-of-the-art models?
   \item \textbf{RQ2} How does AFRL make trade-off between fairness and accuracy as compared to the state-of-the-art models?
   \item \textbf{RQ3} How do the attribute-specific embeddings and the debiased collaborative embedding contribute to the performance of AFRL?
   \item \textbf{RQ4} How do the hyper-parameters affect the performance of AFRL?
   \end{itemize}

\subsection{Experimental Settings}

\subsubsection{Datasets.}
We evaluate AFRL on two datasets:
\begin{itemize}
  \item \textbf{MovieLens-1M$\footnote{\url{ https://grouplens.org/datasets/movielens/}}$ (ML-1M)} This dataset consists of 1,000,000 movie ratings provided by 6,040 users with attributes G (Gender), A (Age) and O (Occupation). For simplicity, we convert the ratings into binary values, where ratings not less than 4 are set to 1, while the rest are set to 0.
  \item \textbf{Taobao Display Ad Click$\footnote{\url{https://tianchi.aliyun.com/dataset/56}}$ (Taobao)} This dataset comprises more than 26,000,000 interactions between 1,140,000 users and 840,000 advertisements on Taobao website. A user in Taobao is associated with attributes G (Gender), A (Age), and C (Consumption level). 
 \end{itemize}
For each dataset, we remove the users with less than 10 instances. We generate interaction sequences for a user with sliding window sizes 100 and 50 on ML-1M and Taobao, respectively. As previous work did \cite{wu2022selective}, for each user, we choose the most recent sequence as a testing instance and the second most recent sequence as a validation instance.

\begin{table}[t]
  \centering
  \caption{Hyper-parameter setting of AFRL}
  \resizebox{\linewidth}{!}{
    \begin{tabular}{l|c|c|c|c}
    \toprule
    \multicolumn{1}{c|}{\multirow{2}{*}{$\bold{Method}$}}& \multicolumn{2}{c|}{\makecell{$\bold{\beta \ is\ selected\ from}$\\ $\bold{\{0.001,0.01,0.1,1,10,100\}}$}} & \multicolumn{2}{c}{\makecell{$\bold{\lambda\ is\ selected\ from}$\\$\bold{\{0.001,0.01,0.1,1,10,100\}}$}} \\
\cline{2-5} & ML-1M & Taobao	 & ML-1M & Taobao	 \\
    \hline
    AFRL+PMF &   0.1    &   1    &   0.1    &  10\\
  \hline
   AFRL+DeepModel &  0.1     &   1    &   0.1    &  1\\
    \hline
    AFRL+SASRec &  1     &     1  &    1   & 10 \\
    \hline
    AFRL+BERT4Rec &  1     &    1   &  1     & 10  \\
    \bottomrule
    \end{tabular}%
    }
  \label{tab:hyperparameter_set}%
\end{table}%
  
 \subsubsection{Baselines}
 We will compare AFRL with following three state-of-the-art models for personalized fairness in recommendations:
\begin{itemize}
  \item \textbf{SM \cite{li2021towards}} To meet personalized fairness in recommendations, SM trains a attribute filter for each possible combination of sensitive attributes.
  \item \textbf{PFRec \cite{wu2022selective}} PFRec builds a set of prompt-based bias eliminators and adapters with predefined attribute-specific prompts to learn fair representations for different attribute combinations. 
  \item \textbf{FFVAE \cite{creager2019flexibly}} FFVAE disentangles an unfair embedding into latent factors, each of which corresponds to an attribute, and excludes relevant sensitive latent factors based on different fairness requirements of users.
\end{itemize}
We will check the performance of AFRL and the baseline models in combination with the following four base recommendation models, including two non-sequential models \textbf{PMF} \cite{mnih2007probabilistic} and \textbf{DeepModel} \cite{cheng2016wide}, and two sequential models \textbf{SASRec} \cite{kang2018self} and \textbf{BERT4Rec} \cite{sun2019bert4rec}.

 \subsubsection{Evaluation Protocols.}
 \label{sec:protocol}
We evaluate the accuracy of top-$N$ recommendations with two popular metrics Normalized Discounted Cumulative Gain (N@$10$) and Hit rate (H@$10$). Following the idea in \cite{kang2018self, sun2019bert4rec, wu2022selective}, we sample 1 and 99 negative examples for a positive example during training and testing, respectively. To evaluate the fairness, following the popular method used in the existing works \cite{li2021towards, hua2023up5}, we first train a surrogate classifier for each attribute using the fair embeddings generated by AFRL and the baseline models, respectively, and report the average AUC over all classifiers as the fairness metric. The closer to 0.5 AUC, the better fairness.

\begin{table*}[t]
\caption{ Fairness (AUC) and Accuracy (N@10, H@10) comparison on ML-1M. A fairness requirement is represented by an attribute combination where each attribute is assumed to be sensitive to users. For instance, G+A+O represents Gender, Age and Occupation are sensitive attributes. The best runs per metric are marked in boldface.}
\resizebox{\textwidth}{!}{
	\begin{tabular}{r|ccc|ccc|ccc|ccc|ccc|ccc|ccc}
     \toprule
   \multicolumn{1}{c|}{\multirow{3}{*}{$\bold{Method}$}} & \multicolumn{21}{c}{$\bold{Fairness\ Requirement}$}\\
\cline{2-22}    \multicolumn{1}{c|}{} & \multicolumn{3}{c|}{$\bold{G}$} & \multicolumn{3}{c|}{$\bold{A}$} & \multicolumn{3}{c|}{$\bold{O}$} & \multicolumn{3}{c|}{$\bold{G+A}$} & \multicolumn{3}{c|}{$\bold{G+O}$} & \multicolumn{3}{c|}{$\bold{A+O}$} & \multicolumn{3}{c}{$\bold{G+A+O}$}\\
\cline{2-22}    \multicolumn{1}{c|}{} 
& AUC$\downarrow$   &{N@10$\uparrow$} & {H@10$\uparrow$} 
& AUC$\downarrow$   &{N@10$\uparrow$} & {H@10$\uparrow$} 
& AUC$\downarrow$   &{N@10$\uparrow$} & {H@10$\uparrow$} 
& AUC$\downarrow$   &{N@10$\uparrow$} & {H@10$\uparrow$} 
& AUC$\downarrow$   &{N@10$\uparrow$} & {H@10$\uparrow$} 
& AUC$\downarrow$   &{N@10$\uparrow$} & {H@10$\uparrow$}
& AUC$\downarrow$   &{N@10$\uparrow$} & {H@10$\uparrow$} \\
    \hline
   PMF  &   0.7890 & 0.4157 & 0.6918 & 0.7908 & 0.4157 & 0.6918 & 0.6703 & 0.4157 & 0.6918 & 0.7863 & 0.4157 & 0.6918 & 0.7211 & 0.4157 & 0.6918 & 0.7220 & 0.4157 & 0.6918 & 0.7438 & 0.4157 & 0.6918 \\
   SM+PMF  &  0.6547 & 0.3733 & 0.6492 & 0.6394 & 0.3739 & 0.6496 & 0.5703 & 0.3690 & 0.6467 & 0.6382 & 0.3703 & 0.6471 & 0.5890 & 0.3711 & 0.6498 & 0.5828 & 0.3682 & 0.6440 & 0.5915 & 0.3699 & 0.6464  \\
   FFVAE+PMF  & 0.6433 & 0.3122 & 0.5625 & 0.6843 & 0.2993 & 0.5409 & 0.6036 & 0.2772 & 0.5048 & 0.7261 & 0.2979 & 0.5418 & 0.6553 & 0.2765 & 0.5034 & 0.6258 & 0.2684 & 0.4874 & 0.6394 & 0.2664 & 0.4844\\
  AFRL+PMF &  \textbf{0.5310} & \textbf{0.3944} & \textbf{0.6707} & \textbf{0.5238} & \textbf{0.3908} & \textbf{0.6655} & \textbf{0.5112} & \textbf{0.3945} & \textbf{0.6729} & \textbf{0.5274} & \textbf{0.3884} & \textbf{0.6622} & \textbf{0.5211} & \textbf{0.3915} & \textbf{0.6669} & \textbf{0.5175} & \textbf{0.3881} & \textbf{0.6601} & \textbf{0.5220} & \textbf{0.3846} & \textbf{0.6555} \\
    \hline
    DeepModel &   0.7656 & 0.3683 & 0.6397 & 0.7365 & 0.3683 & 0.6397 & 0.6542 & 0.3683 & 0.6397 & 0.7589 & 0.3683 & 0.6397 & 0.7113 & 0.3683 & 0.6397 & 0.6876 & 0.3683 & 0.6397 & 0.7202 & 0.3683 & 0.6397 \\
    SM+DeepModel& 0.6529 & 0.3403 & 0.6063 & 0.6319 & 0.3346 & 0.6047 & 0.5709 & 0.3356 & 0.6009 & 0.6495 & 0.3299 & 0.5949 & 0.6169 & 0.3314 & 0.5952 & 0.6024 & 0.3263 & 0.5906 & 0.6158 & 0.3205 & 0.5802  \\   
    FFVAE+DeepModel &    0.6289 & 0.2981 & 0.5204 & 0.6491 & 0.2954 & 0.5137 & \textbf{0.5000} & 0.2955 & 0.5145 & 0.6258 & 0.2957 & 0.5125 & 0.5647 & 0.2925 & 0.5095 & \textbf{0.5000} & 0.2928 & 0.509 & 0.5798 & 0.2902 & 0.5029 \\
    AFRL+DeepModel & \textbf{0.5358} & \textbf{0.3504} & \textbf{0.6151} & \textbf{0.5018} & \textbf{0.3474} & \textbf{0.6103} & 0.5231 & \textbf{0.3525} & \textbf{0.6202} & \textbf{0.5188} & \textbf{0.3430} & \textbf{0.6024} & \textbf{0.5295} & \textbf{0.3471} & \textbf{0.6108} & 0.5125 & \textbf{0.3438} & \textbf{0.6050} & \textbf{0.5203} & \textbf{0.3377} & \textbf{0.5921}\\
    \hline
  SASRec &  0.7554 & 0.5709 & 0.8103 & 0.7259 & 0.5709 & 0.8103 & 0.6143 & 0.5709 & 0.8103 & 0.7284 & 0.5709 & 0.8103 & 0.6879 & 0.5709 & 0.8103 & 0.6634 & 0.5709 & 0.8103 & 0.6883 & 0.5709 & 0.8103  \\
 SM+SASRec & 0.6366 & 0.4808 & 0.7474 & 0.6183 & 0.4792 & 0.7447 & 0.5917 & 0.4803 & 0.7454 & 0.6021 & 0.4790 & 0.7427 & 0.5947 & 0.4813 & 0.7486 & 0.5968 & 0.4773 & 0.7378 & 0.6229 & 0.4746 & 0.7382  \\ 
  PFRec+SASRec & \textbf{ 0.5179} & 0.4784 & 0.7418 & \textbf{0.5159} & 0.4684 & 0.7341 & 0.5288 & 0.4499 & 0.7044 &  \textbf{0.5146} & 0.4420 & 0.7002 &  \textbf{0.5179} & 0.4478 & 0.7069 & \textbf{0.5073} & 0.3798 & 0.5825 & \textbf{0.5173} & 0.4187 & 0.6830\\  
   FFVAE+SASRec & 0.5677 & 0.4190 & 0.6585 & 0.6540 & 0.4035 & 0.6376 & 0.6029 & 0.3728 & 0.5995 & 0.6433 & 0.4027 & 0.6383 & 0.6513 & 0.3727 & 0.5966 & 0.5631 & 0.3598 & 0.5785 & 0.5962 & 0.3614 & 0.5829\\
   AFRL+SASRec & 0.5543 & \textbf{0.5571} & \textbf{0.7980} & 0.5346 & \textbf{0.5543} & \textbf{0.7953} & \textbf{0.5200} & \textbf{0.5595} & \textbf{0.7982} &0.5445 & \textbf{0.5518} & \textbf{0.7945} & 0.5372 & \textbf{0.5559} & \textbf{0.7953} & 0.5273 & \textbf{0.5526} & \textbf{0.7934} & 0.5463 & \textbf{0.5493} & \textbf{0.7903}\\
    \hline
     BERT4Rec & 0.7257 & 0.5657 & 0.8103 & 0.6946 & 0.5657 & 0.8103 & 0.6331 & 0.5657 & 0.8103 & 0.7171 & 0.5657 & 0.8103 & 0.6894 & 0.5657 & 0.8103 & 0.6686 & 0.5657 & 0.8103 & 0.6922 & 0.5657 & 0.8103 \\
 SM+BERT4Rec &0.5485 & 0.4764 & 0.7393 & 0.5674 & 0.4731 & 0.7370 & 0.5668 & 0.4744 & 0.7373 & 0.6200 & 0.4660 & 0.7297 & 0.6007 & 0.4702 & 0.7299 & 0.6112 & 0.4587 & 0.7244 & 0.6063 & 0.4606 & 0.7235  \\
  PFRec+BERT4Rec  & \textbf{0.5258} & 0.4875 & 0.7400 & 0.5147 & 0.4576 & 0.7113 & 0.5277 & 0.4699 & 0.7183 & 0.5351 & 0.4438 & 0.6963 & \textbf{0.5139} & 0.4510 & 0.7037 & 0.5179 & 0.3868 & 0.5936 & 0.5278 & 0.4247 & 0.6765\\
 FFVAE+BERT4Rec & 0.6506 & 0.4229 & 0.6790 & 0.5933 & 0.4090 & 0.6595 & 0.5963 & 0.3528 & 0.5817 & 0.6263 & 0.4080 & 0.6592 & 0.6018 & 0.3549 & 0.5850 & 0.5567 & 0.3431 & 0.5625 & 0.5765 & 0.3434 & 0.5691\\ 
    AFRL+BERT4Rec  & 0.5461 & \textbf{0.5390} & \textbf{0.7906} & \textbf{0.5066} & \textbf{0.5326} & \textbf{0.7830} & \textbf{0.5248} & \textbf{0.5405} & \textbf{0.7903} & \textbf{0.5264} & \textbf{0.5285} & \textbf{0.7798} & 0.5354 & \textbf{0.5381} & \textbf{0.7897} & \textbf{0.5157} & \textbf{0.5310} & \textbf{0.7817} & \textbf{0.5258} & \textbf{0.5207} & \textbf{0.7770}  \\
    \bottomrule
    \end{tabular}%
}
 \label{tab:ml-1m}
\end{table*}

\begin{table*}
\caption{Fairness and Accuracy comparison on Taobao. }
\resizebox{\textwidth}{!}{
	\begin{tabular}{r|ccc|ccc|ccc|ccc|ccc|ccc|ccc}
     \toprule
   \multicolumn{1}{c|}{\multirow{3}{*}{$\bold{Method}$}} & \multicolumn{21}{c}{$\bold{Fairness\ Requirement}$} \\
\cline{2-22}    \multicolumn{1}{c|}{} & \multicolumn{3}{c|}{$\bold{G}$} & \multicolumn{3}{c|}{$\bold{A}$} & \multicolumn{3}{c|}{$\bold{C}$} & \multicolumn{3}{c|}{$\bold{G+A}$} & \multicolumn{3}{c|}{$\bold{G+C}$} & \multicolumn{3}{c|}{$\bold{A+C}$} & \multicolumn{3}{c}{$\bold{G+A+C}$}\\
\cline{2-22}    \multicolumn{1}{c|}{} 
& AUC$\downarrow$   &{N@10$\uparrow$} & {H@10$\uparrow$} 
& AUC$\downarrow$   &{N@10$\uparrow$} & {H@10$\uparrow$} 
& AUC$\downarrow$   &{N@10$\uparrow$} & {H@10$\uparrow$} 
& AUC$\downarrow$   &{N@10$\uparrow$} & {H@10$\uparrow$} 
& AUC$\downarrow$   &{N@10$\uparrow$} & {H@10$\uparrow$} 
& AUC$\downarrow$   &{N@10$\uparrow$} & {H@10$\uparrow$}
& AUC$\downarrow$   &{N@10$\uparrow$} & {H@10$\uparrow$} \\
    \hline
     PMF  &  0.7144 & 0.5329 & 0.7839 & 0.6872 & 0.5329 & 0.7839 & 0.6406 & 0.5329 & 0.7839 & 0.7007 & 0.5329 & 0.7839 & 0.6785 & 0.5329 & 0.7839 & 0.6707 & 0.5329 & 0.7839 & 0.6829 & 0.5329 & 0.7839 \\
    SM+PMF  & 0.5979 & 0.4892 & 0.7420 & \textbf{0.5345} & \textbf{0.5001} & \textbf{0.7546} & 0.5577 & 0.4701 & 0.7270 & 0.5798 & 0.4811 & 0.7364 & 0.6089 & 0.4654 & 0.7271 & 0.5832 & 0.4538 & 0.7169 & 0.5783 & 0.4797 & 0.7313 \\
   FFVAE+PMF & 0.6867&0.4382&0.6939& 0.6212 &0.4209& 0.6734& 0.5360& 0.3806 & 0.6068 & 0.6325 & 0.4168 & 0.6653 & 0.5767& 0.3762& 0.6032 & 0.5456 & 0.3783& 0.6058 & 0.5722 & 0.3720 &0.5983  \\
   AFRL+PMF & \textbf{0.5659} & \textbf{0.4998} & \textbf{0.7529} &  0.5443 & 0.4983 & 0.7521 & \textbf{0.5073} & \textbf{0.5014} & \textbf{0.7550} & \textbf{0.5551} & \textbf{0.4908} & \textbf{0.7445} & \textbf{0.5366} & \textbf{0.4943} & \textbf{0.7473} & \textbf{0.5258} & \textbf{0.4929} & \textbf{0.7469} & \textbf{0.5391} & \textbf{0.4856} & \textbf{0.7492}   \\
      \hline
    DeepModel & 0.6582 & 0.4972 & 0.7485 & 0.6016 & 0.4972 & 0.7485 & 0.5652 & 0.4972 & 0.7485 & 0.6301 & 0.4972 & 0.7485 & 0.6095 & 0.4972 & 0.7485 & 0.5835 & 0.4972 & 0.7485 & 0.6082 & 0.4972 & 0.7485 \\
    SM+DeepModel&0.6057 & 0.4714 & 0.7244 & 0.5541 & 0.4734 & 0.7248 & 0.6066 & 0.4725 & 0.7237 & 0.5800  & 0.4715 & 0.7244 & 0.6010 & 0.4728 & 0.7256 & 0.5803 & 0.4723 & 0.7247 & 0.5845 & 0.4615 & 0.7136\\
    FFVAE+DeepModel & 0.6221 & 0.4185 & 0.6624 & 0.6392& 0.4158 &  0.6573 & 0.5175 & 0.3997& 0.6304& 0.6133& 0.4166 & 0.6582 & 0.5441 & 0.4009 & 0.6331 & 0.5337 & 0.3990 &0.6345& 0.5447 & 0.4011 & 0.6336  \\
    AFRL+DeepModel &   \textbf{0.5237} & \textbf{0.4879} & \textbf{0.7444} & \textbf{0.5335} & \textbf{0.4897} & \textbf{0.7389} & \textbf{0.5085} & \textbf{0.4894} & \textbf{0.7380} & \textbf{0.5310} & \textbf{0.4831} & \textbf{0.7303} & \textbf{0.5166} & \textbf{0.4828} & \textbf{0.7292} & \textbf{0.5204} & \textbf{0.4847} & \textbf{0.7359} & \textbf{0.5209} & \textbf{0.4801} & \textbf{0.7273} \\
        \hline
     SASRec &  0.7080 & 0.5399 & 0.7888 & 0.6656 & 0.5399 & 0.7888 & 0.5591 & 0.5399 & 0.7888 & 0.6832 & 0.5399 & 0.7888 & 0.6264 & 0.5399 & 0.7888 & 0.6171 & 0.5399 & 0.7888 & 0.6418 & 0.5399 & 0.7888  \\
  SM+SASRec &  0.6305 & 0.4827 & 0.7346 &0.5868& 0.4889 & 0.7393 & 0.5762 & 0.4860 & 0.7373 & 0.6121 & 0.4845 & 0.7377 & 0.6042 & 0.4848 & 0.7377 & 0.5826 & 0.4841 & 0.7365 & 0.6011 & 0.4854 & 0.7379  \\
    PFRec+SASRec & 0.6160 & 0.4957 & 0.7478 & 0.6027 & 0.4848 & 0.7373 &0.5373 & 0.4945 & 0.7457 & 0.6086 & 0.4808 & 0.7318 & 0.5703 & 0.4860 & 0.7380 & 0.5655 & 0.4851 & 0.7375 & 0.5789 & 0.4794 & 0.7314 \\   
    FFVAE+SASRec&  0.6506 & 0.4429 & 0.6990 & 0.5933 & 0.4290 & 0.6795 & 0.5500 & 0.3728 & 0.6017 & 0.6000 & 0.4280 & 0.6792 & 0.5700 & 0.3749 & 0.6050 & 0.5567 & 0.3631 & 0.5825 & 0.5765 & 0.3634 & 0.5891 \\
  AFRL+SASRec & \textbf{0.5942} & \textbf{0.5266} & \textbf{0.7767} & \textbf{0.5699} & \textbf{0.5263} & \textbf{0.7765} & \textbf{0.5133} & \textbf{0.5271} & \textbf{0.7773} & \textbf{0.5821} & \textbf{0.5211} & \textbf{0.7712} & \textbf{0.5537} & \textbf{0.5216} & \textbf{0.7717} & \textbf{0.5416} & \textbf{0.5212} & \textbf{0.7714} & \textbf{0.5691} & \textbf{0.5161} & \textbf{0.7663}  \\
    \hline
   BERT4Rec & 0.7119 & 0.5417 & 0.7900 & 0.6752 & 0.5417 & 0.7900 & 0.5863 & 0.5417 & 0.7900 & 0.6895 & 0.5417 & 0.7900 & 0.6331 & 0.5417 & 0.7900 & 0.6282 & 0.5417 & 0.7900 & 0.6543 & 0.5417 & 0.7900 \\
   SM+BERT4Rec &0.6305 & 0.4627 & 0.7246 & 0.5868 & 0.4689 & 0.7293 & 0.5762 & 0.4660 & 0.7273 & 0.6121 & 0.4645 & 0.7277 & 0.6042 & 0.4648 & 0.7277 & 0.5826 & 0.4641 & 0.7265 & 0.6011 & 0.4654 & 0.7279  \\
  PFRec+BERT4Rec  &  0.6020 & 0.4938 & 0.7449 & 0.5810 & 0.4891 & 0.7389 &0.5451 & 0.4964 & 0.7491 &  \textbf{0.5803} & 0.4938 & 0.7451 & 0.5724 & 0.4934 & 0.7456 & 0.5544 & 0.4709 & 0.7206 & 0.5737 & 0.4873 & 0.7323 \\
    FFVAE+BERT4Rec & 0.6638 & 0.4269 &0.6760&0.6519&0.4254 &0.6746 &0.5784 &0.3740 & 0.6045& 0.6462& 0.4174 &0.6644 &0.6027 & 0.3714& 0.6015&0.6014&0.3721& 0.6042&0.5972&0.3660 &0.5945\\
    AFRL+BERT4Rec  & \textbf{0.5879} & \textbf{0.5273} & \textbf{0.7772} & \textbf{0.5799} & \textbf{0.5263} & \textbf{0.7784} & \textbf{0.5082} & \textbf{0.5307} & \textbf{0.7810} &0.5839 & \textbf{0.5212} & \textbf{0.7725} & \textbf{0.5480} & \textbf{0.5220} & \textbf{0.7720} & \textbf{0.5440} & \textbf{0.5213} & \textbf{0.7730} & \textbf{0.5586} & \textbf{0.5111} & \textbf{0.7618}\\
    \bottomrule
    \end{tabular}%
}
 \label{tab:Taobao}
\end{table*}

\begin{figure*}[!t]
  \centering
    \subfigure[ML-1M G]{
    \begin{minipage}[t]{0.16\linewidth}
      \centering
      \includegraphics[width=\linewidth]{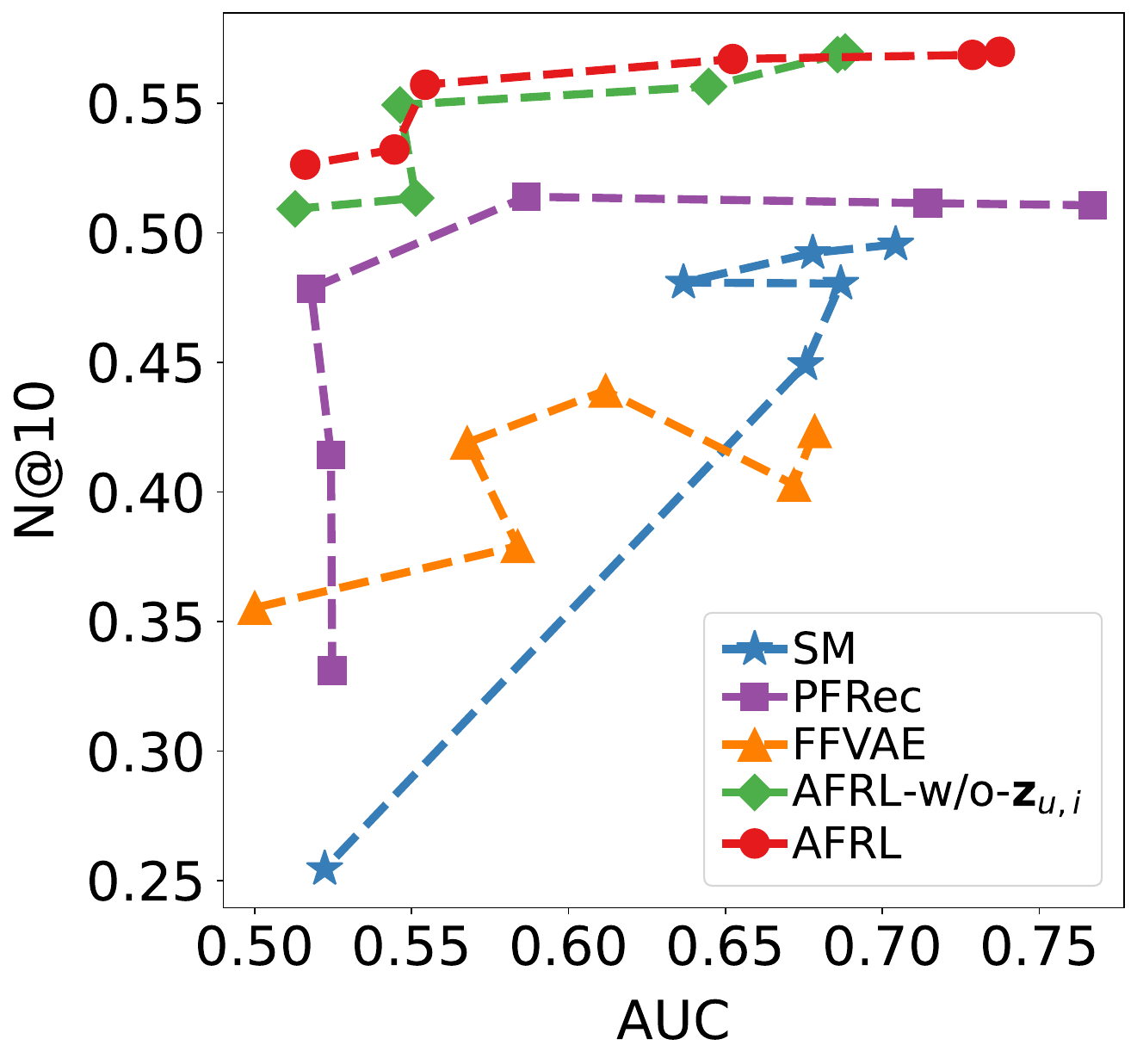}
    \end{minipage}%
  }%
    \subfigure[ML-1M A]{
    \begin{minipage}[t]{0.16\linewidth}
      \centering
      \includegraphics[width=\linewidth]{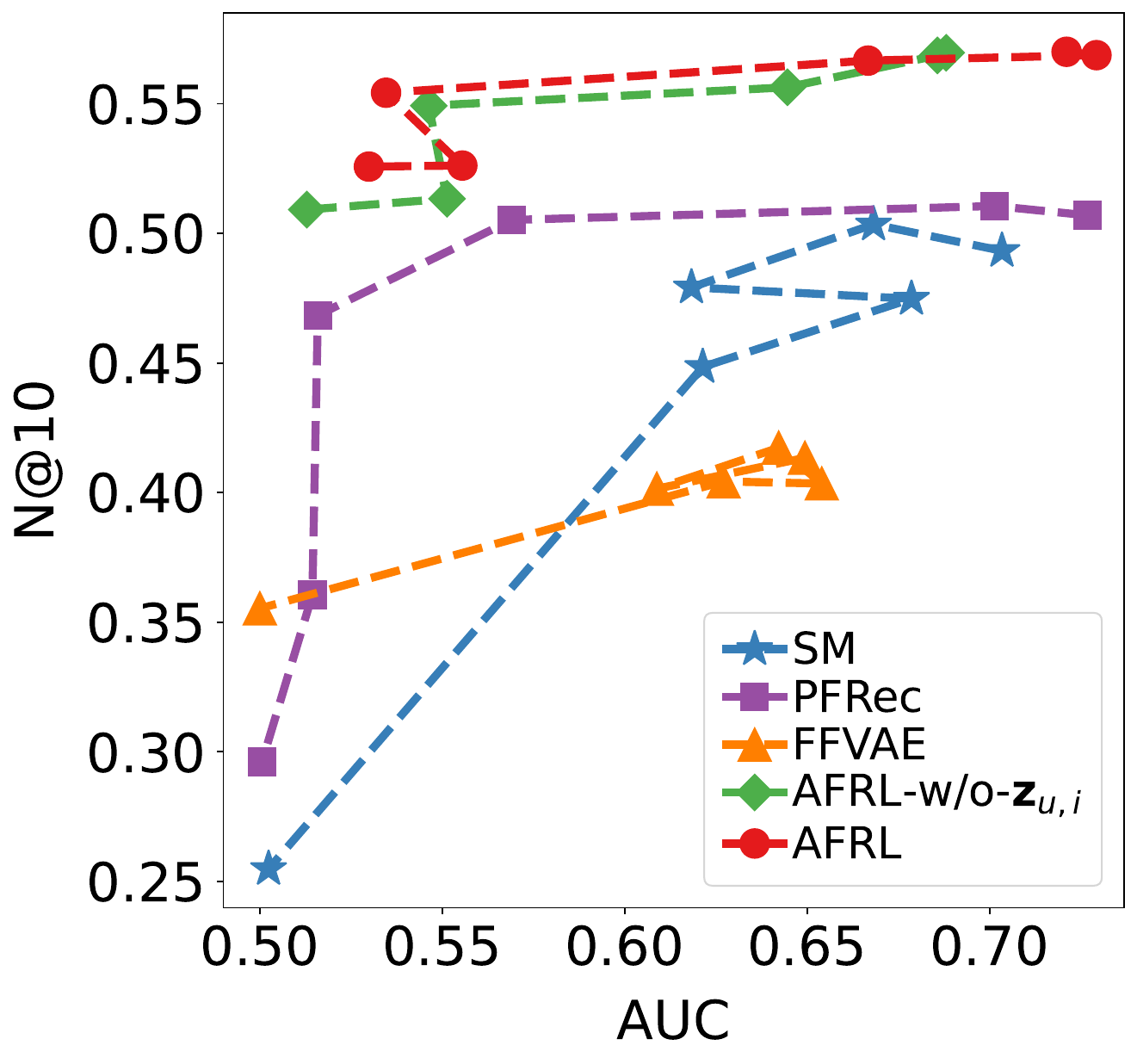}
    \end{minipage}%
  }%
  \subfigure[ML-1M O]{
    \begin{minipage}[t]{0.16\linewidth}
      \centering
      \includegraphics[width=\linewidth]{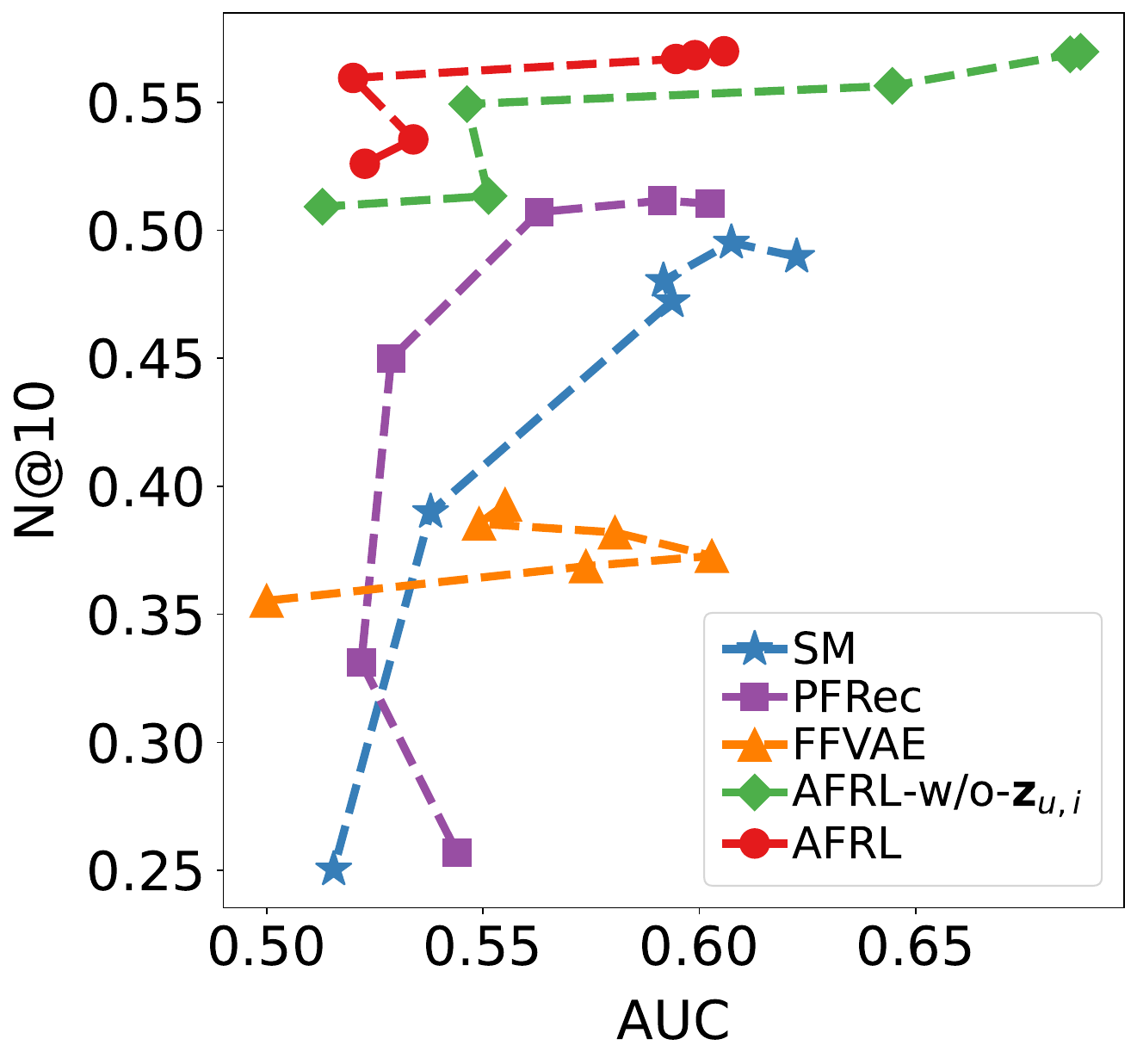}
    \end{minipage}%
  }%
\subfigure[ML-1M G+A]{
    \begin{minipage}[t]{0.16\linewidth}
      \centering
      \includegraphics[width=\linewidth]{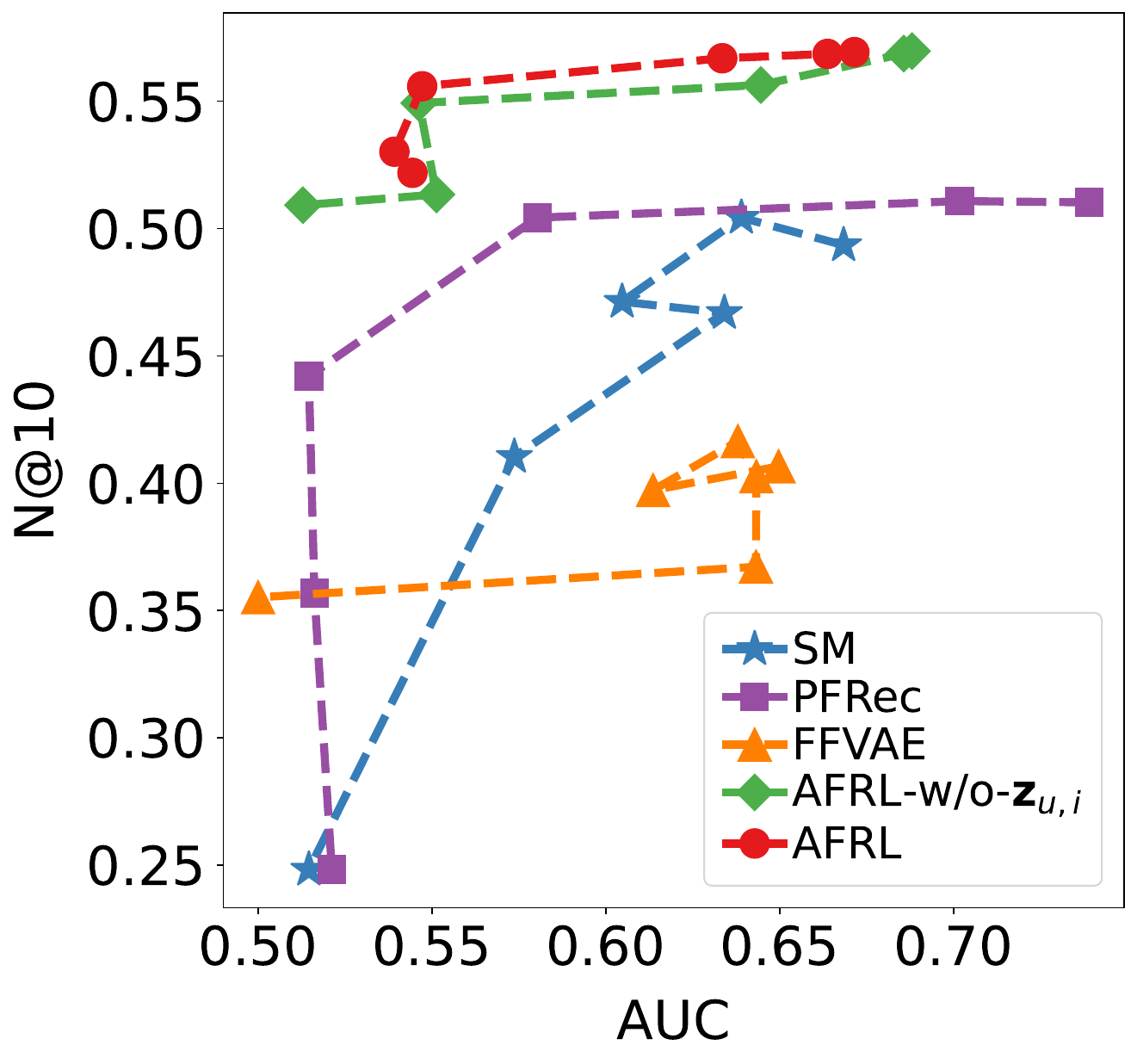}
    \end{minipage}%
  }%
\subfigure[ML-1M G+O]{
    \begin{minipage}[t]{0.16\linewidth}
      \centering
      \includegraphics[width=\linewidth]{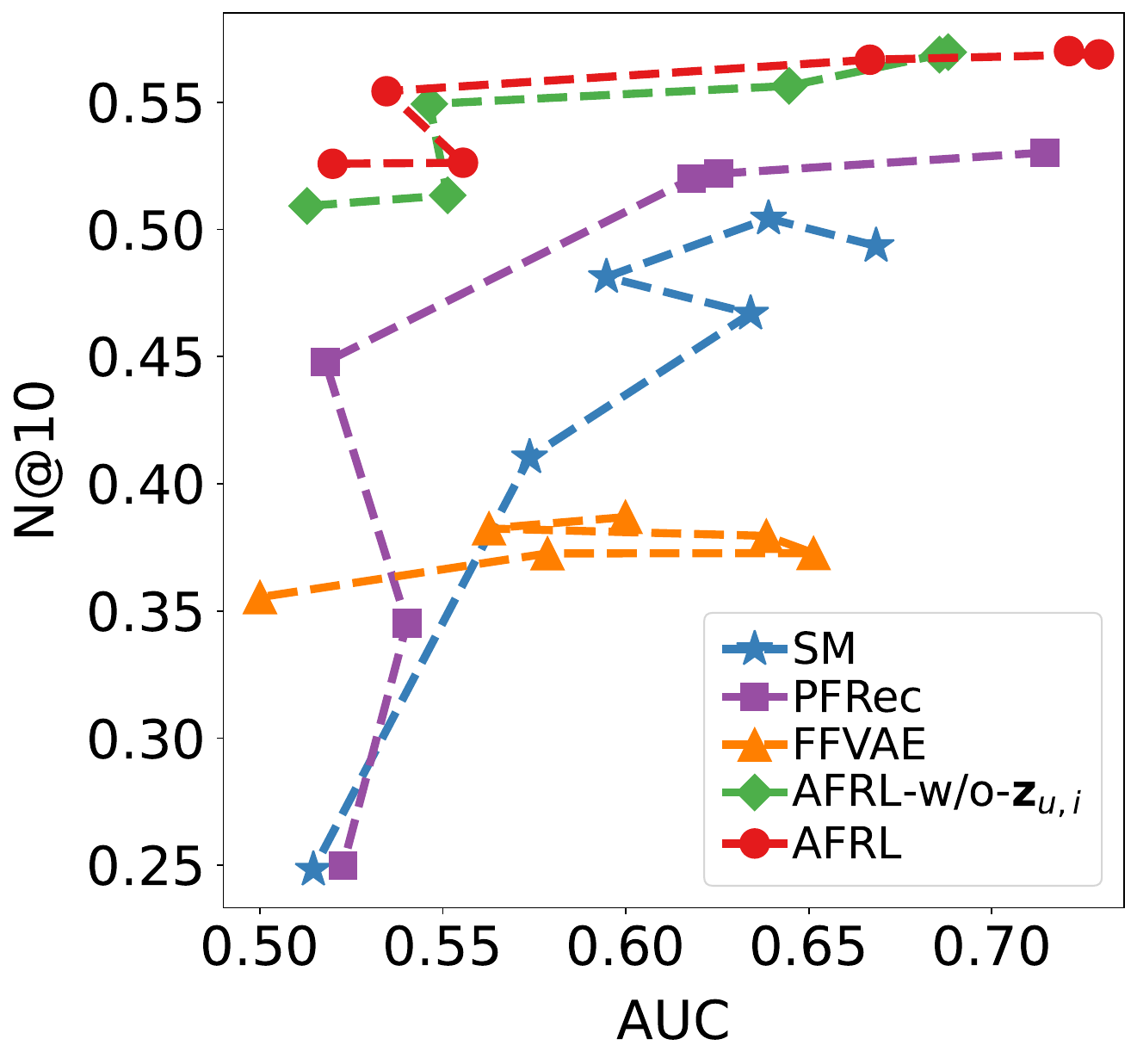}
    \end{minipage}%
  }%
\subfigure[ML-1M A+O]{
    \begin{minipage}[t]{0.16\linewidth}
      \centering
      \includegraphics[width=\linewidth]{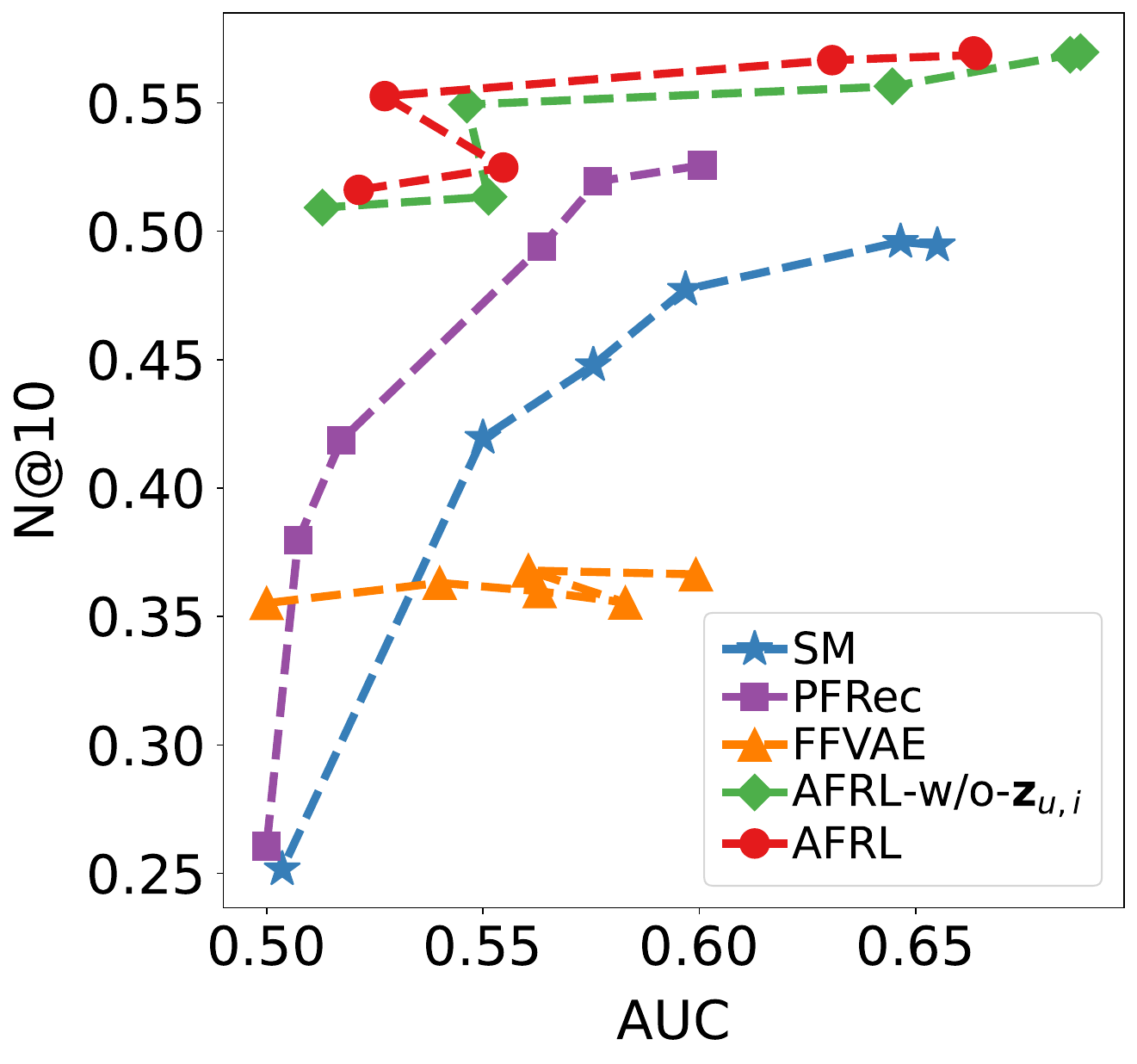}
    \end{minipage}%
  }%
  
  \subfigure[Taobao G]{
    \begin{minipage}[t]{0.16\linewidth}
      \centering
      \includegraphics[width=\linewidth]{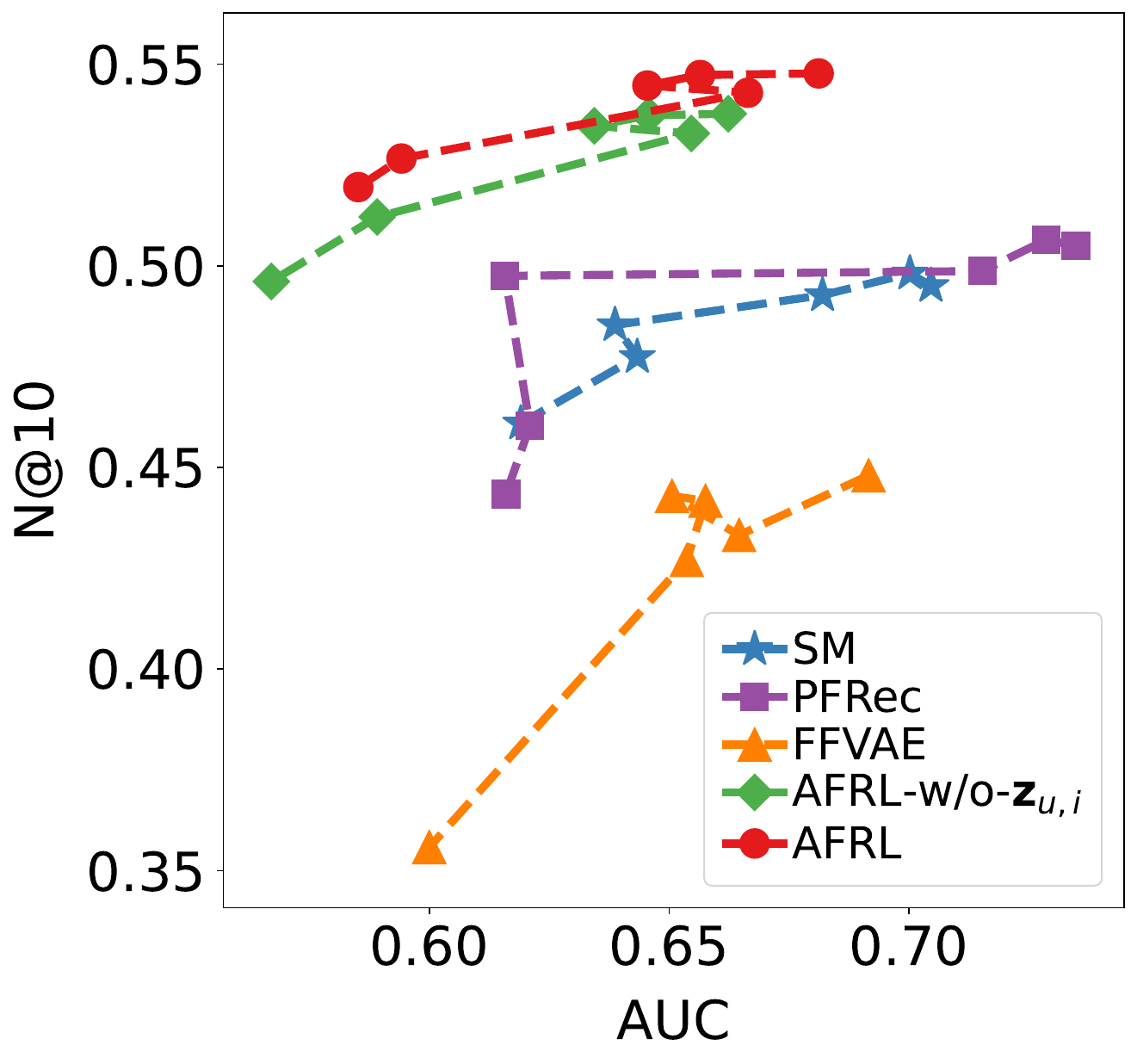}
    \end{minipage}%
  }%
    \subfigure[Taobao A]{
    \begin{minipage}[t]{0.16\linewidth}
      \centering
      \includegraphics[width=\linewidth]{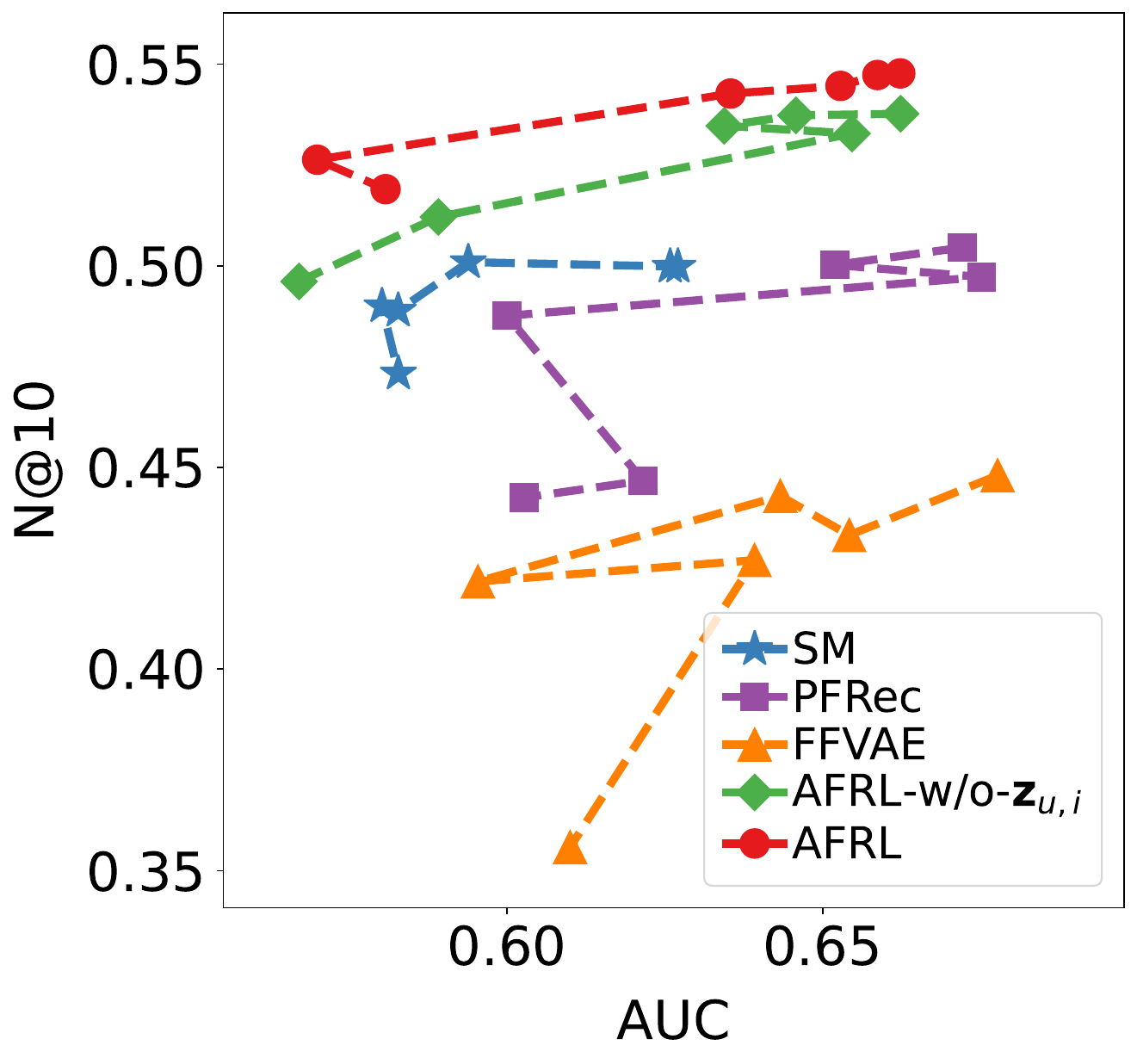}
    \end{minipage}%
  }%
  \subfigure[Taobao C]{
    \begin{minipage}[t]{0.16\linewidth}
      \centering
      \includegraphics[width=\linewidth]{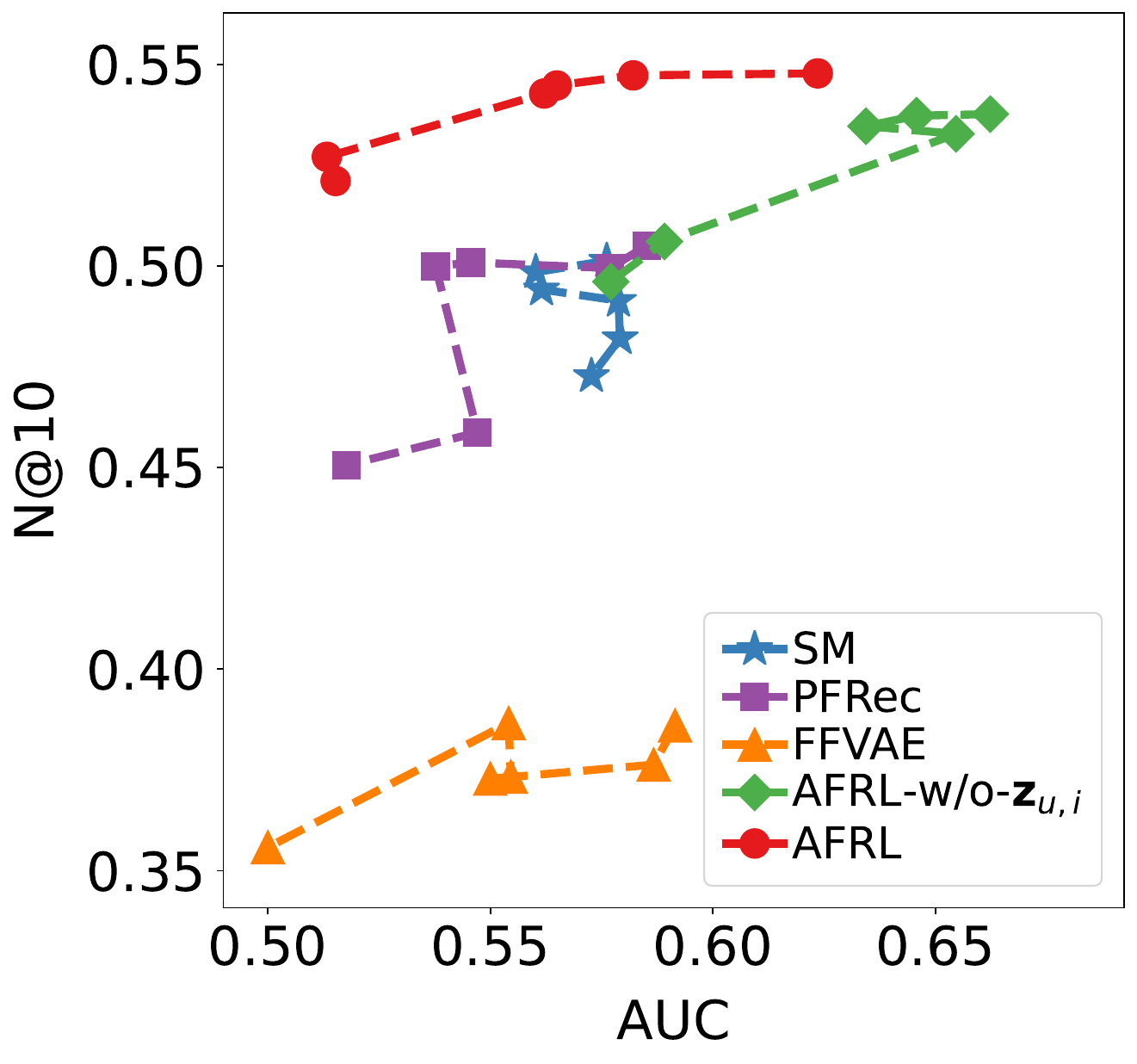}
    \end{minipage}%
  }%
\subfigure[Taobao G+A]{
    \begin{minipage}[t]{0.16\linewidth}
      \centering
      \includegraphics[width=\linewidth]{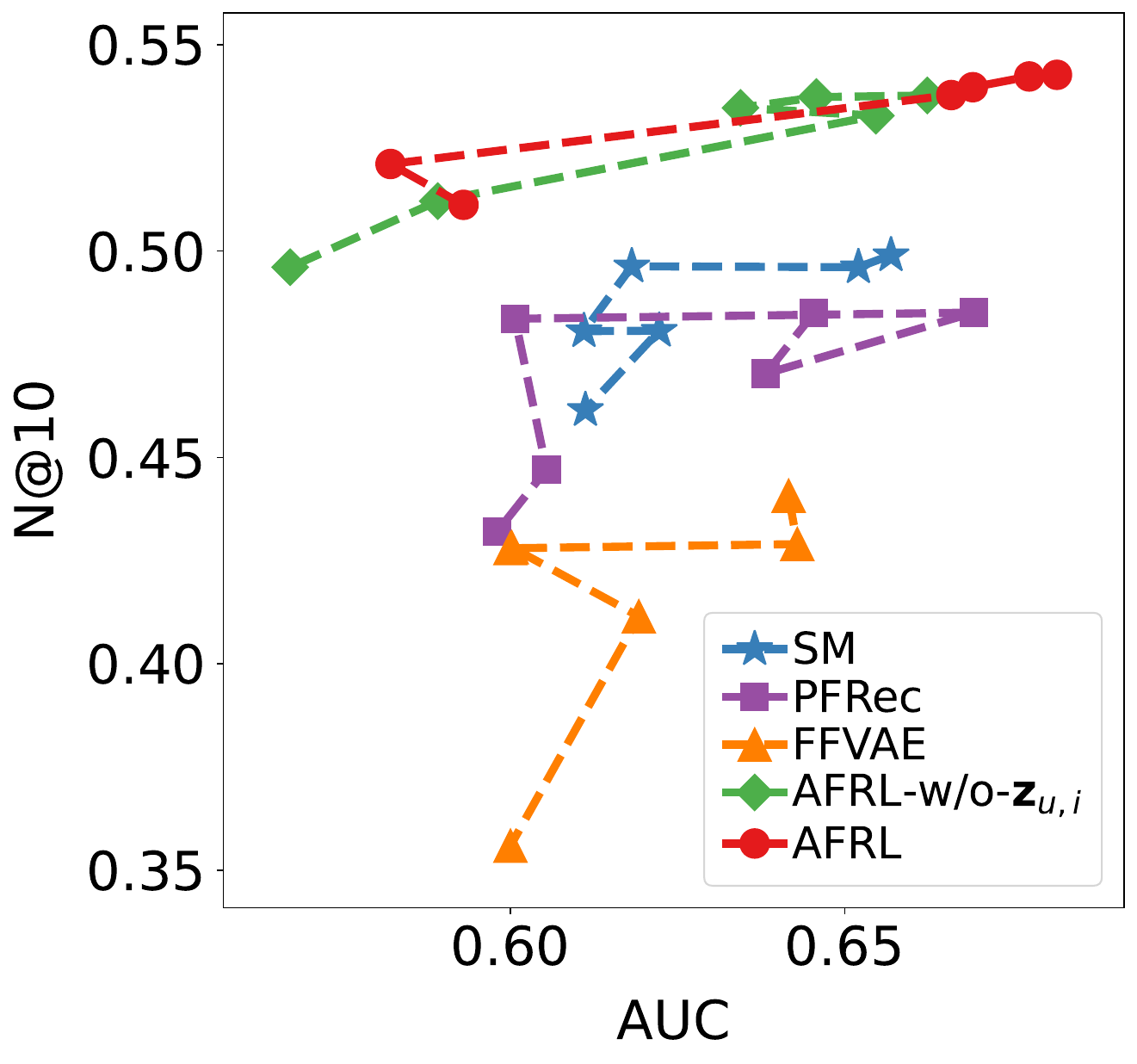}
    \end{minipage}%
  }%
\subfigure[Taobao G+C]{
    \begin{minipage}[t]{0.16\linewidth}
      \centering
      \includegraphics[width=\linewidth]{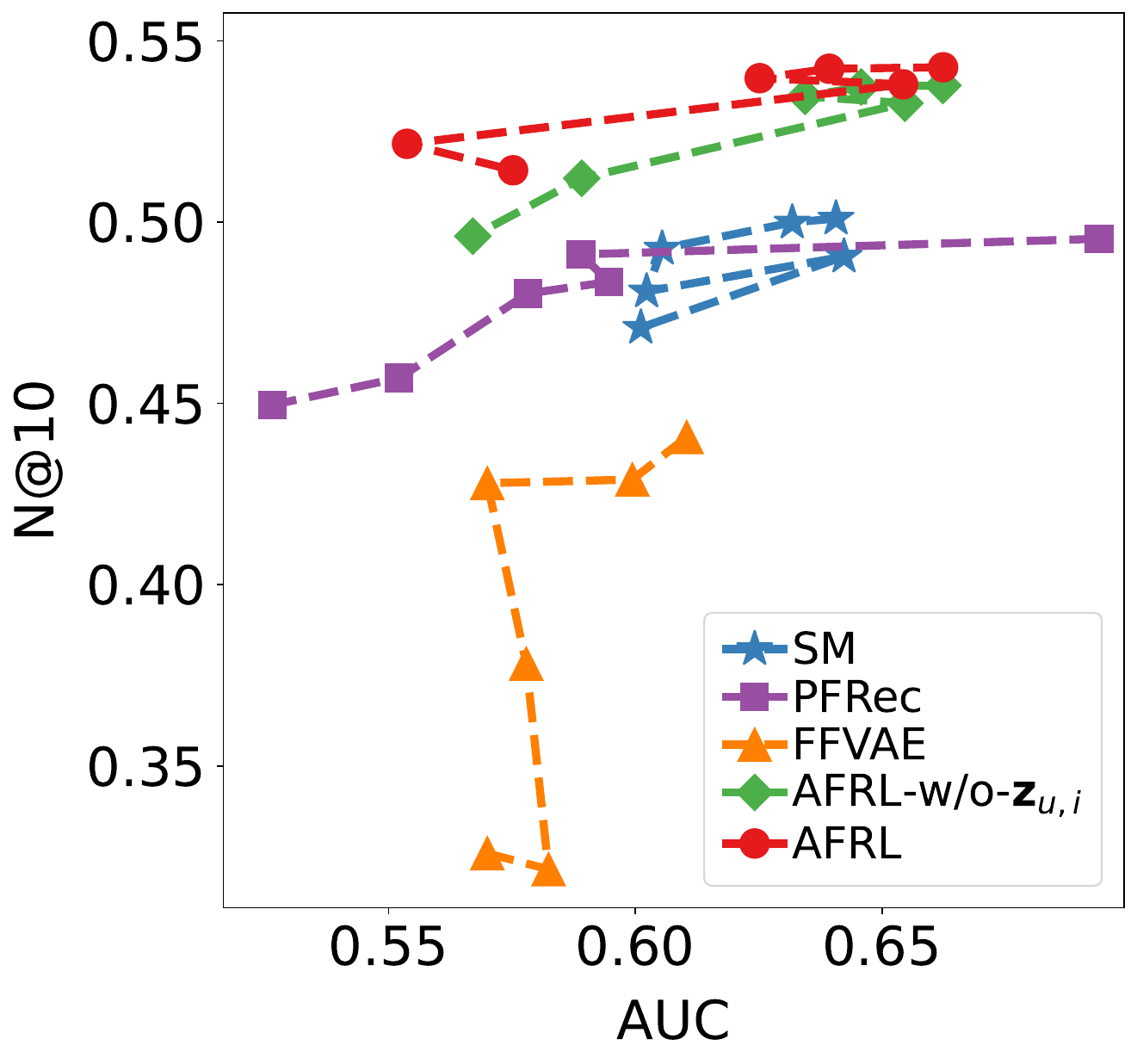}
    \end{minipage}%
  }%
\subfigure[Taobao A+C]{
    \begin{minipage}[t]{0.16\linewidth}
      \centering
      \includegraphics[width=\linewidth]{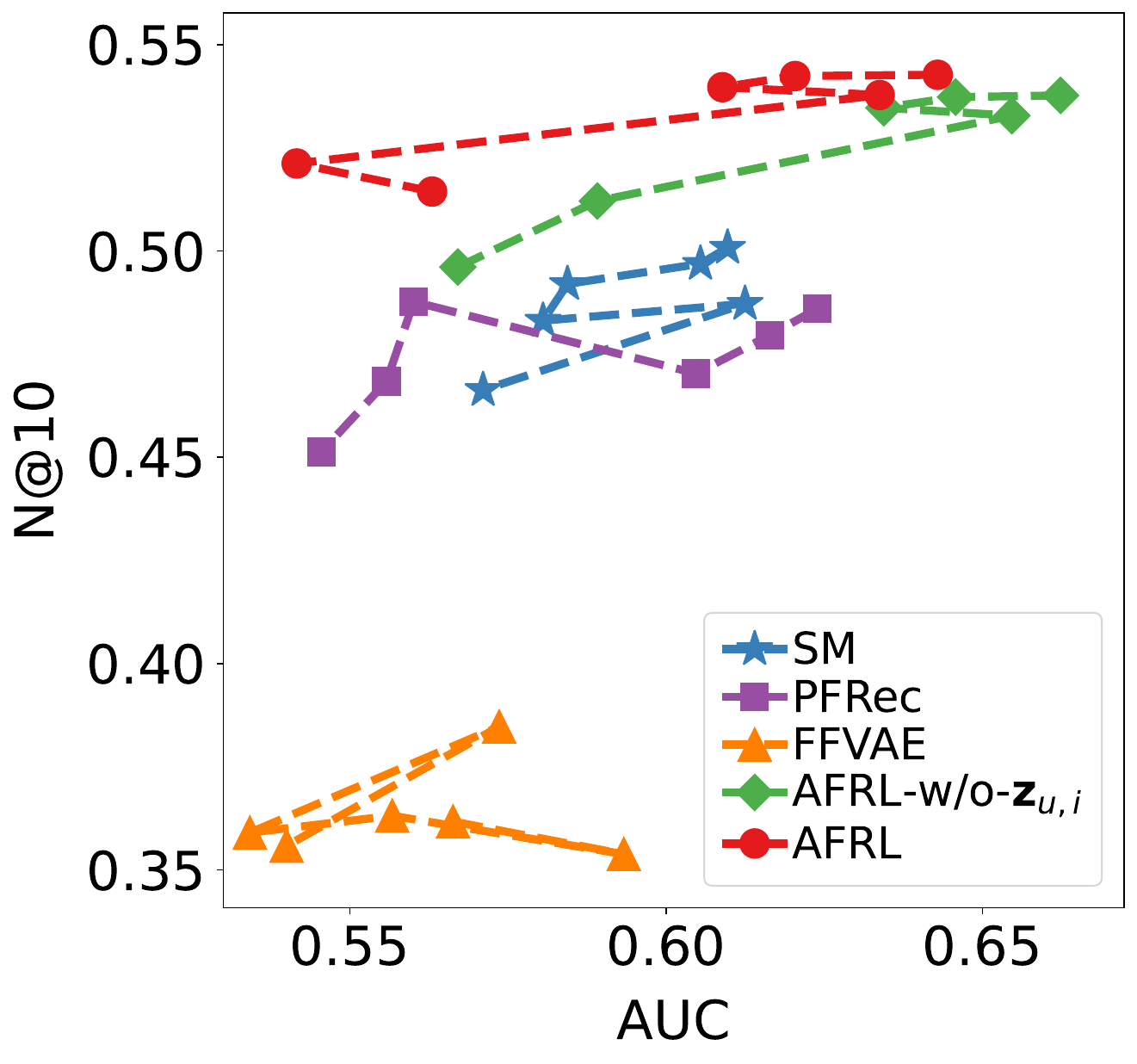}
    \end{minipage}%
  }%
  \centering
  \caption{Pareto front curves under different fairness requirements with SASRec \cite{kang2018self} as base model. In each curve, the points from left to right correspond to $\boldsymbol{\lambda}\in\{100,10,1,0.1,0.01,0.001\}$, respectively.}
  \label{Fig:pareto}
\end{figure*}

\begin{figure*}[t]
  \centering
    \subfigure[ML-1M G]{
    \begin{minipage}[t]{0.16\linewidth}
      \centering
      \includegraphics[width=\linewidth]{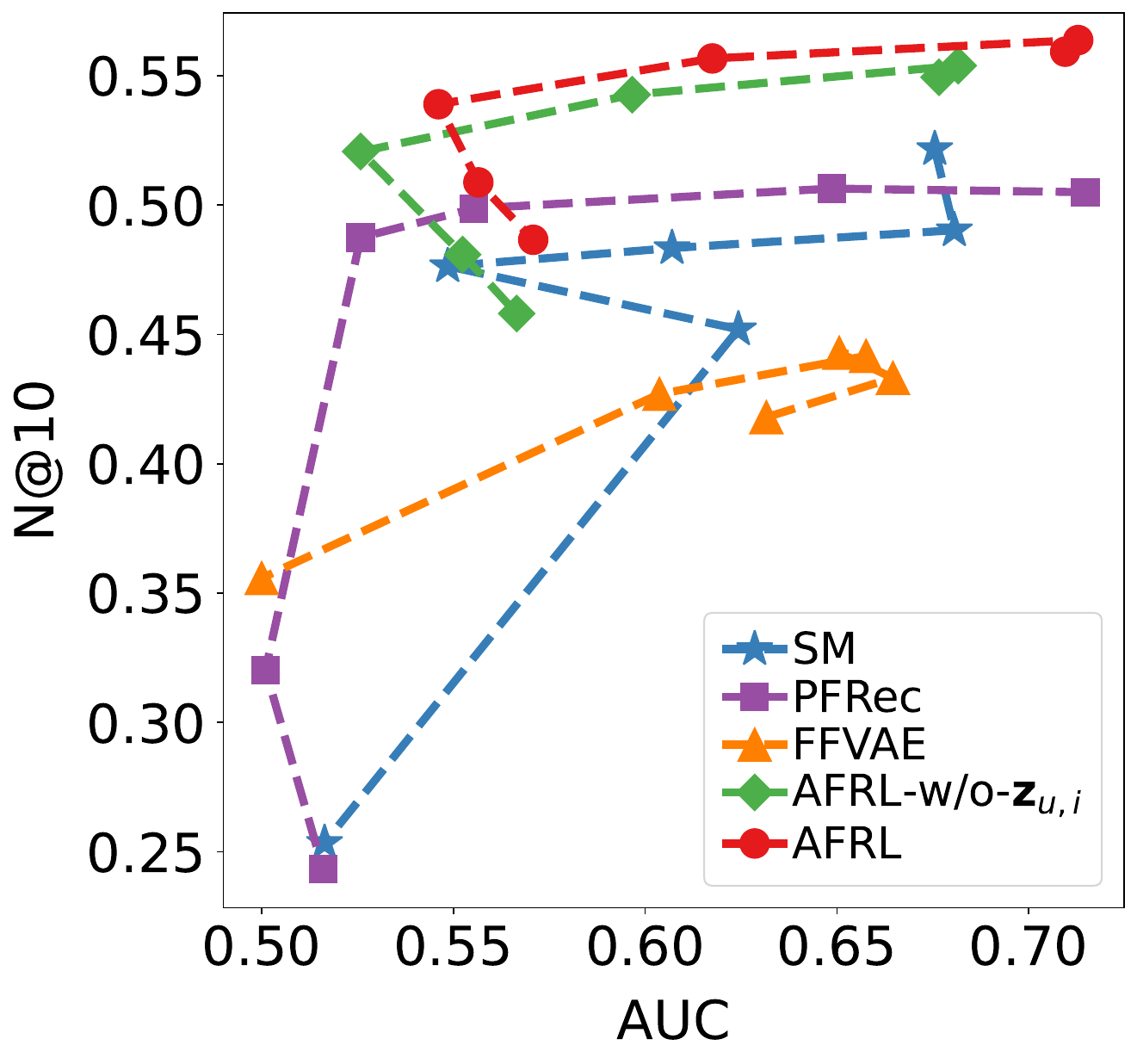}
    \end{minipage}%
  }%
    \subfigure[ML-1M A]{
    \begin{minipage}[t]{0.165\linewidth}
      \centering
      \includegraphics[width=\linewidth]{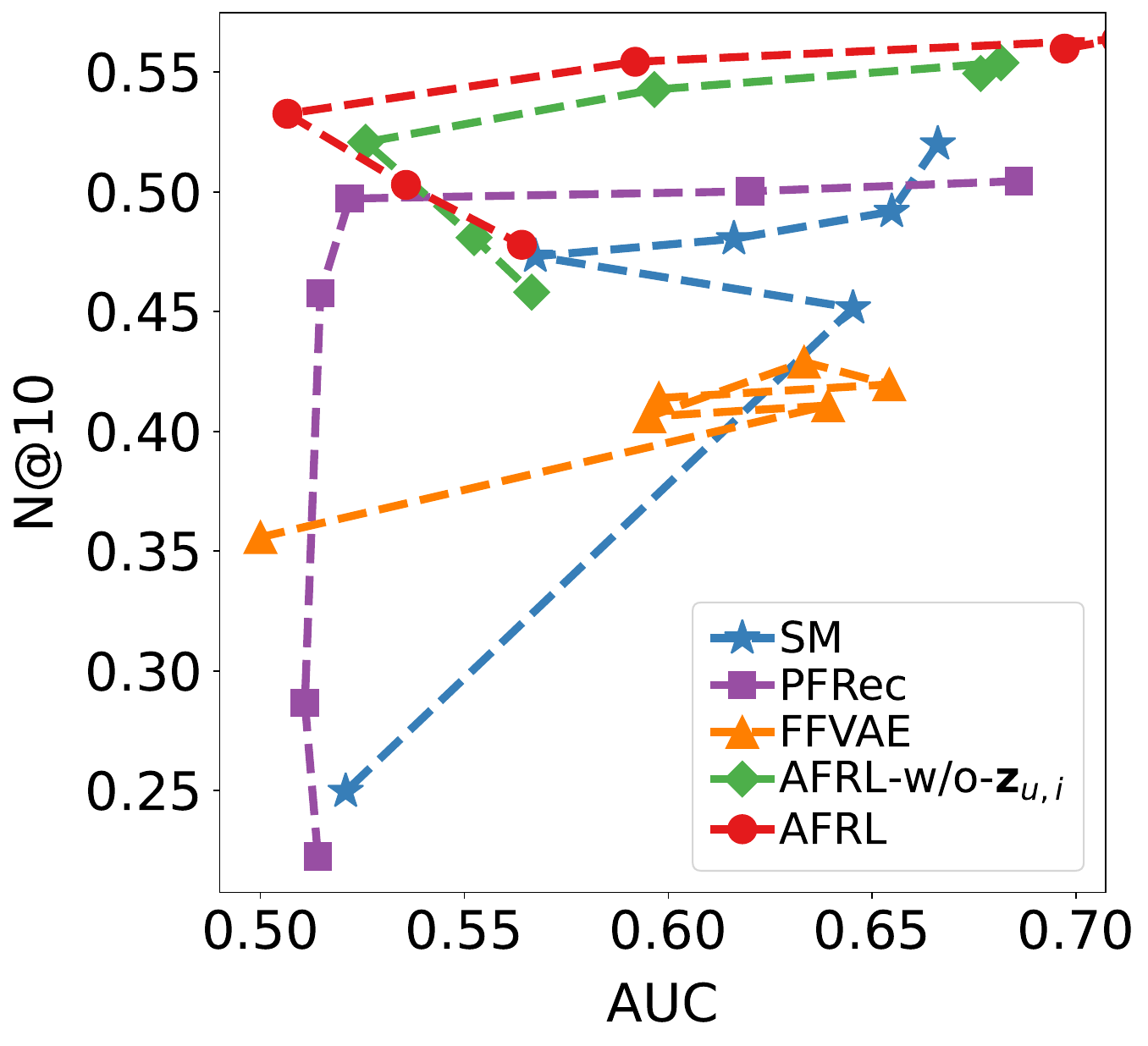}
    \end{minipage}%
  }%
  \subfigure[ML-1M O]{
    \begin{minipage}[t]{0.16\linewidth}
      \centering
      \includegraphics[width=\linewidth]{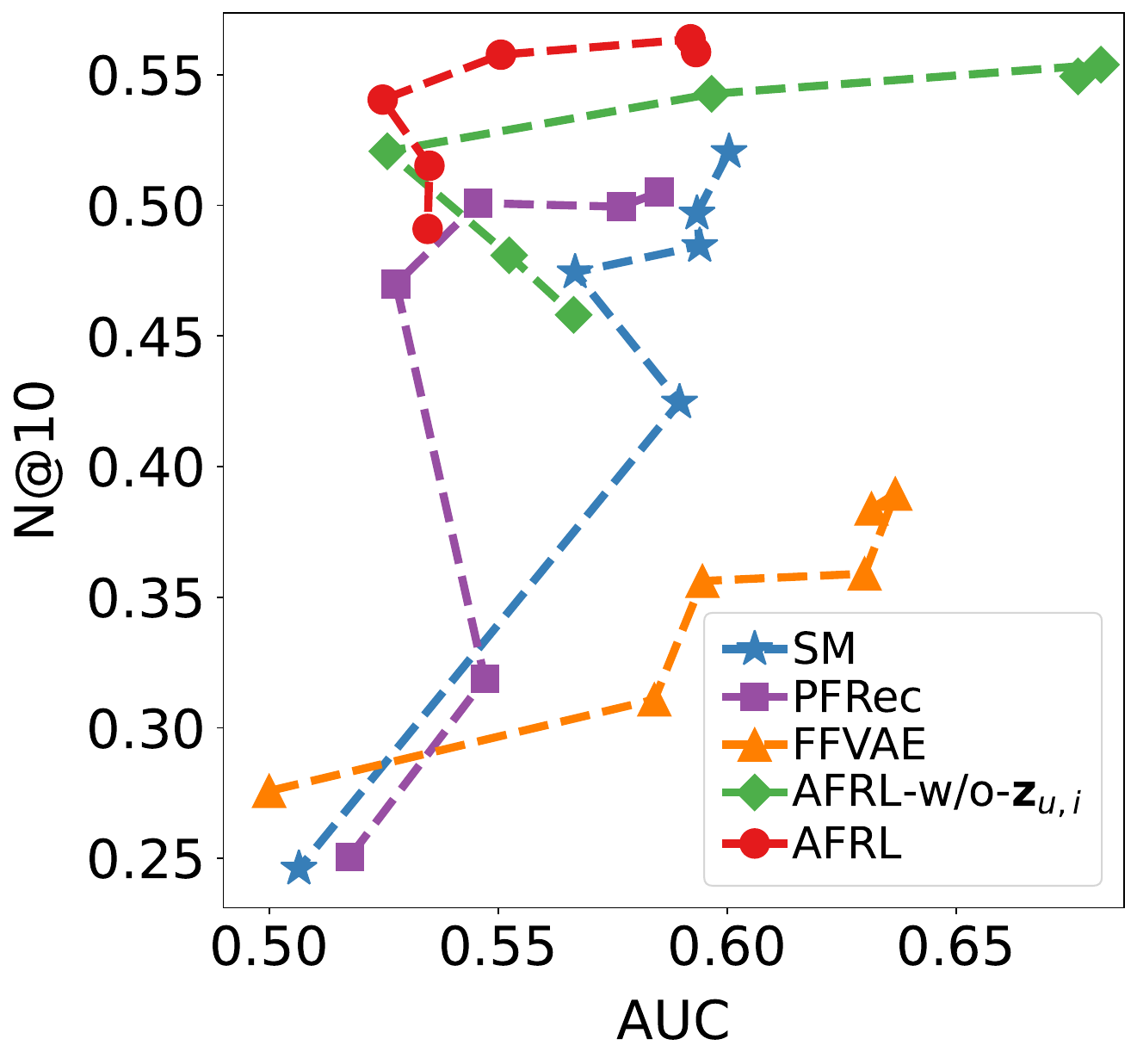}
    \end{minipage}%
  }%
\subfigure[ML-1M G+A]{
    \begin{minipage}[t]{0.16\linewidth}
      \centering
      \includegraphics[width=\linewidth]{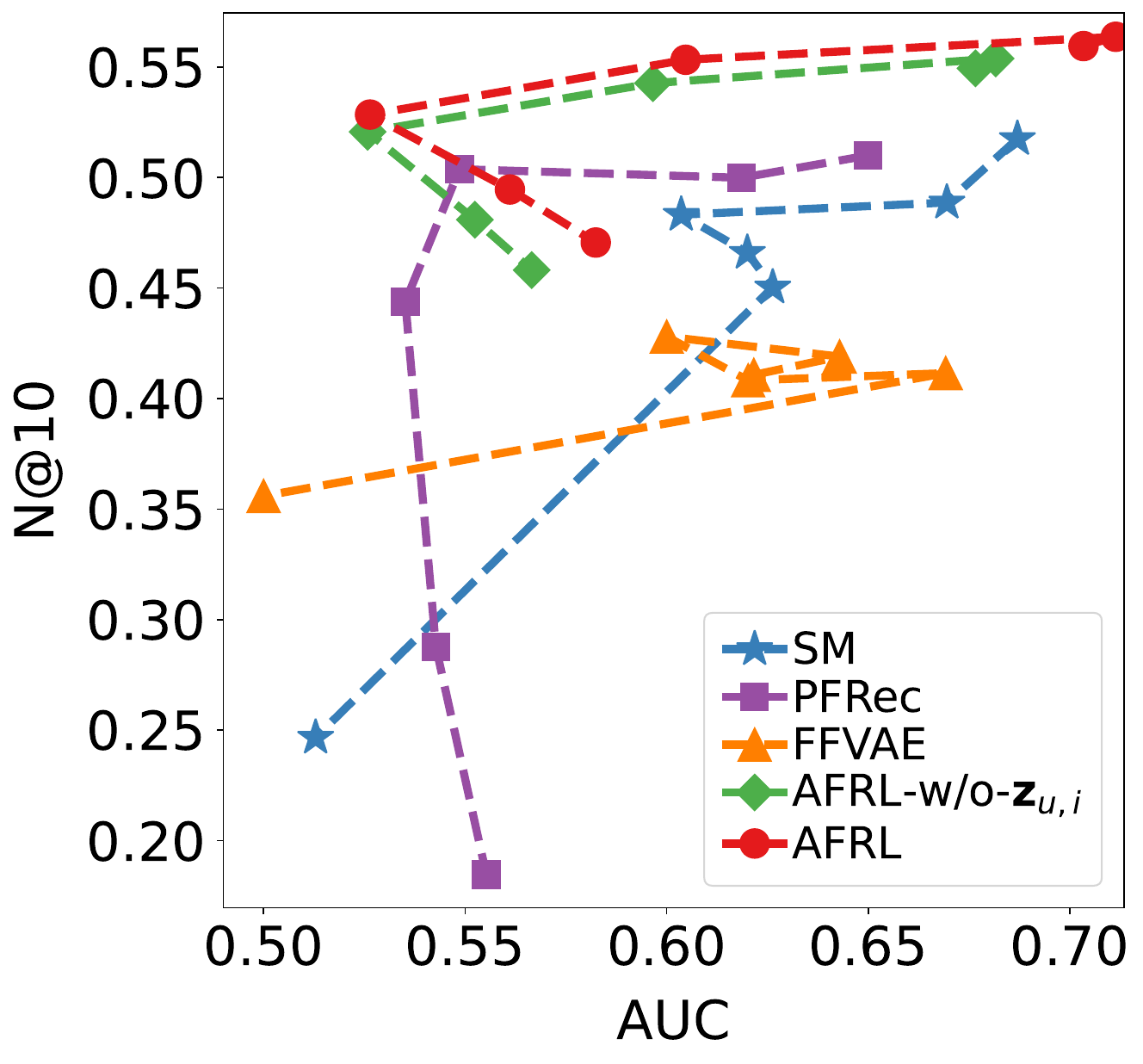}
    \end{minipage}%
  }%
\subfigure[ML-1M G+O]{
    \begin{minipage}[t]{0.16\linewidth}
      \centering
      \includegraphics[width=\linewidth]{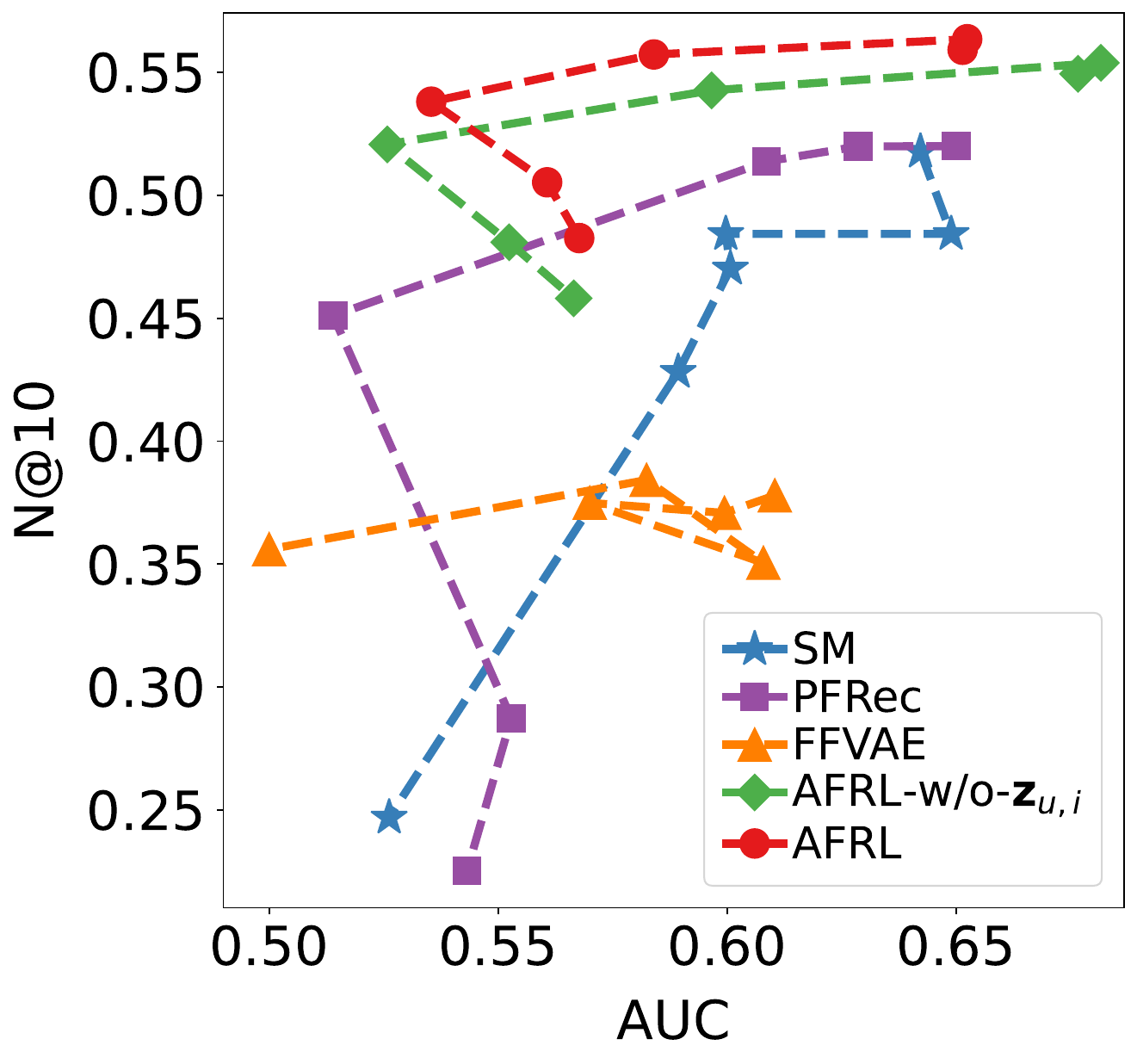}
    \end{minipage}%
  }%
\subfigure[ML-1M A+O]{
    \begin{minipage}[t]{0.16\linewidth}
      \centering
      \includegraphics[width=\linewidth]{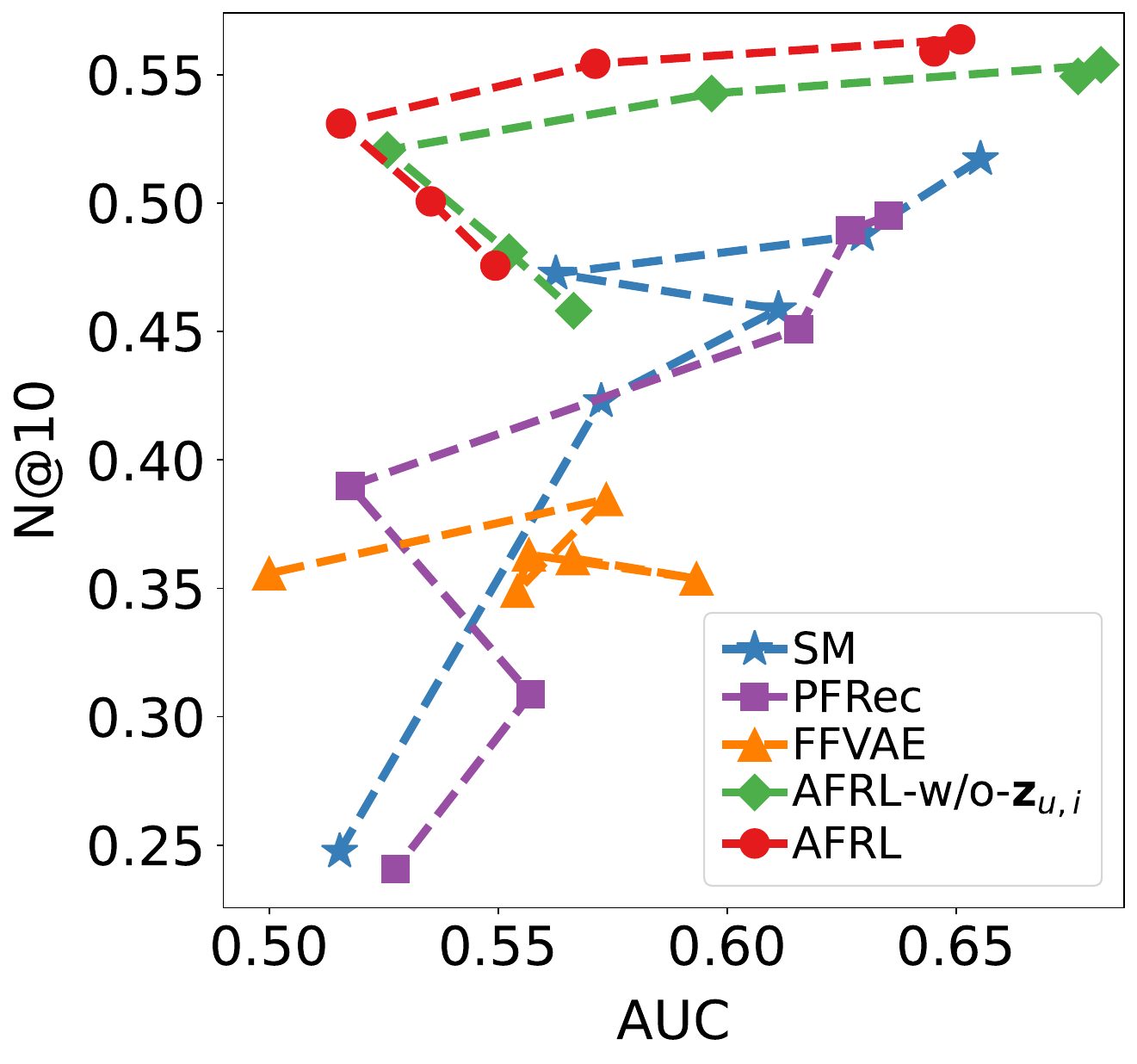}
    \end{minipage}%
  }%
  
  \subfigure[Taobao G]{
    \begin{minipage}[t]{0.16\linewidth}
      \centering
      \includegraphics[width=\linewidth]{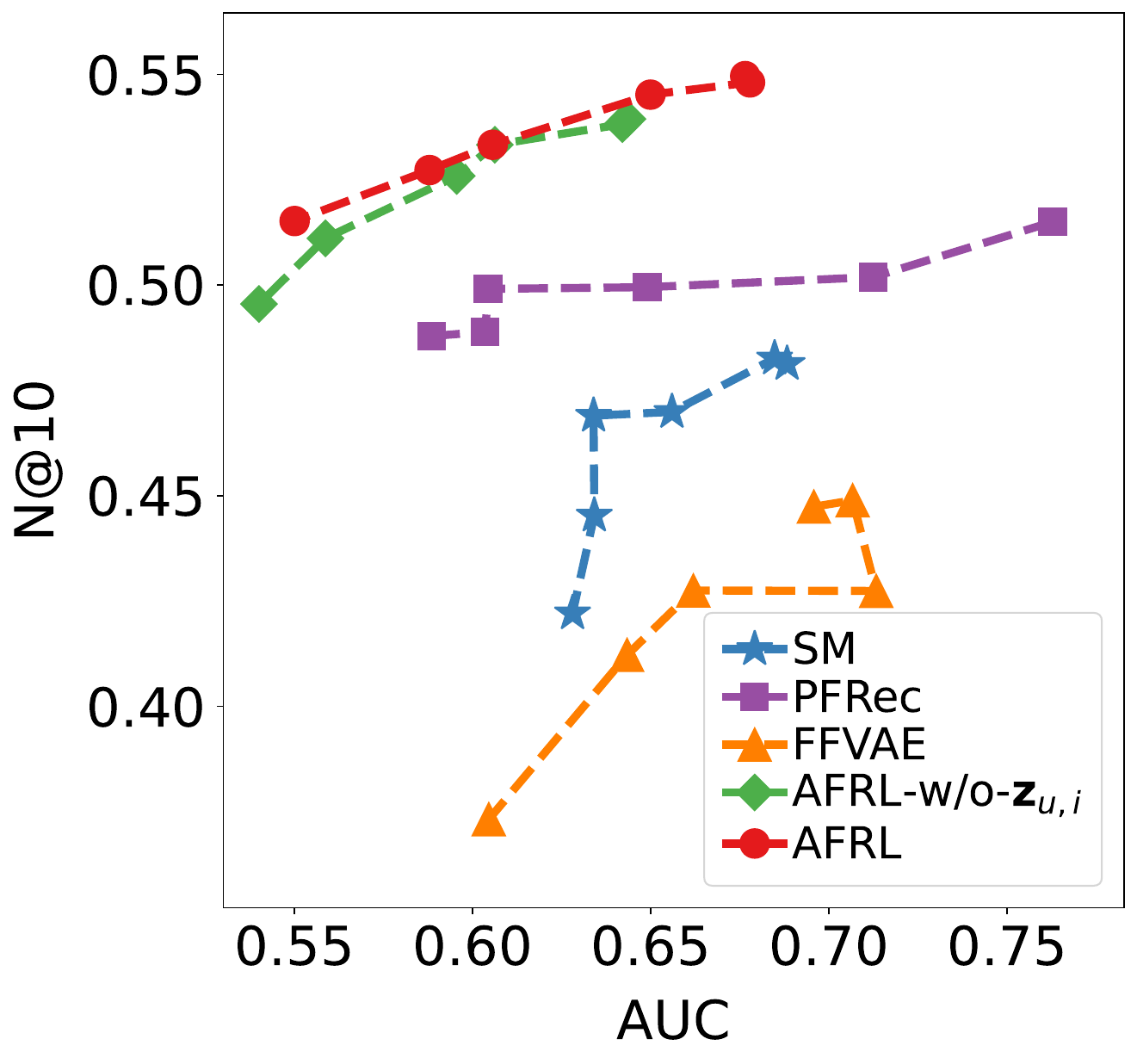}
    \end{minipage}%
  }%
    \subfigure[Taobao A]{
    \begin{minipage}[t]{0.16\linewidth}
      \centering
      \includegraphics[width=\linewidth]{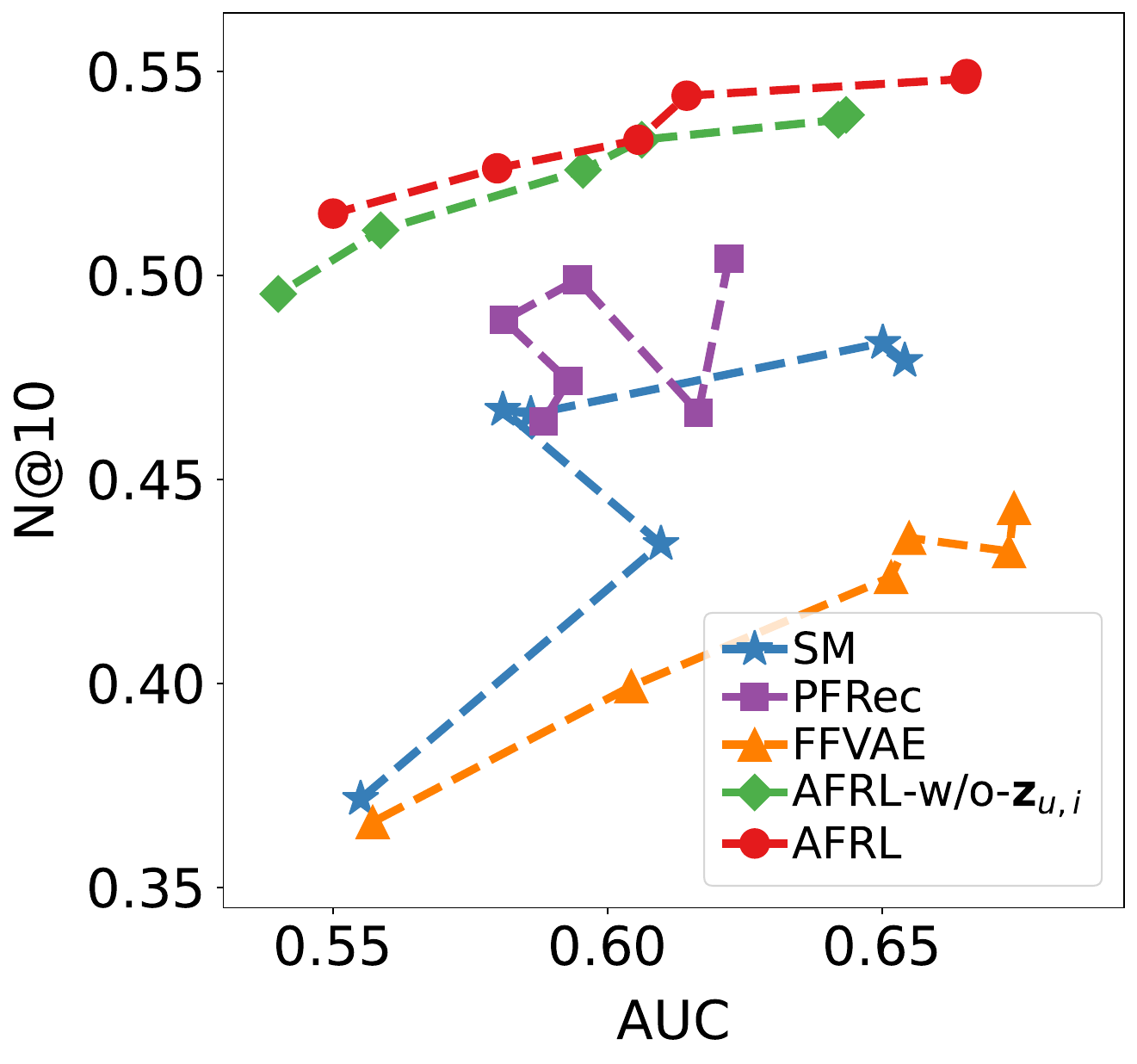}
    \end{minipage}%
  }%
  \subfigure[Taobao C]{
    \begin{minipage}[t]{0.16\linewidth}
      \centering
      \includegraphics[width=\linewidth]{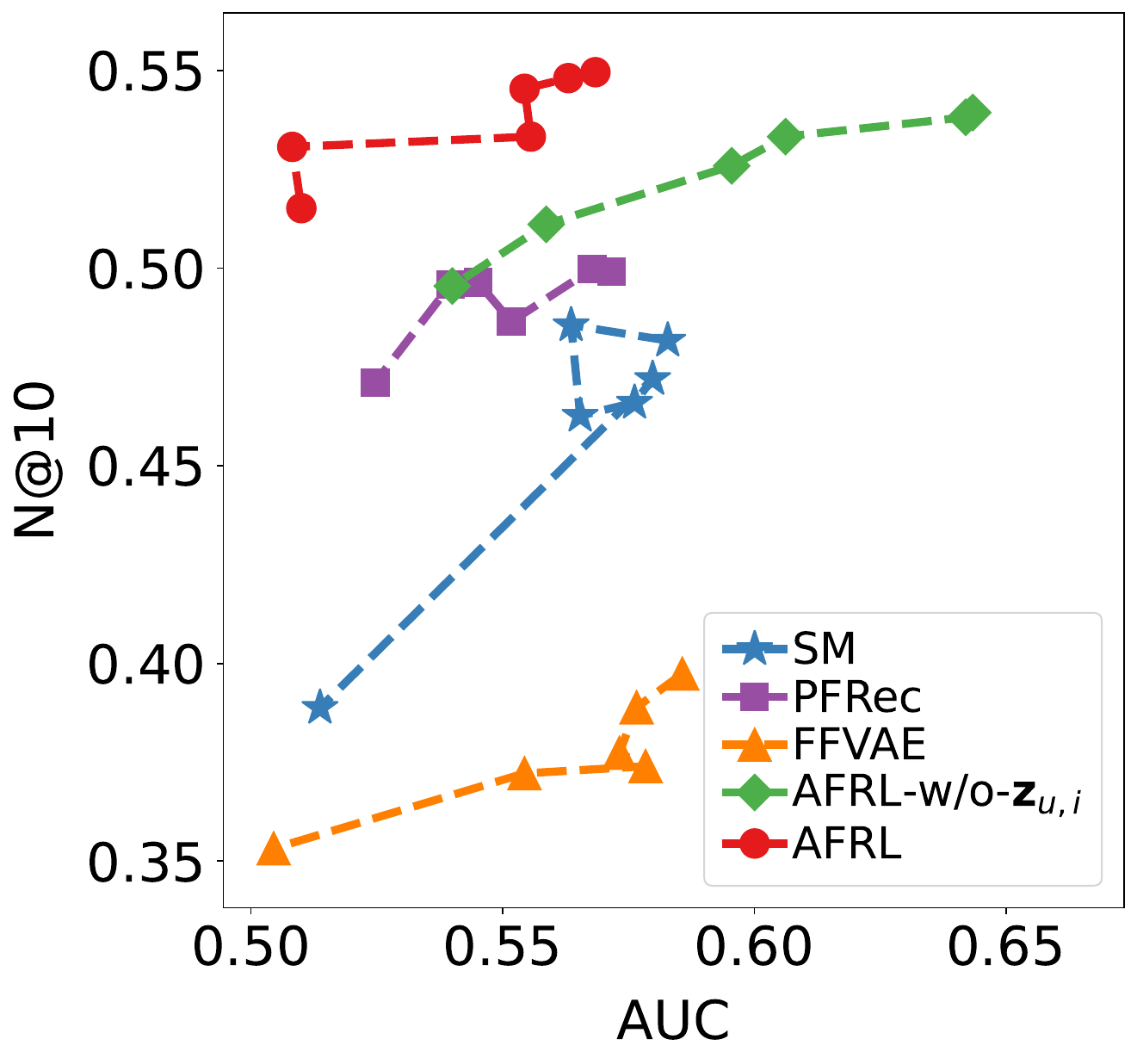}
    \end{minipage}%
  }%
\subfigure[Taobao G+A]{
    \begin{minipage}[t]{0.16\linewidth}
      \centering
      \includegraphics[width=\linewidth]{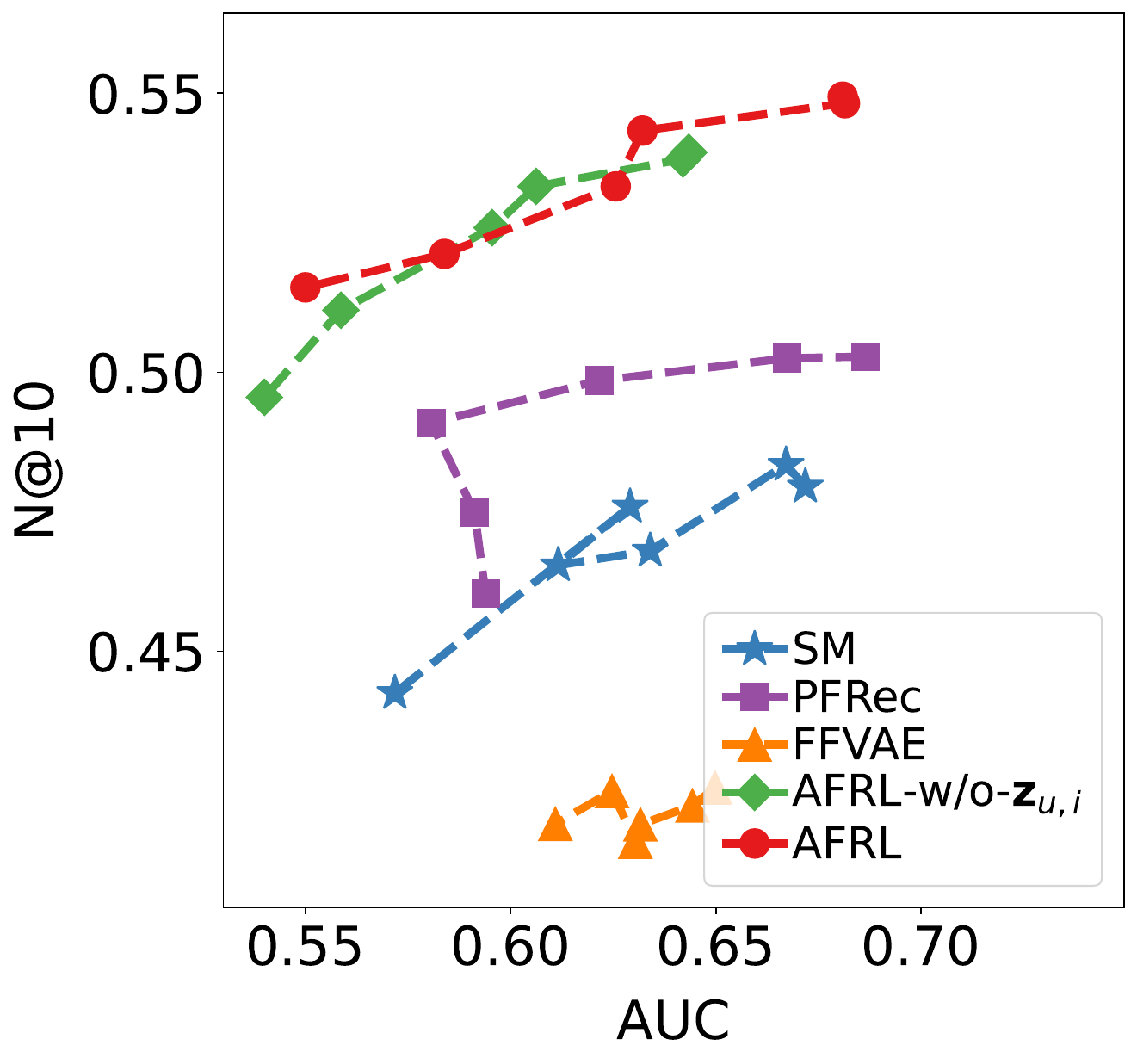}
    \end{minipage}%
  }%
\subfigure[Taobao G+C]{
    \begin{minipage}[t]{0.16\linewidth}
      \centering
      \includegraphics[width=\linewidth]{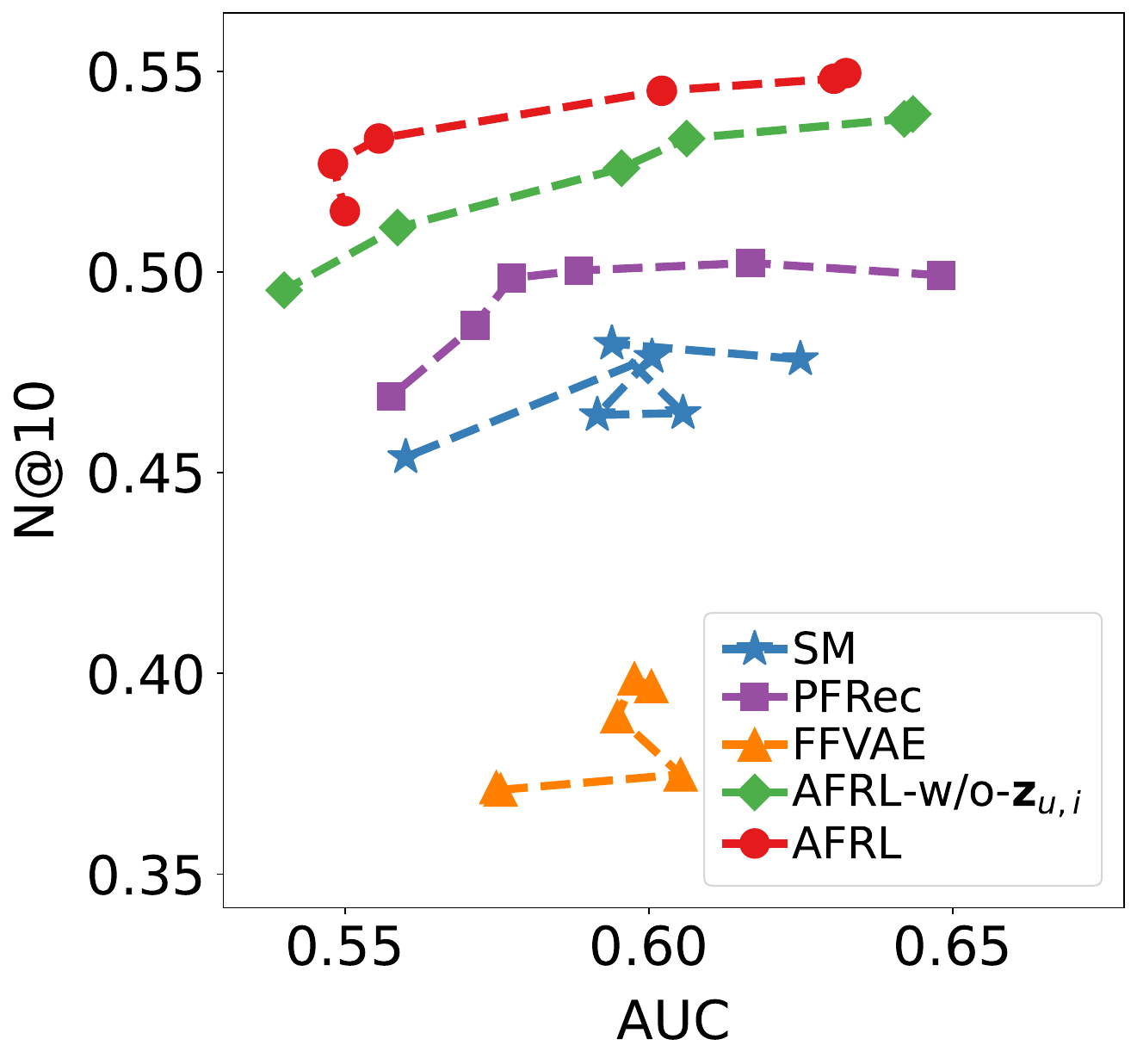}
    \end{minipage}%
  }%
\subfigure[Taobao A+C]{
    \begin{minipage}[t]{0.16\linewidth}
      \centering
      \includegraphics[width=\linewidth]{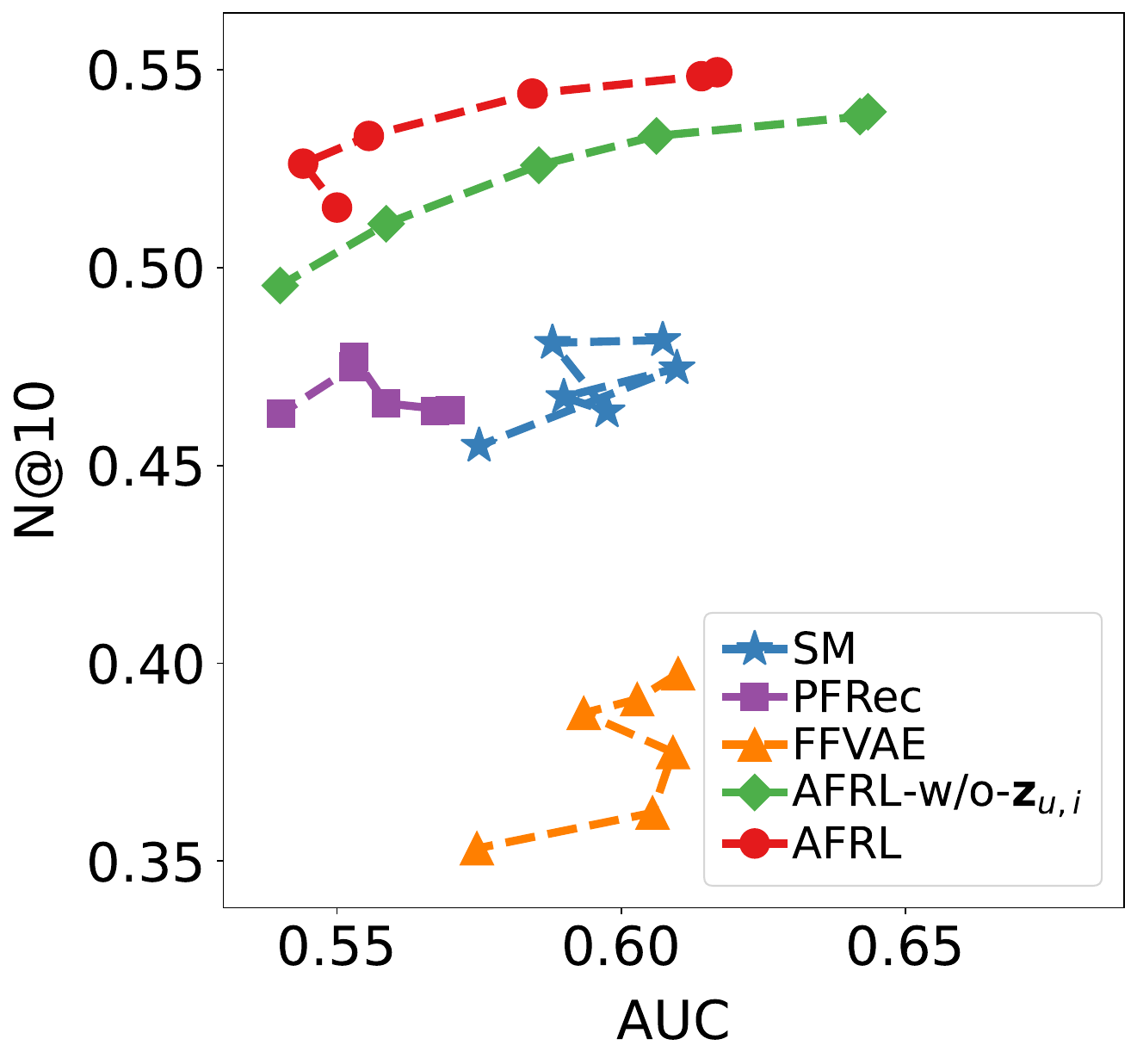}
    \end{minipage}%
  }%
  \centering
  \caption{Pareto front curves under different fairness requirements with BERT4Rec \cite{sun2019bert4rec} as base model. In each curve, the points from left to right correspond to $\boldsymbol{\lambda}\in\{100,10,1,0.1,0.01,0.001\}$, respectively.}
  \label{Fig:pareto_bert}
\end{figure*}

 \subsubsection{Parameter Setting.}
We set the batch size to 256, the initial learning rate to 5e-5, and the embedding dimensionality to 64 on both datasets. We employ early stopping as the regularizer for avoiding overfitting, and Adam algorithm as the optimizer. $F$ and $\{E_i\}$ are implemented as a six-layer MLP using ReLU as the activation function. $\{D_i\}$ and $\{C_i\}$ are implemented as a two-layer MLP using ReLU as the activation function. Table \ref{tab:hyperparameter_set} gives the settings of the hyper-parameters $\beta$ and $\lambda$ in Equations (\ref{Eq:loss_ei}) and (\ref{Eq:Di}) in different scenarios. For fairness, the hyper-parameters of the baseline models are set to their optimal configuration tuned on validation sets.

\subsection{Fairness and Accuracy (RQ1)}
Tables \ref{tab:ml-1m} and \ref{tab:Taobao} show the performaces of AFRL and the baseline models on ML-1M and Taobao, respectively. We can see that on both datasets, compared with the AUC of original model, the AUC after applying any fair model decreases, which means all the fair models can improve the fairness of the original recommendation models. However, the AUC after applying AFRL is closest to 0.5 for most fairness requirements, by which we can conclude that compared with the baseline fair models, AFRL can achieve the best personalized fairness of a recommendation model, due to its capability of learning the attribute-specific embeddings of attributes based on information alignment. 

For accuracy, we can observe that the applying of AFRL and the baseline fair models will reduce the accuracy of the original recommendation models in terms of NDCG and Hit rate, which is reasonable because of the loss of information in the fair embeddings learned by the fair models. However, we see that in most cases, AFRL outperforms the baseline fair models, which indicates that AFRL can achieve less accuracy loss, due to its ability to capture more discriminative information from non-sensitive attributes and collaborative signals in the fair representations learned by it.

\begin{figure}[t]
  \centering
  \subfigure[ML-1M]{
    \begin{minipage}[t]{0.32\linewidth}
      \centering
      \label{Fig:ablation_ml1m}
      \includegraphics[width=\linewidth]{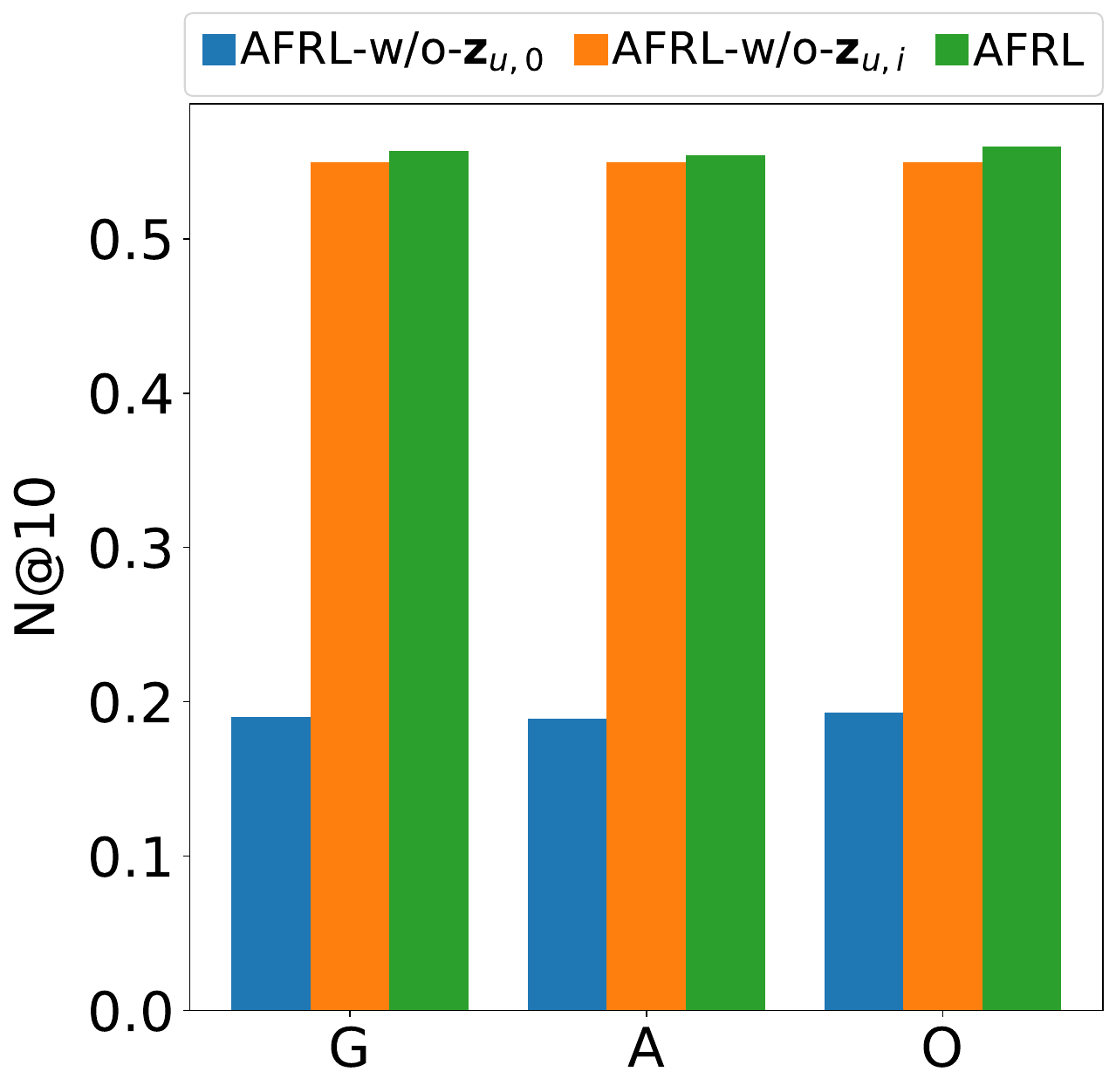}
    \end{minipage}%
  }%
  \subfigure [Taobao]{
    \begin{minipage}[t]{0.32\linewidth}
      \centering
      \label{Fig:ablation_taobao}
      \includegraphics[width=\linewidth]{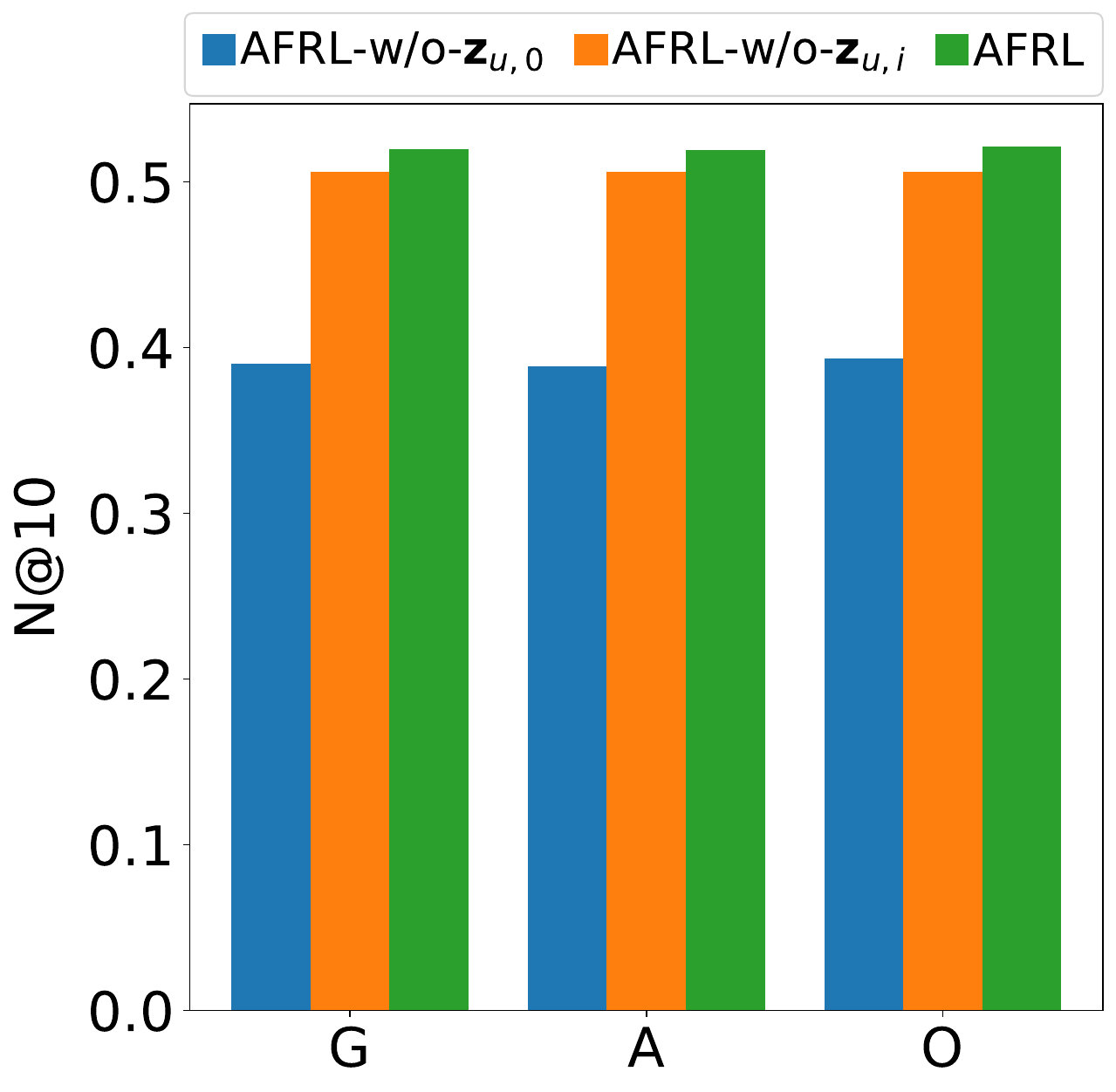}
    \end{minipage}%
  }%
  \centering
  \caption{Contributions to accuracy by $\mathbf{z}_{u, 0}$ and $\mathbf{z}_{u, i}$.}
  \label{Fig:ablation}
\end{figure}

\subsection{Trade-off (RQ2)}
AFRL uses $\lambda$ in Equation (\ref{Eq:Di}) to control the trade-off between fairness and accuracy, similar to the baseline fair models. Figures \ref{Fig:pareto} and \ref{Fig:pareto_bert} show the pareto front curves of AFRL and the baseline methods over different $\lambda$ and different fairness requirements on both datasets, with SASRec \cite{kang2018self} and BERT4Rec \cite{sun2019bert4rec} as the base recommendation models, respectively.

We can first see that in most cases, the point of the curve of AFRL at a specific $\lambda$ basically locates in the uppermost-left position compared to all other points, i.e., there are no points of the curves of the baseline models in its upper-left region. This result confirms the advantage of AFRL to achieve a better trade-off between fairness and accuracy than the baseline fair models. At the same time, we also note that as $\lambda$ is decreasing, the points of the curve of AFRL move to upper-right region, i.e., the accuracy improves while the fairness degrades. This is because in Equation (\ref{Eq:Di}), the smaller $\lambda$ will cause more information from original user embedding to be preserved in the fair embedding, which is favorable to its discriminability but unfavorable to its fairness.

\subsection{Ablation Study (RQ3)}
Now we investigate the contributions of the attribute-specific embeddings $\mathbf{z}_{u, i}$ ($1 \le i \le M$) and the debiased collaborative embedding $\mathbf{z}_{u, 0}$. For this purpose, we compare AFRL with its two variants, AFRL-w/o-$\mathbf{z}_{u, i}$ where the embeddings $\mathbf{z}_{u, i}$ of non-sensitive attributes are removed from the fair user embedding $\mathbf{u}^*$, and AFRL-w/o-$\mathbf{z}_{u, 0}$ where the debiased collaborative embedding $\mathbf{z}_{u, 0}$ is removed from $\mathbf{u}^*$. The results are shown in Figures \ref{Fig:pareto}, \ref{Fig:pareto_bert} and \ref{Fig:ablation}.

\begin{figure}[t]
  \centering
   \subfigure [$\beta$]{
    \begin{minipage}[t]{0.28\linewidth}
      \centering
      \label{Fig:beta}
      \includegraphics[width=\linewidth]{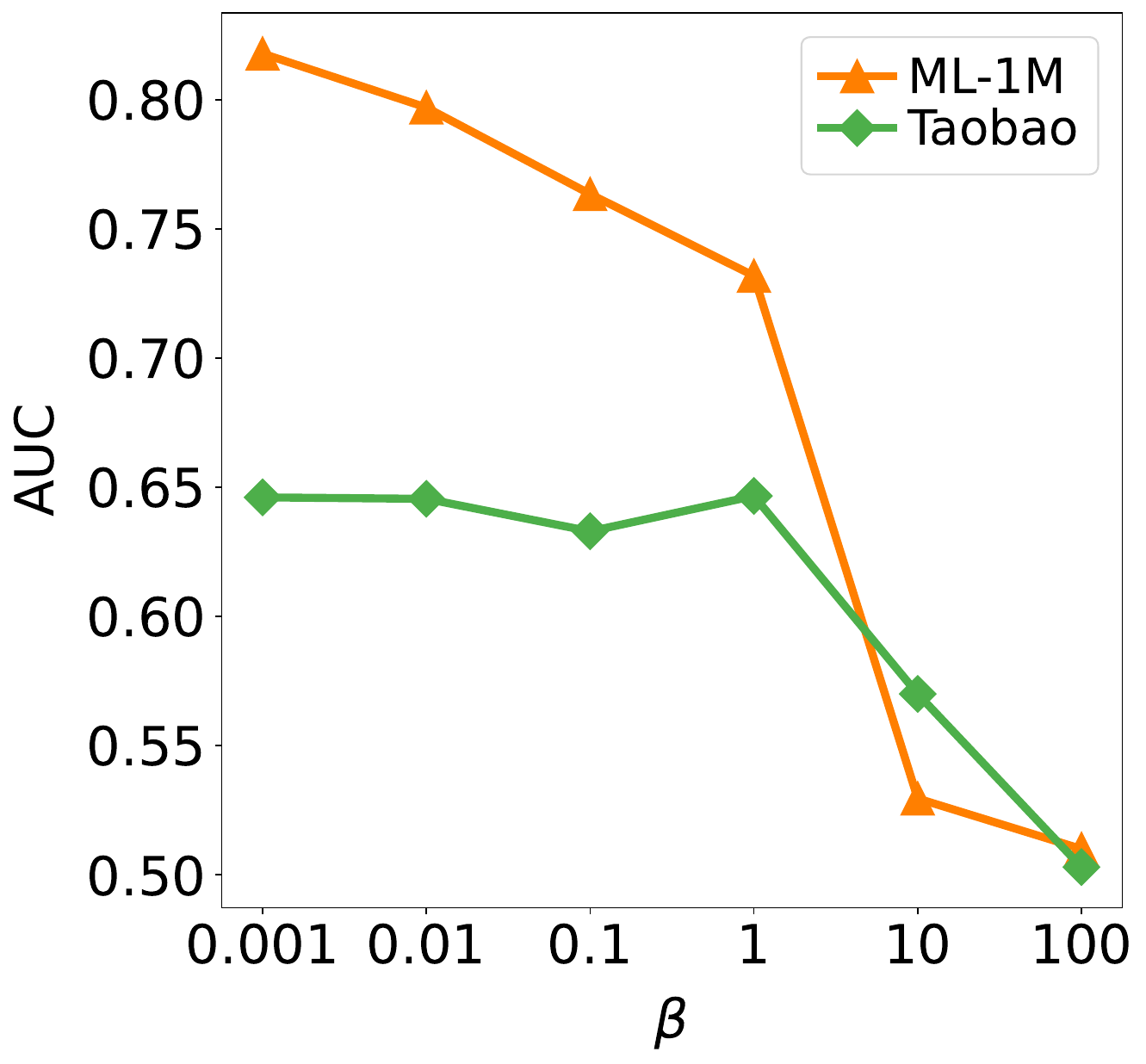}
    \end{minipage}%
  }%
  \subfigure[$\lambda$ on ML-1M]{
    \begin{minipage}[t]{0.33\linewidth}
      \centering
      \label{Fig:lambda_ml1m}
      \includegraphics[width=\linewidth]{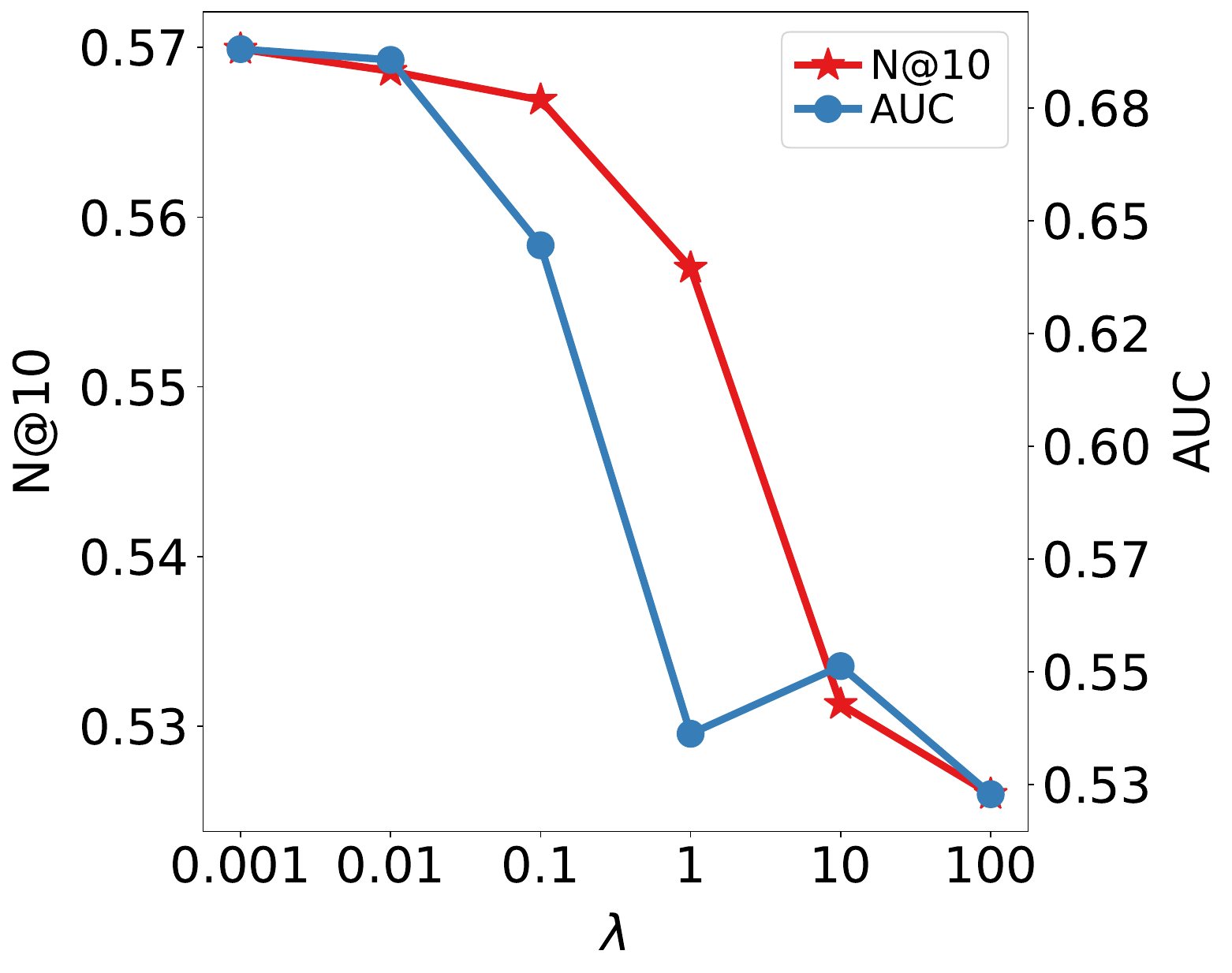}
    \end{minipage}%
  }%
  \subfigure [$\lambda$ on Taobao]{
    \begin{minipage}[t]{0.338\linewidth}
      \centering
      \label{Fig:lambda_ali}
      \includegraphics[width=\linewidth]{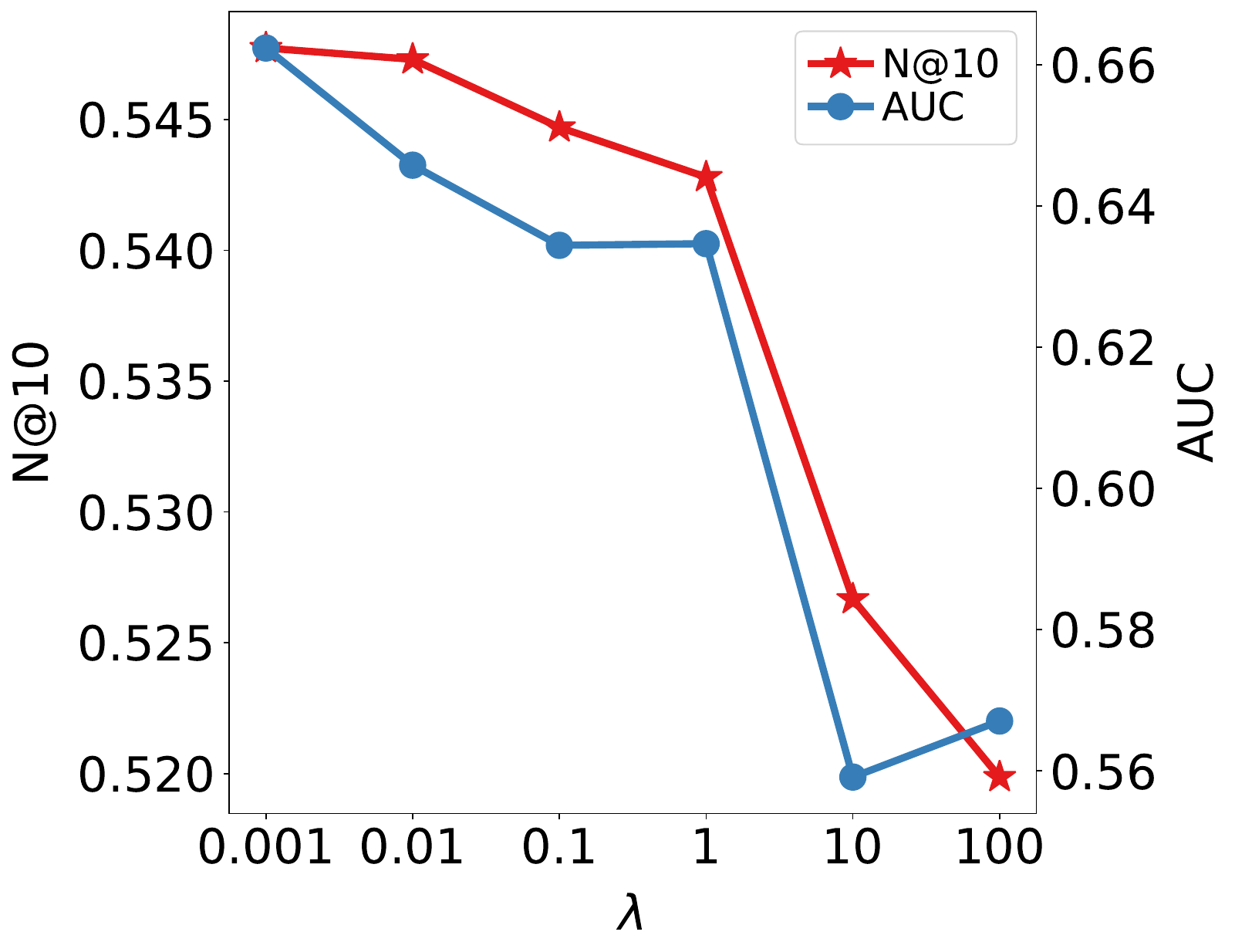}
    \end{minipage}%
  }%
  \centering
  \caption{Tuning of hyper-parameters.}
  \label{Fig:tuning}
\end{figure}

At first, from Figures \ref{Fig:pareto} and \ref{Fig:pareto_bert} we can see that basically at the same $\lambda$, AFRL and AFRL-w/o-$\mathbf{z}_{u, i}$ can achieve almost the same fairness, but AFRL has higher accuracy. 
This indicates that due to the information alignment, the fair embedding learned by AFRL through $\mathbf{z}_{u, i}$ preserves more discriminability compared to the fair embedding learned by AFRL-w/o-$\mathbf{z}_{u, i}$, without sacrificing fairness. At the same time, Fig. \ref{Fig:ablation} shows that AFRL achieves the higher accuracy than both AFRL-w/o-$\mathbf{z}_{u, 0}$ and AFRL-w/o-$\mathbf{z}_{u, i}$, which indicates both $\mathbf{z}_{u, 0}$ and $\mathbf{z}_{u, i}$ contribute to the accuracy because of the discriminative information encoded by them. However, we also note that the accuracy of AFRL-w/o-$\mathbf{z}_{u, i}$ is higher than that of AFRL-w/o-$\mathbf{z}_{u, 0}$, which suggests the debiased collaborative signals encoded by $\mathbf{z}_{u, 0}$ contribute more to the discriminability of the fair embedding.

\subsection{Hyper-parameter Tuning (RQ4)}


To determine the appropriate value of $\beta$, we train a classifier for each attribute $A_i$ using the attribute embeddings $Z_i$ and utilize AUC to evaluate the amount of information of $A_i$ encoded by $Z_i$. Fig. \ref{Fig:beta} shows the average AUC curves over all surrogate classifiers on both datasets. We can see that basically the smaller $\beta$, the higher AUC, and the more information of $A_i$ encoded, which is consistent with our analysis of Fig. \ref{Fig:ialignm} in Section \ref{Sec:AE}. We also note that after $\beta$ reaches 1, AUC begins to drop sharply. Since our objective is information alignment, i.e., making the information encoded by $Z_i$ exactly cover that of attribute $A_i$ rather than maximize or minimize it, we heuristically choose $\beta = 1$ to approximate the ideal case shown in Fig. \ref{Fig:idea_beta}. 

As $\lambda$ controls the trade-off between accuracy and fairness, we need to consider both accuracy and fairness to determine its optimal value. As mentioned in Section \ref{sec:protocol}, we train a set of surrogate classifiers employing the final fair user embedding $\mathbf{u}^*$ as input to predict the sensitive attribute value, with AUC as the metric for fairness performance. Accuracy is evaluated by N@$10$. Fig. \ref{Fig:lambda_ml1m} and Fig. \ref{Fig:lambda_ali} show the impact of $\lambda$ on recommendation accuracy and fairness on ML-1M and Taobao, respectively. We see that the larger $\lambda$, the more the fairness (the smaller the AUC), and the less the accuracy (the smaller the NDCG). We choose $\lambda = 1$ for the ML-1M dataset and $\lambda=10$ for Taobao, because they are a balance point beyond which the accuracy will drastically drop but without significant rise of the fairness.

 \section{related work}
The fairness challenge in recommender systems is complex due to the involvement of multiple stakeholders, generally categorized into item-side fairness and user-side fairness. Item-side fairness seeks that different types of items have an equitable opportunity to be recommended, by minimizing prediction errors for various item categories \cite{rastegarpanah2019fighting} or allocating exposure based on item relevance \cite{biega2018equity}. User-side fairness aims to make fair recommendations to different users \cite{ekstrand2018all, andrus2022demographic, ge2022explainable}, or ensure that similar users receive similar treatments \cite{fu2020fairness, li2023accurate, hua2023up5}.


 A range of approaches have been proposed for user-side fairness, following various technical lines including Pareto optimization \cite{lin2019pareto,yao2017beyond,zhu2018fairness,beutel2019fairness,burke2018balanced}, 
adversarial training \cite{bose2019compositional,wu2021fairness,li2021user,li2021leave}, and disentangled representation learning \cite{ zhao2023fair, park2021learning, gong2020jointly, sarhan2020fairness}. The common idea of them is to filter out the information of user sensitive attributes from the learned fair embeddings, so that recommendations can be made without bias to sensitive attributes. At the same time, some works apply reinforcement learning to fair recommendations, which usually impose fairness constraints to the cumulative return function for long-term and dynamic fairness \cite{ge2021towards, liu2020balancing}. 

The traditional methods often assume all users share the same sensitive attributes, which is not true in real world where different users have personalized fairness requirements. To address this issue, recently, some models have been proposed for personalized fairness in recommendations. For example, Li \textit{et al}. \cite{li2021towards} propose to train a filter for each possible combination of sensitive attributes, while Wu \textit{et al}. \cite{wu2022selective} propose a model PFRec to build a set of prompt-based bias eliminators and adapters with customized attribute-specific prompts to learn fair embeddings for different attribute combinations. However, they  treat a fairness requirement as a hyper-parameter, and train a mode for each combination of attributes, which makes them implement the personalized fairness in a brute force way with unacceptable training cost. Creager \textit{et al}. \cite{creager2019flexibly} propose the FFVAE model, which disentangles an unfair embedding into latent factors, each of which corresponds to an attribute, and excludes relevant sensitive latent factors based on different fairness requirements of users. 
Nevertheless, the existing methods often pursue an extreme fairness by completely removing the information of sensitive attributes from the fair embeddings, which makes them suffer from a suboptimal trade-off between accuracy and fairness. 



\section{Conclusion}
In this paper, we propose a novel model, called Adaptive Fair Representation Learning (AFRL), for personalized fairness in recommendations. AFRL can overcome the challenge of the explosion of attribute combinations and achieve a better trade-off between accuracy and fairness. At first, AFRL treats fairness requirements as inputs and can learn an attribute-specific embedding for each attribute from the unfair user embedding, which endows AFRL with the adaptability during inference phase to dynamically determine the non-sensitive attributes that should be encoded into the fair embedding under the guidance of the user's unique fairness requirement. With the novel information alignment offered by the proposed Information Alignment Module (IAlignM), AFRL can reduce the loss of recommendation accuracy, without loss of fairness, by exactly preserving discriminative information of non-sensitive attributes and incorporating a debiased collaborative embedding into the fair embedding to capture attribute-independent collaborative signals. At last, the extensive experiments conducted on real datasets together with the sound theoretical analysis demonstrate the superiority of AFRL. 

\begin{acks}
This work is supported by National Natural Science Foundation of China under grant 61972270.
\end{acks}


\newpage
\bibliographystyle{ACM-Reference-Format}
\balance
\bibliography{AFRL.bib}


\begin{thebibliography}{42}


\ifx \showCODEN    \undefined \def \showCODEN     #1{\unskip}     \fi
\ifx \showDOI      \undefined \def \showDOI       #1{#1}\fi
\ifx \showISBNx    \undefined \def \showISBNx     #1{\unskip}     \fi
\ifx \showISBNxiii \undefined \def \showISBNxiii  #1{\unskip}     \fi
\ifx \showISSN     \undefined \def \showISSN      #1{\unskip}     \fi
\ifx \showLCCN     \undefined \def \showLCCN      #1{\unskip}     \fi
\ifx \shownote     \undefined \def \shownote      #1{#1}          \fi
\ifx \showarticletitle \undefined \def \showarticletitle #1{#1}   \fi
\ifx \showURL      \undefined \def \showURL       {\relax}        \fi
\providecommand\bibfield[2]{#2}
\providecommand\bibinfo[2]{#2}
\providecommand\natexlab[1]{#1}
\providecommand\showeprint[2][]{arXiv:#2}

\bibitem[Andrus and Villeneuve(2022)]%
        {andrus2022demographic}
\bibfield{author}{\bibinfo{person}{McKane Andrus} {and} \bibinfo{person}{Sarah
  Villeneuve}.} \bibinfo{year}{2022}\natexlab{}.
\newblock \showarticletitle{Demographic-reliant algorithmic fairness:
  Characterizing the risks of demographic data collection in the pursuit of
  fairness}. In \bibinfo{booktitle}{\emph{FAT$^*$}}.
\newblock


\bibitem[Beutel et~al\mbox{.}(2019)]%
        {beutel2019fairness}
\bibfield{author}{\bibinfo{person}{Alex Beutel}, \bibinfo{person}{Jilin Chen},
  \bibinfo{person}{Tulsee Doshi}, \bibinfo{person}{Hai Qian},
  \bibinfo{person}{Li Wei}, \bibinfo{person}{Yi Wu}, \bibinfo{person}{Lukasz
  Heldt}, \bibinfo{person}{Zhe Zhao}, \bibinfo{person}{Lichan Hong},
  \bibinfo{person}{Ed~H Chi}, {et~al\mbox{.}}} \bibinfo{year}{2019}\natexlab{}.
\newblock \showarticletitle{Fairness in recommendation ranking through pairwise
  comparisons}. In \bibinfo{booktitle}{\emph{SIGKDD}}.
\newblock


\bibitem[Biega et~al\mbox{.}(2018)]%
        {biega2018equity}
\bibfield{author}{\bibinfo{person}{Asia~J Biega}, \bibinfo{person}{Krishna~P
  Gummadi}, {and} \bibinfo{person}{Gerhard Weikum}.}
  \bibinfo{year}{2018}\natexlab{}.
\newblock \showarticletitle{Equity of attention: Amortizing individual fairness
  in rankings}. In \bibinfo{booktitle}{\emph{SIGIR}}.
\newblock


\bibitem[Bose and Hamilton(2019)]%
        {bose2019compositional}
\bibfield{author}{\bibinfo{person}{Avishek Bose} {and} \bibinfo{person}{William
  Hamilton}.} \bibinfo{year}{2019}\natexlab{}.
\newblock \showarticletitle{Compositional fairness constraints for graph
  embeddings}. In \bibinfo{booktitle}{\emph{ICML}}.
\newblock


\bibitem[Burke et~al\mbox{.}(2018)]%
        {burke2018balanced}
\bibfield{author}{\bibinfo{person}{Robin Burke}, \bibinfo{person}{Nasim
  Sonboli}, {and} \bibinfo{person}{Aldo Ordonez-Gauger}.}
  \bibinfo{year}{2018}\natexlab{}.
\newblock \showarticletitle{Balanced neighborhoods for multi-sided fairness in
  recommendation}. In \bibinfo{booktitle}{\emph{FAT$^*$}}.
\newblock


\bibitem[Cheng et~al\mbox{.}(2016)]%
        {cheng2016wide}
\bibfield{author}{\bibinfo{person}{Heng-Tze Cheng}, \bibinfo{person}{Levent
  Koc}, \bibinfo{person}{Jeremiah Harmsen}, \bibinfo{person}{Tal Shaked},
  \bibinfo{person}{Tushar Chandra}, \bibinfo{person}{Hrishi Aradhye},
  \bibinfo{person}{Glen Anderson}, \bibinfo{person}{Greg Corrado},
  \bibinfo{person}{Wei Chai}, \bibinfo{person}{Mustafa Ispir}, {et~al\mbox{.}}}
  \bibinfo{year}{2016}\natexlab{}.
\newblock \showarticletitle{Wide \& deep learning for recommender systems}. In
  \bibinfo{booktitle}{\emph{DLRS}}.
\newblock


\bibitem[Creager et~al\mbox{.}(2019)]%
        {creager2019flexibly}
\bibfield{author}{\bibinfo{person}{Elliot Creager}, \bibinfo{person}{David
  Madras}, \bibinfo{person}{J{\"o}rn-Henrik Jacobsen}, \bibinfo{person}{Marissa
  Weis}, \bibinfo{person}{Kevin Swersky}, \bibinfo{person}{Toniann Pitassi},
  {and} \bibinfo{person}{Richard Zemel}.} \bibinfo{year}{2019}\natexlab{}.
\newblock \showarticletitle{Flexibly fair representation learning by
  disentanglement}. In \bibinfo{booktitle}{\emph{International conference on
  machine learning}}. PMLR.
\newblock


\bibitem[Dwork et~al\mbox{.}(2012)]%
        {dwork2012fairness}
\bibfield{author}{\bibinfo{person}{Cynthia Dwork}, \bibinfo{person}{Moritz
  Hardt}, \bibinfo{person}{Toniann Pitassi}, \bibinfo{person}{Omer Reingold},
  {and} \bibinfo{person}{Richard Zemel}.} \bibinfo{year}{2012}\natexlab{}.
\newblock \showarticletitle{Fairness through awareness}. In
  \bibinfo{booktitle}{\emph{ITCS}}.
\newblock


\bibitem[Ekstrand et~al\mbox{.}(2018)]%
        {ekstrand2018all}
\bibfield{author}{\bibinfo{person}{Michael~D Ekstrand}, \bibinfo{person}{Mucun
  Tian}, \bibinfo{person}{Ion~Madrazo Azpiazu}, \bibinfo{person}{Jennifer~D
  Ekstrand}, \bibinfo{person}{Oghenemaro Anuyah}, \bibinfo{person}{David
  McNeill}, {and} \bibinfo{person}{Maria~Soledad Pera}.}
  \bibinfo{year}{2018}\natexlab{}.
\newblock \showarticletitle{All the cool kids, how do they fit in?: Popularity
  and demographic biases in recommender evaluation and effectiveness}. In
  \bibinfo{booktitle}{\emph{FAT$^*$}}.
\newblock


\bibitem[Fu et~al\mbox{.}(2020)]%
        {fu2020fairness}
\bibfield{author}{\bibinfo{person}{Zuohui Fu}, \bibinfo{person}{Yikun Xian},
  \bibinfo{person}{Ruoyuan Gao}, \bibinfo{person}{Jieyu Zhao},
  \bibinfo{person}{Qiaoying Huang}, \bibinfo{person}{Yingqiang Ge},
  \bibinfo{person}{Shuyuan Xu}, \bibinfo{person}{Shijie Geng},
  \bibinfo{person}{Chirag Shah}, \bibinfo{person}{Yongfeng Zhang},
  {et~al\mbox{.}}} \bibinfo{year}{2020}\natexlab{}.
\newblock \showarticletitle{Fairness-aware explainable recommendation over
  knowledge graphs}. In \bibinfo{booktitle}{\emph{SIGIR}}.
\newblock


\bibitem[Ge et~al\mbox{.}(2021)]%
        {ge2021towards}
\bibfield{author}{\bibinfo{person}{Yingqiang Ge}, \bibinfo{person}{Shuchang
  Liu}, \bibinfo{person}{Ruoyuan Gao}, \bibinfo{person}{Yikun Xian},
  \bibinfo{person}{Yunqi Li}, \bibinfo{person}{Xiangyu Zhao},
  \bibinfo{person}{Changhua Pei}, \bibinfo{person}{Fei Sun},
  \bibinfo{person}{Junfeng Ge}, \bibinfo{person}{Wenwu Ou}, {et~al\mbox{.}}}
  \bibinfo{year}{2021}\natexlab{}.
\newblock \showarticletitle{Towards long-term fairness in recommendation}. In
  \bibinfo{booktitle}{\emph{WSDM}}.
\newblock


\bibitem[Ge et~al\mbox{.}(2022)]%
        {ge2022explainable}
\bibfield{author}{\bibinfo{person}{Yingqiang Ge}, \bibinfo{person}{Juntao Tan},
  \bibinfo{person}{Yan Zhu}, \bibinfo{person}{Yinglong Xia},
  \bibinfo{person}{Jiebo Luo}, \bibinfo{person}{Shuchang Liu},
  \bibinfo{person}{Zuohui Fu}, \bibinfo{person}{Shijie Geng},
  \bibinfo{person}{Zelong Li}, {and} \bibinfo{person}{Yongfeng Zhang}.}
  \bibinfo{year}{2022}\natexlab{}.
\newblock \showarticletitle{Explainable fairness in recommendation}. In
  \bibinfo{booktitle}{\emph{SIGIR}}.
\newblock


\bibitem[Gong et~al\mbox{.}(2020)]%
        {gong2020jointly}
\bibfield{author}{\bibinfo{person}{Sixue Gong}, \bibinfo{person}{Xiaoming Liu},
  {and} \bibinfo{person}{Anil~K Jain}.} \bibinfo{year}{2020}\natexlab{}.
\newblock \showarticletitle{Jointly de-biasing face recognition and demographic
  attribute estimation}. In \bibinfo{booktitle}{\emph{ECCV}}.
\newblock


\bibitem[Gupta et~al\mbox{.}(2021)]%
        {gupta2021controllable}
\bibfield{author}{\bibinfo{person}{Umang Gupta}, \bibinfo{person}{Aaron~M
  Ferber}, \bibinfo{person}{Bistra Dilkina}, {and} \bibinfo{person}{Greg
  Ver~Steeg}.} \bibinfo{year}{2021}\natexlab{}.
\newblock \showarticletitle{Controllable guarantees for fair outcomes via
  contrastive information estimation}. In \bibinfo{booktitle}{\emph{AAAI}}.
\newblock


\bibitem[He et~al\mbox{.}(2017)]%
        {he2017neural}
\bibfield{author}{\bibinfo{person}{Xiangnan He}, \bibinfo{person}{Lizi Liao},
  \bibinfo{person}{Hanwang Zhang}, \bibinfo{person}{Liqiang Nie},
  \bibinfo{person}{Xia Hu}, {and} \bibinfo{person}{Tat-Seng Chua}.}
  \bibinfo{year}{2017}\natexlab{}.
\newblock \showarticletitle{Neural collaborative filtering}. In
  \bibinfo{booktitle}{\emph{WWW}}.
\newblock


\bibitem[Hua et~al\mbox{.}(2023)]%
        {hua2023up5}
\bibfield{author}{\bibinfo{person}{Wenyue Hua}, \bibinfo{person}{Yingqiang Ge},
  \bibinfo{person}{Shuyuan Xu}, \bibinfo{person}{Jianchao Ji}, {and}
  \bibinfo{person}{Yongfeng Zhang}.} \bibinfo{year}{2023}\natexlab{}.
\newblock \showarticletitle{UP5: Unbiased Foundation Model for Fairness-aware
  Recommendation}.
\newblock \bibinfo{journal}{\emph{arXiv preprint arXiv:2305.12090}}
  (\bibinfo{year}{2023}).
\newblock


\bibitem[Kang and McAuley(2018)]%
        {kang2018self}
\bibfield{author}{\bibinfo{person}{Wang-Cheng Kang} {and}
  \bibinfo{person}{Julian McAuley}.} \bibinfo{year}{2018}\natexlab{}.
\newblock \showarticletitle{Self-attentive sequential recommendation}. In
  \bibinfo{booktitle}{\emph{ICDM}}.
\newblock


\bibitem[Kearns et~al\mbox{.}(2019)]%
        {kearns2019empirical}
\bibfield{author}{\bibinfo{person}{Michael Kearns}, \bibinfo{person}{Seth
  Neel}, \bibinfo{person}{Aaron Roth}, {and} \bibinfo{person}{Zhiwei~Steven
  Wu}.} \bibinfo{year}{2019}\natexlab{}.
\newblock \showarticletitle{An empirical study of rich subgroup fairness for
  machine learning}. In \bibinfo{booktitle}{\emph{FAT$^*$}}.
\newblock


\bibitem[Kleinberg et~al\mbox{.}(2016)]%
        {kleinberg2016inherent}
\bibfield{author}{\bibinfo{person}{Jon Kleinberg}, \bibinfo{person}{Sendhil
  Mullainathan}, {and} \bibinfo{person}{Manish Raghavan}.}
  \bibinfo{year}{2016}\natexlab{}.
\newblock \showarticletitle{Inherent trade-offs in the fair determination of
  risk scores}.
\newblock \bibinfo{journal}{\emph{arXiv preprint arXiv:1609.05807}}
  (\bibinfo{year}{2016}).
\newblock


\bibitem[Kusner et~al\mbox{.}(2017)]%
        {kusner2017counterfactual}
\bibfield{author}{\bibinfo{person}{Matt~J Kusner}, \bibinfo{person}{Joshua
  Loftus}, \bibinfo{person}{Chris Russell}, {and} \bibinfo{person}{Ricardo
  Silva}.} \bibinfo{year}{2017}\natexlab{}.
\newblock \showarticletitle{Counterfactual fairness}.
\newblock \bibinfo{journal}{\emph{Advances in neural information processing
  systems}}  \bibinfo{volume}{30} (\bibinfo{year}{2017}).
\newblock


\bibitem[Li et~al\mbox{.}(2021c)]%
        {li2021leave}
\bibfield{author}{\bibinfo{person}{Roger~Zhe Li}, \bibinfo{person}{Juli{\'a}n
  Urbano}, {and} \bibinfo{person}{Alan Hanjalic}.}
  \bibinfo{year}{2021}\natexlab{c}.
\newblock \showarticletitle{Leave no user behind: Towards improving the utility
  of recommender systems for non-mainstream users}. In
  \bibinfo{booktitle}{\emph{WSDM}}.
\newblock


\bibitem[Li et~al\mbox{.}(2023)]%
        {li2023accurate}
\bibfield{author}{\bibinfo{person}{Xuran Li}, \bibinfo{person}{Peng Wu}, {and}
  \bibinfo{person}{Jing Su}.} \bibinfo{year}{2023}\natexlab{}.
\newblock \showarticletitle{Accurate fairness: Improving individual fairness
  without trading accuracy}. In \bibinfo{booktitle}{\emph{AAAI}}.
\newblock


\bibitem[Li et~al\mbox{.}(2021a)]%
        {li2021user}
\bibfield{author}{\bibinfo{person}{Yunqi Li}, \bibinfo{person}{Hanxiong Chen},
  \bibinfo{person}{Zuohui Fu}, \bibinfo{person}{Yingqiang Ge}, {and}
  \bibinfo{person}{Yongfeng Zhang}.} \bibinfo{year}{2021}\natexlab{a}.
\newblock \showarticletitle{User-oriented fairness in recommendation}. In
  \bibinfo{booktitle}{\emph{WWW}}.
\newblock


\bibitem[Li et~al\mbox{.}(2021b)]%
        {li2021towards}
\bibfield{author}{\bibinfo{person}{Yunqi Li}, \bibinfo{person}{Hanxiong Chen},
  \bibinfo{person}{Shuyuan Xu}, \bibinfo{person}{Yingqiang Ge}, {and}
  \bibinfo{person}{Yongfeng Zhang}.} \bibinfo{year}{2021}\natexlab{b}.
\newblock \showarticletitle{Towards personalized fairness based on causal
  notion}. In \bibinfo{booktitle}{\emph{SIGIR}}.
\newblock


\bibitem[Liang et~al\mbox{.}(2018)]%
        {liang2018variational}
\bibfield{author}{\bibinfo{person}{Dawen Liang}, \bibinfo{person}{Rahul~G
  Krishnan}, \bibinfo{person}{Matthew~D Hoffman}, {and} \bibinfo{person}{Tony
  Jebara}.} \bibinfo{year}{2018}\natexlab{}.
\newblock \showarticletitle{Variational autoencoders for collaborative
  filtering}. In \bibinfo{booktitle}{\emph{WWW}}.
\newblock


\bibitem[Lin et~al\mbox{.}(2019)]%
        {lin2019pareto}
\bibfield{author}{\bibinfo{person}{Xiao Lin}, \bibinfo{person}{Hongjie Chen},
  \bibinfo{person}{Changhua Pei}, \bibinfo{person}{Fei Sun},
  \bibinfo{person}{Xuanji Xiao}, \bibinfo{person}{Hanxiao Sun},
  \bibinfo{person}{Yongfeng Zhang}, \bibinfo{person}{Wenwu Ou}, {and}
  \bibinfo{person}{Peng Jiang}.} \bibinfo{year}{2019}\natexlab{}.
\newblock \showarticletitle{A pareto-efficient algorithm for multiple objective
  optimization in e-commerce recommendation}. In
  \bibinfo{booktitle}{\emph{Recsys}}.
\newblock


\bibitem[Liu et~al\mbox{.}(2021)]%
        {liu2021mitigating}
\bibfield{author}{\bibinfo{person}{Dugang Liu}, \bibinfo{person}{Pengxiang
  Cheng}, \bibinfo{person}{Hong Zhu}, \bibinfo{person}{Zhenhua Dong},
  \bibinfo{person}{Xiuqiang He}, \bibinfo{person}{Weike Pan}, {and}
  \bibinfo{person}{Zhong Ming}.} \bibinfo{year}{2021}\natexlab{}.
\newblock \showarticletitle{Mitigating confounding bias in recommendation via
  information bottleneck}. In \bibinfo{booktitle}{\emph{Recsys}}.
\newblock


\bibitem[Liu et~al\mbox{.}(2020)]%
        {liu2020balancing}
\bibfield{author}{\bibinfo{person}{Weiwen Liu}, \bibinfo{person}{Feng Liu},
  \bibinfo{person}{Ruiming Tang}, \bibinfo{person}{Ben Liao},
  \bibinfo{person}{Guangyong Chen}, {and} \bibinfo{person}{Pheng~Ann Heng}.}
  \bibinfo{year}{2020}\natexlab{}.
\newblock \showarticletitle{Balancing between accuracy and fairness for
  interactive recommendation with reinforcement learning}. In
  \bibinfo{booktitle}{\emph{PAKDD}}.
\newblock


\bibitem[Mnih and Salakhutdinov(2007)]%
        {mnih2007probabilistic}
\bibfield{author}{\bibinfo{person}{Andriy Mnih} {and} \bibinfo{person}{Russ~R
  Salakhutdinov}.} \bibinfo{year}{2007}\natexlab{}.
\newblock \showarticletitle{Probabilistic matrix factorization}. In
  \bibinfo{booktitle}{\emph{NeurIPS}}.
\newblock


\bibitem[Park et~al\mbox{.}(2021)]%
        {park2021learning}
\bibfield{author}{\bibinfo{person}{Sungho Park}, \bibinfo{person}{Sunhee
  Hwang}, \bibinfo{person}{Dohyung Kim}, {and} \bibinfo{person}{Hyeran Byun}.}
  \bibinfo{year}{2021}\natexlab{}.
\newblock \showarticletitle{Learning disentangled representation for fair
  facial attribute classification via fairness-aware information alignment}. In
  \bibinfo{booktitle}{\emph{AAAI}}.
\newblock


\bibitem[Pessach and Shmueli(2022)]%
        {pessach2022review}
\bibfield{author}{\bibinfo{person}{Dana Pessach} {and} \bibinfo{person}{Erez
  Shmueli}.} \bibinfo{year}{2022}\natexlab{}.
\newblock \showarticletitle{A review on fairness in machine learning}.
\newblock \bibinfo{journal}{\emph{ACM Computing Surveys (CSUR)}}
  \bibinfo{volume}{55}, \bibinfo{number}{3} (\bibinfo{year}{2022}),
  \bibinfo{pages}{1--44}.
\newblock


\bibitem[Rastegarpanah et~al\mbox{.}(2019)]%
        {rastegarpanah2019fighting}
\bibfield{author}{\bibinfo{person}{Bashir Rastegarpanah},
  \bibinfo{person}{Krishna~P Gummadi}, {and} \bibinfo{person}{Mark Crovella}.}
  \bibinfo{year}{2019}\natexlab{}.
\newblock \showarticletitle{Fighting fire with fire: Using antidote data to
  improve polarization and fairness of recommender systems}. In
  \bibinfo{booktitle}{\emph{WSDM}}.
\newblock


\bibitem[Sarhan et~al\mbox{.}(2020)]%
        {sarhan2020fairness}
\bibfield{author}{\bibinfo{person}{Mhd~Hasan Sarhan}, \bibinfo{person}{Nassir
  Navab}, \bibinfo{person}{Abouzar Eslami}, {and} \bibinfo{person}{Shadi
  Albarqouni}.} \bibinfo{year}{2020}\natexlab{}.
\newblock \showarticletitle{Fairness by learning orthogonal disentangled
  representations}. In \bibinfo{booktitle}{\emph{Computer Vision--ECCV 2020:
  16th European Conference, Glasgow, UK, August 23--28, 2020, Proceedings, Part
  XXIX 16}}. Springer.
\newblock


\bibitem[Shwartz-Ziv and Tishby(2017)]%
        {shwartz2017opening}
\bibfield{author}{\bibinfo{person}{Ravid Shwartz-Ziv} {and}
  \bibinfo{person}{Naftali Tishby}.} \bibinfo{year}{2017}\natexlab{}.
\newblock \showarticletitle{Opening the black box of deep neural networks via
  information}.
\newblock \bibinfo{journal}{\emph{arXiv preprint arXiv:1703.00810}}
  (\bibinfo{year}{2017}).
\newblock


\bibitem[Sun et~al\mbox{.}(2019)]%
        {sun2019bert4rec}
\bibfield{author}{\bibinfo{person}{Fei Sun}, \bibinfo{person}{Jun Liu},
  \bibinfo{person}{Jian Wu}, \bibinfo{person}{Changhua Pei},
  \bibinfo{person}{Xiao Lin}, \bibinfo{person}{Wenwu Ou}, {and}
  \bibinfo{person}{Peng Jiang}.} \bibinfo{year}{2019}\natexlab{}.
\newblock \showarticletitle{BERT4Rec: Sequential recommendation with
  bidirectional encoder representations from transformer}. In
  \bibinfo{booktitle}{\emph{CIKM}}.
\newblock


\bibitem[Wu et~al\mbox{.}(2021)]%
        {wu2021fairness}
\bibfield{author}{\bibinfo{person}{Chuhan Wu}, \bibinfo{person}{Fangzhao Wu},
  \bibinfo{person}{Xiting Wang}, \bibinfo{person}{Yongfeng Huang}, {and}
  \bibinfo{person}{Xing Xie}.} \bibinfo{year}{2021}\natexlab{}.
\newblock \showarticletitle{Fairness-aware news recommendation with decomposed
  adversarial learning}. In \bibinfo{booktitle}{\emph{AAAI}}.
\newblock


\bibitem[Wu et~al\mbox{.}(2022)]%
        {wu2022selective}
\bibfield{author}{\bibinfo{person}{Yiqing Wu}, \bibinfo{person}{Ruobing Xie},
  \bibinfo{person}{Yongchun Zhu}, \bibinfo{person}{Fuzhen Zhuang},
  \bibinfo{person}{Ao Xiang}, \bibinfo{person}{Xu Zhang}, \bibinfo{person}{Leyu
  Lin}, {and} \bibinfo{person}{Qing He}.} \bibinfo{year}{2022}\natexlab{}.
\newblock \showarticletitle{Selective fairness in recommendation via prompts}.
  In \bibinfo{booktitle}{\emph{SIGIR}}.
\newblock


\bibitem[Xiao et~al\mbox{.}(2017)]%
        {xiao2017fairness}
\bibfield{author}{\bibinfo{person}{Lin Xiao}, \bibinfo{person}{Zhang Min},
  \bibinfo{person}{Zhang Yongfeng}, \bibinfo{person}{Gu Zhaoquan},
  \bibinfo{person}{Liu Yiqun}, {and} \bibinfo{person}{Ma Shaoping}.}
  \bibinfo{year}{2017}\natexlab{}.
\newblock \showarticletitle{Fairness-aware group recommendation with
  pareto-efficiency}. In \bibinfo{booktitle}{\emph{Recsys}}.
\newblock


\bibitem[Yao and Huang(2017)]%
        {yao2017beyond}
\bibfield{author}{\bibinfo{person}{Sirui Yao} {and} \bibinfo{person}{Bert
  Huang}.} \bibinfo{year}{2017}\natexlab{}.
\newblock \showarticletitle{Beyond parity: Fairness objectives for
  collaborative filtering}. In \bibinfo{booktitle}{\emph{NeurIPS}}.
\newblock


\bibitem[Zhao et~al\mbox{.}(2023)]%
        {zhao2023fair}
\bibfield{author}{\bibinfo{person}{Chen Zhao}, \bibinfo{person}{Le Wu},
  \bibinfo{person}{Pengyang Shao}, \bibinfo{person}{Kun Zhang},
  \bibinfo{person}{Richang Hong}, {and} \bibinfo{person}{Meng Wang}.}
  \bibinfo{year}{2023}\natexlab{}.
\newblock \showarticletitle{Fair representation learning for recommendation: a
  mutual information perspective}. In \bibinfo{booktitle}{\emph{AAAI}}.
\newblock


\bibitem[Zhu et~al\mbox{.}(2018)]%
        {zhu2018fairness}
\bibfield{author}{\bibinfo{person}{Ziwei Zhu}, \bibinfo{person}{Xia Hu}, {and}
  \bibinfo{person}{James Caverlee}.} \bibinfo{year}{2018}\natexlab{}.
\newblock \showarticletitle{Fairness-aware tensor-based recommendation}. In
  \bibinfo{booktitle}{\emph{CIKM}}.
\newblock


\bibitem[Zhu et~al\mbox{.}(2021)]%
        {ZhuZiwei2021FaNI}
\bibfield{author}{\bibinfo{person}{Ziwei Zhu}, \bibinfo{person}{Jingu Kim},
  \bibinfo{person}{Trung Nguyen}, \bibinfo{person}{Aish Fenton}, {and}
  \bibinfo{person}{James Caverlee}.} \bibinfo{year}{2021}\natexlab{}.
\newblock \showarticletitle{Fairness among New Items in Cold Start Recommender
  Systems}. In \bibinfo{booktitle}{\emph{SIGIR}}.
\newblock
\showISBNx{1450380379}


\end{thebibliography}

\end{document}